\documentclass[a4paper,11pt]{article}
\usepackage{jheppub} 

\usepackage{lineno}
\usepackage{braket}
\usepackage{booktabs}
\usepackage{xcolor}

\usepackage{booktabs}
\usepackage{tabularx}
\usepackage{array}
\newcolumntype{Y}{>{\raggedright\arraybackslash}X}

\arxivnumber{} 

\title{\boldmath Detecting Topological Transitions and Anisotropy through Multipartite Entanglement in Holographic Weyl Semimetals}

\author[a,b]{Xiantong Chen,}
\author[a,b]{Xuanting Ji,}
\author[a]{Wen-Peng Li,}
\author[b]{Ya-Wen Sun}
\affiliation[a]{Department of Applied Physics, College of Science, \\
    China Agricultural University, Beijing 100083, China    
}
\affiliation[b]{
    School of Physical Sciences,\\ 
    University of Chinese Academy of Sciences, Beijing 100049, China
}

\emailAdd{{chenxiantong23@mails.ucas.ac.cn}, {jixuanting@cau.edu.cn},{liwenpengsci@cau.edu.cn}, {yawen.sun@ucas.ac.cn}}

\abstract{
We study multipartite entanglement structures in the zero-temperature holographic Weyl semimetal, focusing on tripartite and four-partite structures. For strip regions, we compute the conditional mutual information, the entanglement wedge cross section, tripartite measures \(\kappa\) and the Markov gap, multi-EWCS, and two multi-EWCS based four-partite signals $\Delta$ and $g$. These quantities are studied as functions of the strip width \(l\) and the tuning parameter across the topological transition. At large \(l\), their $l$ dependence takes a power-law form governed by the IR scaling of the system. At fixed large \(l\), all these entanglement quantities develop clear features near the critical point, showing that tripartite and four-partite entanglement structures can diagnose the topological quantum phase transition. We further study strips pointing in different directions to probe the anisotropy of the system. The anisotropic large \(l\) behavior distinguishes the nontrivial phase from the trivial phase. These results establish multipartite holographic entanglement as a sensitive, nonlocal probe of topological phase transitions and anisotropic IR physics.}

\begin{document}
\raggedbottom
\setlength{\parskip}{0pt}
\setlength{\parindent}{2em}
\maketitle
\flushbottom

\section{Introduction}
Conventional phase transitions in Landau paradigm are often characterized by local order parameters associated with symmetry breaking. Topological phases are different: their distinction is tied instead to global properties of the quantum state, anomalous responses, topological invariants, and long-distance patterns of correlation~\cite{TopologicalPhaseMatterWitten:2015aoa,Wen:2012hm}. This makes gapless topological matter, such as Weyl semimetals, a natural setting in which to ask how phase structure is encoded when no local Landau order parameter is available. In a Weyl semimetal, the low-energy excitations are Weyl fermions, and the topologically nontrivial regime is characterized by separated Weyl nodes and anomalous transport, most notably the anomalous Hall conductivity~\cite{HolographicWeylSemimetalsLandsteiner:2015lsa,HolographicWeylSemimetalsLandsteiner_2016,HolographicWeylSemimetalsLandsteiner:2016stv,HolographicSemimetalsLandsteiner:2019kxb}. Across a topological transition, these nodal and anomalous structures are removed, while the critical point itself remains gapless. The central question is then how this change is reflected in real-space nonlocal quantum correlations.

At weak coupling, the topology of a Weyl semimetal can often be described in terms of quasiparticles, band crossings, Berry curvature, and topological charges associated with the Weyl nodes. At strong coupling, however, the quasiparticle picture and the band-theoretic description are no longer reliable. One would therefore like to identify nonlocal quantities that remain sensitive to the topological structure and to the way quantum correlations are reorganized across the transition. Entanglement provides a natural language for this purpose, since it captures information that is not visible in local observables alone and can distinguish different organizations of many-body correlations~\cite{QuantumInformationCondensedMatterZeng_2019,Wen:2012hm}.

Entanglement entropy can be computed by the replica trick, but this method is technically difficult, especially for interacting many-body systems. In strongly coupled systems with a holographic dual, the Ryu--Takayanagi prescription gives a geometric alternative: the entanglement entropy of a boundary region is proportional to the area of a bulk extremal surface homologous to that region~\cite{RyuTakayanagiFormulaRyu_2006,Ryu_2006}. This turns a difficult quantum many-body problem into a classical geometric problem in the bulk. For strongly coupled topological semimetals, holography is therefore especially useful: it allows one to compute nonlocal entanglement quantities directly from bulk extremal surfaces, entanglement wedges, and related geometric constructions~\cite{HolographicCMTZaanen:2015oix,HolographicSemimetalsLandsteiner:2019kxb}.

In this work we focus on the zero-temperature holographic Weyl semimetal. In this model the competition between the axial source $b$, which sets the Weyl node separation scale, and the scalar mass deformation $M$, which corresponds to the mass of the fermions, is controlled by the dimensionless ratio $M/b$. As $M/b$ is varied, the system passes from a topologically nontrivial phase to a critical solution and then to a topologically trivial phase. The axial source selects a spatial direction so that the topological transition is intrinsically anisotropic, and the direction of the entangling region becomes part of the physical information carried by the nonlocal quantity,
such as topological invariants \cite{HolographicTopologicalInvariantLiu_2018,Chen2025,AnotherChen2025} and entanglement quantities \cite{chen2026multipartiteentanglementcharacterizingtopological}.

Previous work has shown that holographic entanglement entropy and the entropic \(c\)-function can be used to characterize topological quantum phase transitions in holographic semimetals~\cite{Baggioli:2020cld,CFunctionOrderParameterBaggioli_2023}. However, entanglement entropy captures only part of the nonlocal structure of a quantum state. In a strongly coupled gapless system, the entanglement structure is expected to be richer than what is captured by the entanglement entropy alone. In particular, long-distance correlations may involve several spatial regions in an essentially multipartite way. This motivates us to study not only the entropy and the entropic \(c\)-function, but also multipartite and mixed-state entanglement quantities in the holographic Weyl semimetal.

A number of such quantities have been developed in holography in recent years. The tripartite information \(I_3\) and the conditional mutual information(CMI) are finite combinations of entanglement entropies for three regions and could provide information on the tripartite structure of the subsystems~\cite{Hayden_2013,ConditionalMutualInformationJu:2024kuc,ConditionalMutualInformationJu_2024,ConditionalMutualInformationJi_2025,ju2024holographicmultipartiteentanglementupper,Ju_2025}. The entanglement wedge cross section (EWCS) is the geometric quantity dual to the entanglement of purification or, equivalently, to one half of the reflected entropy in holographic states~\cite{Umemoto2018,ReflectedEntropydutta2019canonicalpurificationentanglementwedge}. Related EWCS-based combinations have also been used to define
triangle-information-like quantities and to connect tripartite wedge data with
holographic entanglement of assistance~\cite{ju2025entanglementwedgecrosssection}. The Markov gap, defined by \(S_R-I\), gives a sharper tripartite measure: for a tripartite pure state, its vanishing is tied to the triangle-state or sum-of-triangle-states structure~\cite{MarkovGapHayden_2021,SumOfTrangleStateZou2021}. The multi-entropy provides another multipartite quantity with a holographic network prescription~\cite{Gadde_2022,Gadde_2023,MultiEntropyHarper_2024,gadde2025multiinvariantsbulkreplicasymmetry,iizuka2026junctionlawmultipartiteentanglement,balasubramanian2026constraintsfourpartyentanglementholography}, and the combination \(\kappa\) constructed from it subtracts the reducible pairwise entropy contributions in order to detect the remaining genuine tripartite entanglement. Other recent proposals, such as the \(L\)-entropy, also aim to characterize genuine multipartite entanglement beyond entropy data~\cite{EWCSbasak2025newgenuinemultipartiteentanglement,ahn2026probinghierarchygenuinemultipartite}. Related work on genuine multi-entropy and holographic constraints on GHZ-like structures provides further context for multipartite entanglement in holography~\cite{GenuineMulipartiteiizuka2025genuinemultientropyholography,GenuineMulipartiteiizuka2025genuinemultientropy,Balasubramanian2026, jiang2025holographicdualghzstate,jiang2026divingbookletwormholes}. For higher partitions, the multipartite entanglement wedge cross section (multi-EWCS) gives a geometric multipartite wedge construction and can be used to build four-partite signals such as \(\Delta\) and \(g\)~\cite{Bao2019,AnotherBao2019,AAnotherBao2019,MultiEntanglementUpperBoundju2024holographicmultipartiteentanglementupper,MarkovGapju2025holographicmultipartiteentanglementstructures,bao2026tripartitecorrelationsignalmultipartite}.

These developments provide the set of tools used in this work. In this work, we focus on the holographic \(c\)-function, CMI, EWCS, \(\kappa\), the Markov gap, multi-EWCS, and the multi-EWCS-based signals \(\Delta\) and \(g\). The goal is to determine whether the holographic Weyl semimetal contains nontrivial tripartite and four-partite entanglement structures, and whether these structures reorganize in a way that detects the topological quantum phase transition.

We use strip-shaped boundary subregions as entangling regions. The strip width \(l\) gives a controllable real-space length scale. Thus the large \(l\) behavior provides a direct way to study the long-range multipartite correlations of the boundary state. Since the holographic Weyl semimetal is anisotropic, we also vary the direction in which the strip has a finite width. We first compare strips finite along the transverse and the axial directions, and then consider strips whose finite direction makes an angle \(\theta\) in the plane~\cite{LIU2019155}. This allows us to ask whether the tripartite and four-partite entanglement structures not only detect the topological transition, but also encode the anisotropy associated with the axial source. Viewed together with anomalous transport, these tripartite and four-partite entanglement data provide a more complete characterization of topological semimetals and of the reorganization of quantum correlations across their phase transitions~\cite{PhysRevD.100.126013,HolographicWeylSemimetalsLandsteiner_2016,HolographicWeylSemimetalsLandsteiner:2016stv,HolographicTopologicalInvariantLiu_2018}.

The rest of this paper is organized as follows. In Section~\ref{Section2}, we review the holographic Weyl semimetal background, the strip geometry, and the entanglement quantities and measures used in the analysis. In Section~\ref{Section3}, we first examine how the entanglement quantities vary with the strip width \(l\), and then isolate their large \(l\) scaling behavior. We next tune \(M/b\) at fixed strip width to identify the topological quantum phase transition. In Section~\ref{Section4}, we analyze the angular dependence of the entanglement measures for strips pointing to different directions to detect the anisotropy of the system. In Section~\ref{Section5}, we extract and compare the angular coefficients in the noncritical phases and discuss the special behavior of the critical solution. Section~\ref{Section6} contains our conclusions and possible extensions.

\section{Review of the holographic Weyl semimetal and multipartite entanglement quantities}
    \label{Section2}
    This section reviews the background material needed in the rest of the paper. We first summarize the holographic Weyl semimetal model, whose solutions at zero temperature contain the topological nontrivial phase, the critical phase, and the topological trivial phase. We then introduce Ryu-Takayanagi (RT) surfaces corresponding to strip subregions and the associated holographic entanglement entropy. From the entanglement entropy we introduce the holographic \(c\)-function as a bipartite entanglement quantity. We then move to tripartite quantities, including conditional mutual information(CMI), \(\kappa\) constructed from the multi-entropy, the entanglement wedge cross section (EWCS), and the Markov gap. Finally, we introduce the multi-EWCS, which will be used below to construct four-partite entanglement quantity.
    \subsection{The holographic model for the Weyl semimetal}
        \label{Subsection2.1}
        We begin by recalling the weakly coupled field theory model for a Weyl semimetal that motivates the holographic construction. A minimal description of a Weyl semimetal that breaks time reversal symmetry is a Dirac theory deformed by a mass term and an axial vector source~\cite{HolographicWeylSemimetalsLandsteiner:2015lsa,HolographicWeylSemimetalsLandsteiner_2016},       \begin{equation}            \label{WeakCouplingWeylSemimetal}         \mathcal{L}_{\rm eff}=\bar{\psi}\left(i\gamma^\mu D_\mu-M-\gamma^\mu\gamma^5 A_\mu\right)\psi.
        \end{equation}
        Here \(D_\mu=\partial_\mu-iV_\mu\) and \(V_\mu\) is the vector gauge field. The parameter \(M\) is the Dirac mass deformation, and \(A_\mu\) is an axial vector field. Without loss of generality, we set \(A_\mu dx^\mu=b\,dz\) to explicitly break Lorentz invariance and time reversal symmetry. Physically, \(b\) sets the separation scale of the Weyl nodes in the momentum space, whereas the mass deformation \(M\) tends to couple the two chiralities and drive the system toward a topologically trivial regime. The competition between these two deformations is therefore naturally characterized by the competition of $M$ and $b$.

        The holographic model can be viewed as a strongly coupled realization of {the symmetry breaking pattern introduced in \eqref{WeakCouplingWeylSemimetal}}. The conserved vector current is dual to a bulk vector gauge field \(V_a\), the axial current is dual to a bulk axial gauge field \(A_a\), and the mass operator is represented by a charged scalar field \(\Phi\). The boundary value of \(A_z\) gives the axial source \(b\), while the source coefficient of the scalar field near the boundary gives the mass deformation \(M\). The Chern-Simons coupling encodes the vector and axial anomaly structure, which is responsible for the anomalous Hall conductivity in the topological nontrivial phase. With these ingredients, the bulk action for the holographic Weyl semimetal is~\cite{HolographicWeylSemimetalsLandsteiner:2015lsa,HolographicWeylSemimetalsLandsteiner_2016,HolographicSemimetalsLandsteiner:2019kxb}
        \begin{equation}
            \label{HolographicWeylSemiMetal}
            \begin{aligned}
                S_{\text{Weyl}}&= \int d^5x \sqrt{-\det g}\left[\frac{1}{2\kappa^2}\left(R+\frac{12}{L^2}\right)-\frac{{F_V}^2}{4}-\frac{{F_A}^2}{4}\right. \\
                &\qquad \left.+ \frac{\alpha}{3}\epsilon^{abcde} A_a\left(3 F_{Vbc} F_{Vde} + F_{Abc} F_{Ade}\right)- (D_a\Phi)^* D^a\Phi- m^2 |\Phi|^2-\frac{\lambda}{2} |\Phi|^4\right].
            \end{aligned}
        \end{equation}
        Here \(g\) is the metric of the bulk spacetime in five dimensions and \(R\) is the corresponding Ricci scalar. \(V\) and \(A\) are the vector and axial gauge fields, with field strengths \(F_V\) and \(F_A\). The scalar field \(\Phi\) provides the effective mass deformation for the holographic Weyl semimetal. The parameter \(\kappa\) is the gravitational constant in five dimensions, and \(L\) is the AdS radius. The parameter \(\alpha\) is the Chern-Simons coupling, while \(m\) and \(\lambda\) enter the scalar potential. The covariant derivative is
        \begin{equation}
            D_a=\partial_a-iqA_a ,
        \end{equation}
        where \(q\) is the corresponding gauge coupling.

        We set \(L=1\), \(2\kappa^2=1\), and \(m^2L^2=-3\), so that the dual scalar operator has dimension three and its source has dimension one, as appropriate for a mass deformation. The Chern-Simons coupling \(\alpha\) controls the normalization of anomalous transport. It does not affect the background solutions considered below, because the vector gauge field is set to zero in the homogeneous ansatz at zero temperature. The corresponding homogeneous solutions are obtained with
        \begin{equation}
            \label{ZeroTemperatureSolutionAnsatz}
            \begin{aligned}
                g &= u(-dt^2 + dx^2 + dy^2) + h\,dz^2 + \frac{dr^2}{u}, \\
                A &= A_z dz, \qquad \Phi = \phi ,
            \end{aligned}
        \end{equation}
        where \(u\), \(h\), \(A_z\), and \(\phi\) are real functions of the radial coordinate \(r\), and the vector field \(V\) is set to zero. With this ansatz, the UV boundary behavior is
        \begin{equation}
            \label{UltraVioletBoundaryBehaviour}
            \begin{aligned}
                &\lim_{r\to\infty}\frac{u(r)}{r^2}= \lim_{r\to\infty}\frac{h(r)}{r^2}=1,\\
                &\lim_{r\to\infty}A_z(r) = b,~\lim_{r\to\infty} r\,\phi(r) = M .
            \end{aligned}
        \end{equation}

        Before writing out the explicit IR solutions, let us recall the corresponding phase structure. In the field theory description, the axial source \(b\) favours a Weyl semimetal with separated Weyl nodes and a finite anomalous Hall conductivity, whereas the mass deformation \(M\) tends to remove this topological structure by coupling the two chiral sectors. In the holographic model, the same competition is encoded geometrically in the radial RG flow. Depending on \(M/b\), the solution at zero temperature approaches one of three IR endpoints. In the numerical analysis we take \(q=1\) and \(\lambda=1/10\), for which the critical value is$(M/b)_c \simeq 0.744.$~\cite{HolographicWeylSemimetalsLandsteiner_2016,HolographicWeylSemimetalsLandsteiner:2016stv} Different values of \(M/b\) are classified into three phases, each associated with a distinct IR asymptotic solution of the same bulk equations of motion.

        \noindent\textbf{\(\bullet\) \emph{The topologically nontrivial phase}.}
            For \(M/b<(M/b)_c\), the bulk solution flows to a topological nontrivial IR fixed point. In this phase, the axial gauge field approaches a nonzero constant in the deep IR, \(A_z(0)\neq 0\), while the scalar field is exponentially suppressed. The nonzero IR value of \(A_z\) gives a finite anomalous Hall conductivity, and this provides the holographic signature of the topological nontrivial semimetal phase. With the ansatz \eqref{ZeroTemperatureSolutionAnsatz}, the IR geometry of this phase takes the form
            \begin{equation}
                \label{NonTrivialInfraRedGeometry}
                \begin{aligned}
                    u   &= r^2, \\
                    h   &= r^2, \\
                    A_z &= a_1 + \frac{\pi a_1^2 \phi_1^2}{16 r} e^{-\frac{2 a_1 q}{r}}, \\
                    \phi &= \sqrt{\pi}\,\phi_1\left( \frac{a_1 q}{2r} \right)^{\frac{3}{2}}e^{-\frac{a_1 q}{r}},
                \end{aligned}
            \end{equation}
            where \(a_1\) and \(\phi_1\) are shooting parameters. Starting from the IR geometry \eqref{NonTrivialInfraRedGeometry}, one integrates the equations of motion for the holographic model \eqref{HolographicWeylSemiMetal} toward the UV boundary and tunes the shooting parameters such that the UV asymptotics match \eqref{UltraVioletBoundaryBehaviour}. This class of solutions exists only for \(M/b<(M/b)_c\).

        \noindent\textbf{\(\bullet\) \emph{The critical phase}.}
            At \(M/b=(M/b)_c\), the system reaches a quantum critical point at zero temperature. The corresponding IR solution is an anisotropic geometry of Lifshitz type, with a nontrivial scaling exponent in the \(z\) direction. The anomalous Hall conductivity vanishes at the critical point, while the geometry retains an anisotropic structure with scale invariance. The IR geometry at the critical point is
            \begin{equation}
                \label{CriticalInfraRedGeometry}
                \begin{aligned}
                    u   &= u_0 r^2 (1 + \delta u\, r^\alpha), \\
                    h   &= h_0 r^{2\beta} (1 + \delta h\, r^\alpha), \\
                    A_z &= r^\beta (1 + \delta a\, r^\alpha), \\
                    \phi &= \phi_0 (1 + \delta \phi\, r^\alpha),
                \end{aligned}
            \end{equation}
            where $\delta\phi = -1, \{\delta u,\delta h,\delta a\} = \{0.369,-2.797,0.137\}\delta\phi,$ and $u_0 = 1.468, h_0 = 0.344,\phi_0 = 0.947,\alpha = 1.315,\beta = 0.407.$ This geometry of Lifshitz type governs the behavior with scale invariance at criticality.
            
        \noindent\textbf{\(\bullet\) \emph{The topologically trivial phase}.}
            For \(M/b>(M/b)_c\), the solution flows to a topological trivial IR endpoint. In this case, the axial gauge field vanishes as a power of the radial coordinate in the far IR, whereas the scalar field approaches a nonzero constant. This corresponds to the restoration of invariance under time reversal at the endpoint of the holographic RG flow, in the sense that the axial structure that breaks time reversal is removed in the deep IR. In the holographic model this phase should be regarded as a topological trivial semimetal rather than necessarily as a fully gapped insulator, since some of the UV degrees of freedom are removed along the RG flow.

            The IR geometry of the topological trivial phase is
            \begin{equation}
                \label{TrivialInfraRedGeometry}
                \begin{aligned}
                    u   &= \left(1 + \frac{3}{8\lambda}\right) r^2, \\
                    h   &= r^2, \\
                    A_z &= a_1 r^{\beta_1}, \\
                    \phi &= \sqrt{\frac{3}{\lambda}}+\phi_1 r^{\beta_2},
                \end{aligned}
            \end{equation}
            where $\lambda = \frac{1}{10}, \beta_1 = \sqrt{\frac{259}{19}} - 1, \beta_2 = \frac{10}{\sqrt{19}} - 2,$ and \(a_1,\phi_1\) are shooting parameters. One can numerically obtain the bulk profiles \(u\), \(h\), \(A_z\), and \(\phi\), for which \(M/b>(M/b)_c\), as expected.

        The low energy, large scale behavior of the system is governed by the IR geometry. Consequently, the large scale scaling behavior is determined by the scaling behavior of the IR geometry. Under a scale transformation, the IR geometry exhibits anisotropic scaling symmetry. Specifically, given bulk coordinates \(\{t,x^1,x^2,\cdots,x^n,r\}\), when we perform the transformation $r\to s^{-1}r$, the IR geometry transforms as
        \begin{equation}
            \label{InfraRedGeometryScaling}
            x^i \to s^{\alpha_i} x^i, ~t \to s t,
        \end{equation}
        {where the scaling exponents $\alpha_i$ measure the anisotropy of the ground state for the system and can be read from the corresponding IR geometry explicitly. For the topological nontrivial phase \eqref{NonTrivialInfraRedGeometry} and trivial phase \eqref{TrivialInfraRedGeometry}, corresponding $\alpha_x=\alpha_y=\alpha_z=1$. In contrast, the critical phase \eqref{CriticalInfraRedGeometry} has $\alpha_x=\alpha_y=1,~\alpha_z=\beta=0.407$.} With the scaling exponent $\alpha_i$, under the transformation \eqref{InfraRedGeometryScaling} the volume element in the deep IR region then transforms as
        \begin{equation}
            dx^1 dx^2 \cdots dx^n \to s^{\alpha_V} dx^1 dx^2 \cdots dx^n, 
        \end{equation}
        where $\alpha_V = \sum_{j=1}^n \alpha_j$.
    \subsection{The RT surface for a strip and the corresponding holographic \texorpdfstring{\(c\)}{c}-function}
        \label{Subsection2.2}
        {After introducing the holographic model for the Weyl semimetal, we review the calculation for the entanglement entropy via the RT formula and the definition of the corresponding $c$-function. To render the extremal surface problem tractable, we consider an infinite strip of finite width on the boundary. This choice is standard in holographic analyses for two key reason. First, with the boundary filed theory is equipped with translational symmetry, the action for corresponding RT surface is completely integrable. Second, the entanglement entropy of such a strip is a well-studied observable in boundary field theories, capturing non-local correlations at a scale set by its width.}
    
        Without loss of generality, we work on a slice at constant time with metric $g_{xx}dx^2 + g_{yy}dy^2 + g_{zz}dz^2 + g_{rr}dr^2$ and denote the coordinates by \((x,y,z,r)=(x^1,x^2,x^3,r)\). {To keep the derivation general, we temporarily do not substitute the specific metric components but use the symbolic form \(g_{ii}\).} Considering a strip oriented along the \(x^i\) direction with width \(l_i\), namely \(x^i\in[-l_i/2,l_i/2]\), the RT surface homologous to this strip can then be parametrized by the boundary coordinates \(x^1,x^2,x^3\), with embedding \(r=r(x^i)\). The corresponding area functional reduces to
        \begin{equation}
            \label{Action}
            A=L^2\int_{-l_i/2}^{l_i/2} \sqrt{\prod_{j\ne i}g_{jj}\left(g_{rr}r'^2+g_{ii}\right)}\,dx^i,
        \end{equation}
        where \(L\) is a regulator length for the transverse directions \(x^j\) with \(j\ne i\). Translational invariance along the \(x^i\) direction implies, via Noether's theorem, the conserved quantity
        \begin{equation}
            \label{ConservedQuantity}
            C_i=g_{ii}\sqrt{\frac{\prod_{j\ne i}g_{jj}}{g_{ii}+g_{rr}r'^2}}.
        \end{equation}

        The reduced system therefore admits the conserved quantity \eqref{ConservedQuantity} and is fully integrable. This constant of motion is directly related to the derivative of the entanglement entropy
        \begin{equation}
            \label{DifferentialEntropy}
            \frac{4G}{L^2}\frac{dS_i}{dl_i}=\frac{C_i}{2},
        \end{equation}
        where \(S_i\) is the holographic entanglement entropy for a strip whose finite width lies along the \(x^i\) direction. This relation motivates the definition of the \(c\)-function~\cite{DifferentialEntropyMyers_2012,DifferentialEntropyLiu_2014,Baggioli:2020cld}
        \begin{equation}
            \label{CentralFunctionDefinition}
            c_i=\frac{C_i}{2}l_i^3 .
        \end{equation}

        Before turning to the explicit computation, we analyze the behavior at large \(l_i\) of the \(c\)-function. The turning point \(r_*\), defined by \(r'=0\), determines the maximal bulk depth reached by the RT surface. To probe behavior at long distances and low energies, we require \(r_*\) to lie sufficiently close to the horizon. The conserved quantity \(C_i\) defined in \eqref{ConservedQuantity} can equivalently be evaluated at the turning point as
        \[C_i=\sqrt{\prod_j g_{jj}(r_*)}.\]
        For a given turning point \(r_*\), the corresponding strip width can be obtained by integrating from \(r_*\) to infinity
        \begin{equation}
            \label{StripWidth}
            l_i=2\int_{r_*}^{\infty}\sqrt{\frac{g_{rr}C_i^2} {g_{ii}\left(\prod_j g_{jj}-C_i^2\right)}}\,dr.
        \end{equation}

        In the regime of large strip width, this integral is dominated by the region near \(r_*\). Assuming that the RT surface penetrates deeply into the IR region, we may use the leading IR geometry to determine the relation, governed by a power law, between \(l_i\) and the corresponding turning point \(r_*\). Define
        \[\widetilde{C}_i=\sqrt{\prod_j g_{jj}(s r_*)},\]
        and let
        \begin{equation}
            \label{ScaledStripWidth}
            \widetilde{l}_i=2\int_{s r_*}^{\infty}\sqrt{\frac{g_{rr}\widetilde{C}_i^2}{g_{ii}\left(\prod_j g_{jj}-\widetilde{C}_i^2\right)}}\,dr
        \end{equation}
        denote the strip width associated with an RT surface whose turning point is located at \(s r_*\). When $l_i$ is large, $r_*$ approaches the horizon, and the scaling behavior for \eqref{StripWidth} of the system at such scales is mainly determined by the IR geometry. Therefore, we could get the large $l$ scaling behavior explicitly from the IR geometry as follows. We have 
        \[\widetilde{l}_i(s r_*)=s^{-\alpha_i}l_i(r_*),\]
        which implies
        \begin{equation}
            \label{TurningPointScaling}
            r_*\propto l_i^{-\frac{1}{\alpha_i}} .
        \end{equation}
        Here \(\alpha_i\) is the IR scaling exponent along the \(x^i\) direction. This relation in turn yields the behavior at large distances of the conserved quantity \(C_i\) and the corresponding \(c\)-function
        \begin{equation}
            \label{CentralFunctionScalingBehavior}
            C_i\propto l_i^{-\frac{\alpha_V}{\alpha_i}},~ c_i\propto l_i^{3-\frac{\alpha_V}{\alpha_i}},
        \end{equation}
        where \(\alpha_V\) is the scaling exponent of the volume element in the deep IR region. Similar IR counting of scaling powers also organizes the behavior at large strip width of the CMI, EWCS, \(\kappa\), Markov gap, and multi-EWCS considered below, which will be determined explicitly later.
    \subsection{The holographic conditional mutual information}
        The CMI $I(A:B|E)$ is defined as the information shared between two regions $A$ and $B$ conditioned on a third region $E$:
        \begin{equation}
            \label{CMIDefinition}
            I(A:B|E)=S(A\cup E)+S(B\cup E)-S(A\cup B\cup E)-S(E).
        \end{equation}
        A particularly useful configuration is obtained by taking \(A\) and \(B\) to be infinitesimal regions separated by a strip \(E\) of width \(l_i\) aligned along the \(x^i\) direction~\cite{ConditionalMutualInformationJi_2025,ConditionalMutualInformationJu_2024}. In this case, the CMI is directly related to the second derivative of the entanglement entropy:
        \begin{equation}
            \label{CMISecondDerivative}
            I(A:B|E)=-\frac{d^2 S_i}{d l_i^2} ,
        \end{equation}
        where \(S_i\) is the entanglement entropy of the strip. Here we have already divided the ordinary CMI by the sizes \(l_A\) and \(l_B\) of the two infinitesimal regions, so these factors do not appear explicitly in the expression above. The quantity denoted by \(I(A:B|E)\) here is therefore a CMI density with dimension \(L^{-2}\). Using the holographic relation \eqref{DifferentialEntropy} between the entropy derivative and the conserved quantity \(C_i\), one further obtains
        \begin{equation}
            \label{CMIConservedQuantity}
            I(A:B|E)=-\frac{L^2}{8G}\frac{d C_i}{d l_i}.
        \end{equation}

        The behavior of CMI at large \(l_i\) is controlled by the deep IR geometry. The power law dependence for the CMI can be obtained by \eqref{CentralFunctionScalingBehavior} as
        \begin{equation}
            I(A:B|E)\propto l_i^{-1-\frac{\alpha_V}{\alpha_i}},
        \end{equation}
        where $\alpha_V$ and $\alpha_i$ is presented above. Therefore, the decay of CMI governed by this power provides a sharp nonlocal probe of the IR fixed point and is sensitive to the quantum phase transition.
    \subsection{The holographic entanglement measures based on the multi-entropy}
        The multi-entropy generalizes entanglement entropy to multipartite systems. In field theory, it arises from a construction based on twist operators of higher genus and extends the replica trick to configurations involving multiple disjoint regions~\cite{Gadde_2022,Gadde_2023,MultiEntropyHarper_2024,gadde2025multiinvariantsbulkreplicasymmetry}. Holographically, the multi-entropy is represented by a Steiner tree, namely a network of minimal surfaces that partitions the bulk into regions individually homologous to the corresponding boundary subregions. For a tripartite pure state with boundary regions \(A\), \(B\), and \(C\), the Steiner tree consists of three minimal surfaces meeting at a common junction. Together, these surfaces divide the bulk into three disconnected components, each of which is homologous to one boundary region. The holographic multi-entropy is then defined by the minimal total area of this network:
        \begin{equation}
            \label{MultiEntropyDefinition}
            S^{(3)}(A:B:C)=\frac{1}{4G}\min_{\text{Steiner tree}}\left[\operatorname{Area}(\Gamma_A)+\operatorname{Area}(\Gamma_B)+\operatorname{Area}(\Gamma_C)\right],
        \end{equation}
        where \(\Gamma_A\), \(\Gamma_B\), and \(\Gamma_C\) denote the three minimal
        surfaces meeting at the junction. At the junction, the equilibrium condition requires the three surfaces to meet at mutual angles of \(2\pi/3\), ensuring that the total area is extremal.

        For three subsystems \(A\), \(B\), and \(C\), the multi-entropy \(S^{(3)}(A:B:C)\) contains both genuine tripartite and reducible bipartite contributions. To isolate the irreducible tripartite component, we define
        \begin{equation}
            \label{KappaDefinition}
            \kappa(A:B:C)=S^{(3)}(A:B:C)-\frac{1}{2}\bigl(S_{AB}+S_{BC}+S_{CA}\bigr),
        \end{equation}
        where \(S_{AB}\equiv S(A\cup B)\), \(S_{BC}\equiv S(B\cup C)\), and \(S_{CA}\equiv S(C\cup A)\). This combination cancels the UV divergent contributions and isolates the genuine tripartite entanglement that cannot be reduced to bipartite correlations. In particular, \(\kappa\) vanishes for separable states as well as for triangle states, whose tripartite entanglement can be reduced to bipartite entanglement among smaller subsystems, while it is nonzero for genuinely tripartite entangled states such as the GHZ state. Hence, \(\kappa\) provides a useful measure of genuine multipartite entanglement in holographic systems.
    \subsection{The holographic entanglement measures based on entanglement wedge cross section}
        The EWCS is a fundamental geometric object in holography~\cite{Umemoto2018,ReflectedEntropydutta2019canonicalpurificationentanglementwedge}. For two boundary subregions \(A\) and \(B\) whose entanglement wedge is connected, the EWCS is defined as the surface of minimal area that separates the entanglement wedge into two parts, each homologous to \(A\) and \(B\), respectively
        \begin{equation}
        \label{EWCSDefinition}
            E_W(A:B) = \frac{1}{4G}\, \min_{\gamma_{A:B}} \operatorname{Area}(\gamma_{A:B}),
        \end{equation}
        where \(\gamma_{A:B}\) is a surface within the entanglement wedge that partitions it into two regions. Holographically, the EWCS is dual to two entanglement measures for mixed-states: the entanglement of purification \(E_P(A:B)\) and the reflected entropy \(S_R(A:B)\), satisfying
        \begin{equation}
            S_R(A:B) = 2E_P(A:B) = 2E_W(A:B).
        \end{equation}
        Thus, the EWCS captures quantum correlations in mixed-states and can be used to construct multipartite entanglement quantities.

        A refined tripartite quantity derived from the EWCS is the Markov gap, defined as~\cite{MarkovGapHayden_2021}
        \begin{equation}
        \label{MarkovGapDefinition}
            h(A:B) = S_R(A:B) - I(A:B) = 2E_W(A:B) - I(A:B),
        \end{equation}
        where \(I(A:B)=S_A+S_B-S_{A\cup B}\) is the mutual information. By construction, the Markov gap is non-negative. Its vanishing indicates that the tripartite pure state \(|\psi\rangle_{ABC}\), with \(C\) being the complement of \(A\cup B\), is a sum of triangle states (SOTS) up to local unitary transformations~\cite{SumOfTrangleStateZou2021}. Triangle states are those whose tripartite entanglement reduces to bipartite entanglement among smaller subsystems, while SOTS states are convex mixtures of such triangle states. In contrast, a nonzero Markov gap signals the presence of genuine tripartite entanglement that cannot be decomposed into SOTS.

        Together with \(\kappa\), which vanishes precisely for triangle states, the Markov gap provides a complementary characterization of tripartite entanglement: \(\kappa=0\) indicates that the state is a triangle state, while \(h=0\) indicates that it is a sum of triangle states. Their combined behavior helps characterize the irreducible multipartite entanglement structure in holographic systems.
    \subsection{The four-partite entanglement signals based on multi-EWCS}
        Having introduced the tripartite entanglement quantities based on the multi-entropy and the bipartite EWCS, we now generalize the EWCS to a tripartite entanglement quantity \(E_W(A:B:C)\) in order to construct more partite entanglement signals. We choose three finite boundary regions \(A\), \(B\), and \(C\), and regard the complement \(D=\overline{A\cup B\cup C}\) as the fourth subsystem. The multipartite EWCS \(E_W(A:B:C)\) is therefore used as a geometric quantity of the partition into four parties \(A:B:C:D\). In this notation, the displayed arguments \(A:B:C\) specify the finite regions whose entanglement wedge is constructed, while the complementary region \(D\) enters implicitly as the purifier of the mixed state on \(A\cup B\cup C\).

        The multi-EWCS naturally generalizes the bipartite EWCS to multipartite configurations. Consider \(n\) boundary subregions \(A_i\) \((i=1,\dots,n)\) and let \(\gamma\) denote the RT surface corresponding to \(\bigcup_{i=1}^n A_i\) while \(M\) be the bulk region enclosed by \(\gamma\) together with the boundary region \(\bigcup_{i=1}^n A_i\). This region \(M\) is the entanglement wedge of \(\bigcup_{i=1}^n A_i\). We partition the boundary of \(M\) into \(n\) parts \(\widetilde{A}_i\) such that \(A_i\subseteq \widetilde{A}_i\). For each \(\widetilde{A}_i\), we introduce a cross section \(\Gamma_i\) within \(M\) that is homologous to \(\widetilde{A}_i\). The multi-EWCS is then defined as the minimal total area of these cross sections:
        \begin{equation}
        \label{MultiEWCSDefinition}
            E_W(A_1:A_2:\cdots:A_n)=\frac{1}{4G}\,\min_{\{\Gamma_i\}}\operatorname{Area}\left(\bigcup_{i=1}^n\Gamma_i\right).
        \end{equation}
        \begin{figure}[htbp]
            \centering
            \includegraphics[width=0.75\linewidth]{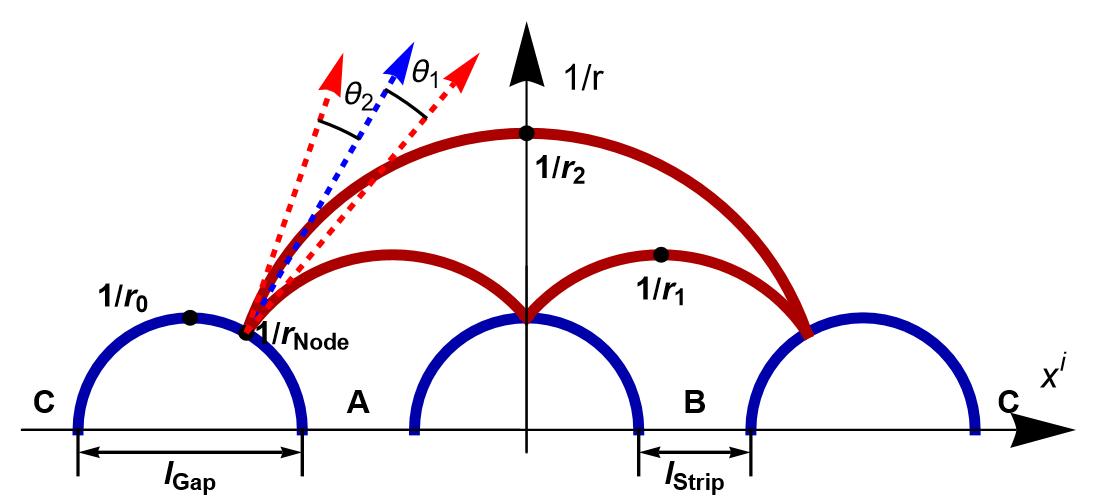}
            \caption{Illustration of the multi-EWCS configuration. The blue curves denote the RT surfaces that bound the entanglement wedge of \(A\cup B\cup C\), while the red curves denote the candidate multipartite cross section inside this wedge. The junction point \(r_{\rm node}\) is varied to minimize the total area of the red network; in the symmetric configuration this gives \(\theta_1=\theta_2\). The boundary parameters \(l_{\rm Strip}\) and \(l_{\rm Gap}\) denote the widths of the finite strips and the gap regions between adjacent regions.}
            \label{FigureMultiEWCSSketch}
        \end{figure}

        For simplicity, we consider the three strip configuration depicted in Fig.~\ref{FigureMultiEWCSSketch}: three strips $A,B,C$ of equal width $l_{\text{Strip}}$ placed in a row, with adjacent strips separated by a gap of width $l_{\text{Gap}}$. In this symmetric setup, the multi-EWCS simplifies considerably. To minimize the total area, it suffices to adjust the position of the junction point \(r_{\rm node}\) such that the angles satisfy \(\theta_1=\theta_2\) as shown in  Fig.\,\ref{FigureMultiEWCSSketch}. The resulting minimal cross section, highlighted in red in the figure, directly yields the multi-EWCS. With the convention introduced above, this quantity could be used to construct four partite entanglement quantities for the partition \(A:B:C:D\), where \(D=\overline{A\cup B\cup C}\).

        Holographically, the multi-EWCS is dual to the multi-entanglement of purification \(E_p(A:B:C)\) in the boundary field theory, where the multi-EoP $E_p(A_1:A_2:\cdots:A_n)$ is defined as \[E_p(A_1:A_2:\cdots:A_n)=\min_{\ket{\psi}_{A_1A_1'\cdots A_nA_n'}}\sum_{i=1}^nS_{A_iA_i'}.\] The tripartite multi-EoP satisfies two key inequalities~\cite{Bao2019,MultiEntanglementUpperBoundju2024holographicmultipartiteentanglementupper,MarkovGapju2025holographicmultipartiteentanglementstructures,bao2026tripartitecorrelationsignalmultipartite}:
        \begin{equation}
            E_p(A:B:C) \ge I(A:B:C),
        \end{equation}
        \begin{equation}
            E_p(A:B:C) \ge E_p(A:BC) + E_p(B:AC) + E_p(C:AB),
        \end{equation}
        where \(I(A:B:C)\) denotes the tripartite mutual information. These inequalities motivate the definition of two multipartite entanglement measures. The first is the tripartite entanglement signal
        \begin{equation}
            \label{MultiPartiteEntanglementSignal}
            \Delta(A:B:C)=E_p(A:B:C)- E_p(A:BC)- E_p(B:AC)- E_p(C:AB),
        \end{equation}
        which isolates the irreducible contribution not captured by bipartite partitions. {The quantity $\Delta(A:B:C)$ vanishes when the state $\rho_{ABC}$ contains only classical correlations and bipartite entanglement among $A$, $B$ and $C$, or when $\rho_{ABC}$ is pure.} The second is
        \begin{equation}
            \label{MultiPartiteEntanglementSignalNew}
            g(A:B:C) = E_p(A:B:C) - I(A:B:C),
        \end{equation}
        which {is a natural generalization for the Markov gap $h(A:B)$. Similar to the Markov gap, \(g(A:B:C)=0\) if and only if the pure state \(\ket{\psi}_{ABCD}\) is a quadrangle state $\ket{\psi}_{A_1D_1}\ket{\psi}_{B_1D_2}\ket{\psi}_{C_1D_3}\ket{\psi}_{A_2B_2C_2}$ up to a local unitary transformation, where \(D\) is the complement of \(ABC\), and the state spaces of the subsystems are appropriately partitioned as $\mathcal{H}_{\alpha} = \mathcal{H}_{\alpha_1} \otimes \mathcal{H}_{\alpha_2},(\alpha = A,B,C)~\mathcal{H}_{D} = \mathcal{H}_{D_1} \otimes \mathcal{H}_{D_2} \otimes \mathcal{H}_{D_3}.$}

        Although these quantities are written with the displayed arguments \(A:B:C\), in the present strip construction they should be understood as quantities defined with respect to the partition into four parties \(A:B:C:D\), with \(D=\overline{A\cup B\cup C}\) implicit through the purification.

\section{Entanglement quantities characterizing the topological phases transition}
\label{Section3}

We now evaluate the entanglement quantities introduced in Section~\ref{Section2} in the holographic Weyl semimetal at zero temperature. We use strip regions because their finite width \(l\) provides a controllable boundary length scale: as \(l\) is increased, the associated RT surfaces, entanglement wedges, and multipartite wedge networks reach deeper into the bulk and become sensitive to the IR geometry. In the holographic Weyl semimetal, the axial source separates the Weyl nodes along the \(z\) direction, while the \(x\) and \(y\) directions remain transverse. A strip with finite width along \(x\) therefore gives the entanglement quantities in transverse directions, whereas a strip with finite width along \(z\) probes the entanglement behavior in the distinguished direction. We first restrict to these two directions; finite-angle strips in the \(x\)-\(z\) plane as a probe of anisotropy will be studied in Section~\ref{Section4}.

The analysis starts from the dependence of the holographic \(c\)-function, CMI, EWCS, \(\kappa\), the Markov gap, multi-EWCS, \(\Delta\) and \(g\) on the strip width \(l\). As \(l\) becomes large, these quantities enter the regime controlled by the IR fixed point of the corresponding zero temperature background. Since the system is gapless, correlations can extend over long distances and decay algebraically; the entanglement structure at large strip width  \(l\) is therefore governed by scaling behavior. We therefore extract the large \(l\) powers of all entanglement quantities in the \(x\) and \(z\) directions and compare them with the expectations from the IR scaling data. We then fix \(l\) and vary \(M/b\) across the transition, in order to test how these entanglement quantities detect the quantum topological phase transition. Finally, we compare the behavior of these quantities of near-critical phases on the two sides of the critical point.

\subsection{The \texorpdfstring{$l$}{l} dependence of various entanglement quantities and multipartite measures}
\label{Subsection3.1}
We begin with the calculation of the entanglement quantities in all three phases of the holographic Weyl semimetal, utilizing the methods presented in Section~\ref{Section2}. The quantities considered here are the holographic \(c\)-function, CMI, EWCS, \(\kappa\), the Markov gap, multi-EWCS, \(\Delta\) and \(g\). We display them as functions of \(l_x\) and \(l_z\), corresponding to strips whose finite width is taken along the transverse $x$ direction and along the axial $z$ direction, respectively, for all three phases of the holographic Weyl semimetal. This gives a scale dependent view of the entanglement quantities before the large width regime is isolated in the next subsection.

For the nontrivial and trivial phases, we use the representative parameter values of \(M/b=(M/b)_c\mp0.1\). This choice is sufficiently far from the critical point to show the noncritical behavior clearly, and compares the two phases at the same parameter distance from the transition. The behavior of the critical solution and the pure AdS case is also shown as reference backgrounds. The common feature of the curves is that the phase dependence is weak at small strip width and becomes more visible as the strip width is increased.
\begin{figure}[htbp]
\centering
\begin{minipage}[t]{0.49\textwidth}
\centering
\includegraphics[width=\linewidth]{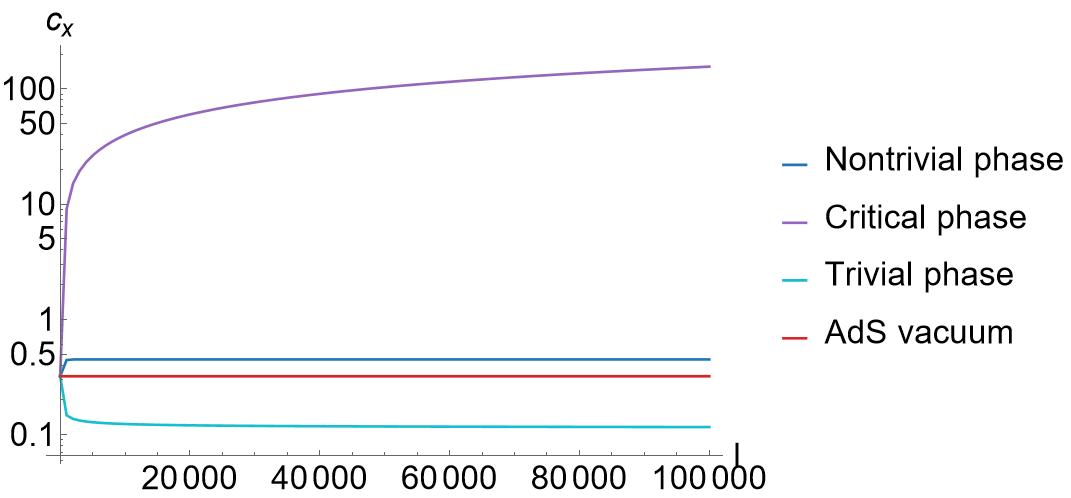}
\end{minipage}
\hfill
\begin{minipage}[t]{0.49\textwidth}
\centering
\includegraphics[width=\linewidth]{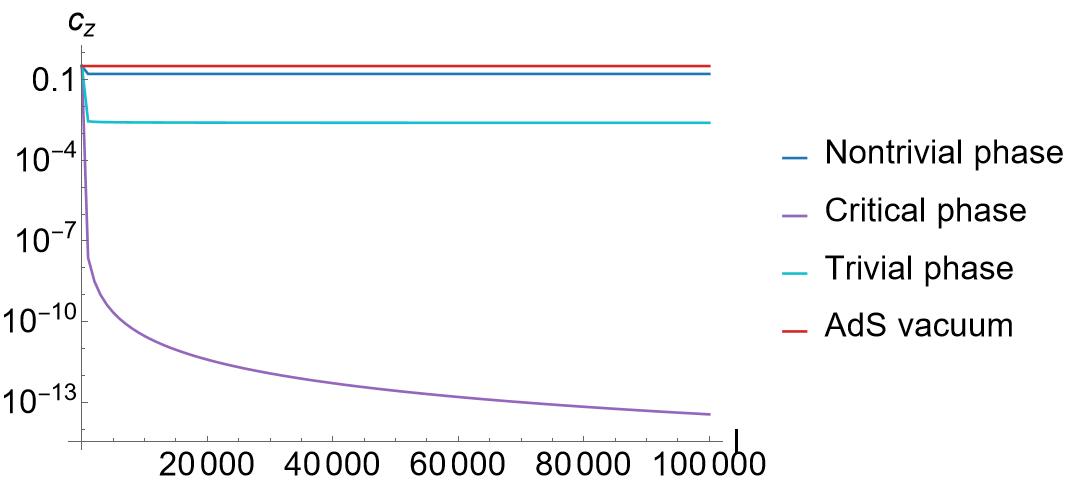}
\end{minipage}
\caption[Holographic \(c\)-function as a function of strip width]{
The holographic \(c\)-function as a function of strip width for four
backgrounds: the topologically nontrivial phase, the critical phase, the
topologically trivial phase, and the pure AdS background. The parameters for the nontrivial and
trivial phases are \(M/b=(M/b)_c\mp0.1\). Left: \(c_x(l_x)\).
Right: \(c_z(l_z)\).
}
\label{fig:b0p1_cfunction_fixed_direction}
\end{figure}

Figure~\ref{fig:b0p1_cfunction_fixed_direction} shows the representative behavior of the holographic \(c\)-function. At small strip width, the curves are close to one another, reflecting the common UV asymptotics of the backgrounds. At larger strip width, the nontrivial and trivial phases approach finite plateaus in both directions. The critical solution instead remains running over the accessible large width range: it grows in the transverse \(x\) direction and is suppressed in the \(z\) direction.

\begin{figure}[htbp]
    \centering

    \begin{minipage}[b]{0.48\textwidth}
        \centering
        \includegraphics[width=\linewidth]{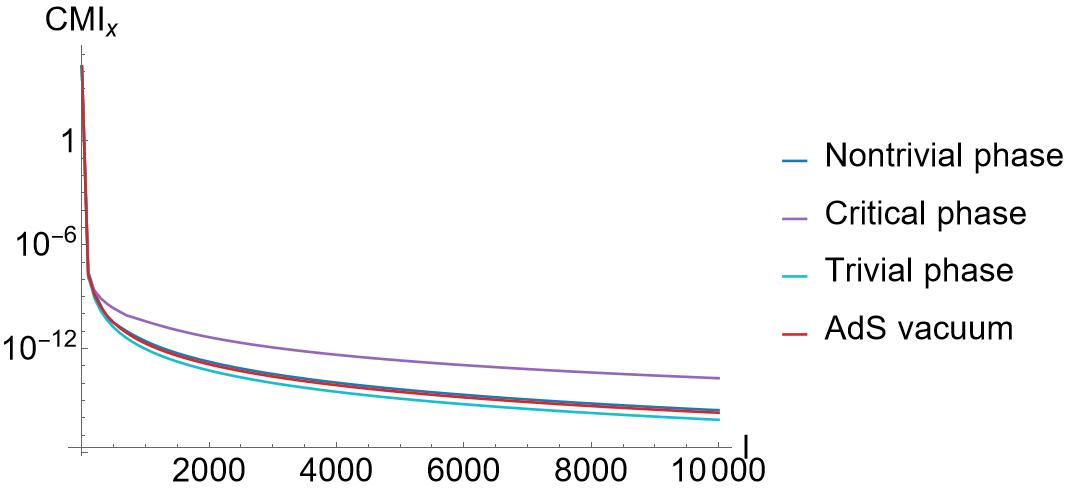}
        \small (a) CMI\(_x(l_x)\)
    \end{minipage}
    \hfill
    \begin{minipage}[b]{0.48\textwidth}
        \centering
        \includegraphics[width=\linewidth]{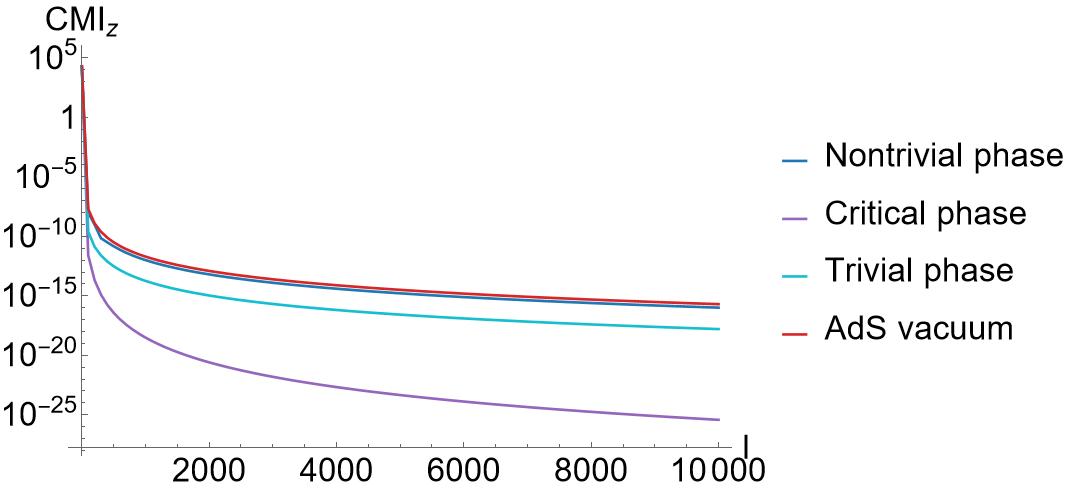}
        \small (b) CMI\(_z(l_z)\)
    \end{minipage}

   \caption[CMI as a function of strip width]{
The CMI as a function of strip width for four backgrounds: the topologically nontrivial phase, the critical phase, the topologically trivial phase, and the pure AdS background. The parameters for the nontrivial and trivial phases are \(M/b=(M/b)_c\mp0.1\). Left: CMI\(_x(l_x)\). Right: CMI\(_z(l_z)\).}
    \label{fig:fixed_l_cmi_comparison}
\end{figure}

\begin{figure}[htbp]
    \centering

    \begin{minipage}[b]{0.48\textwidth}
        \centering
        \includegraphics[width=\linewidth]{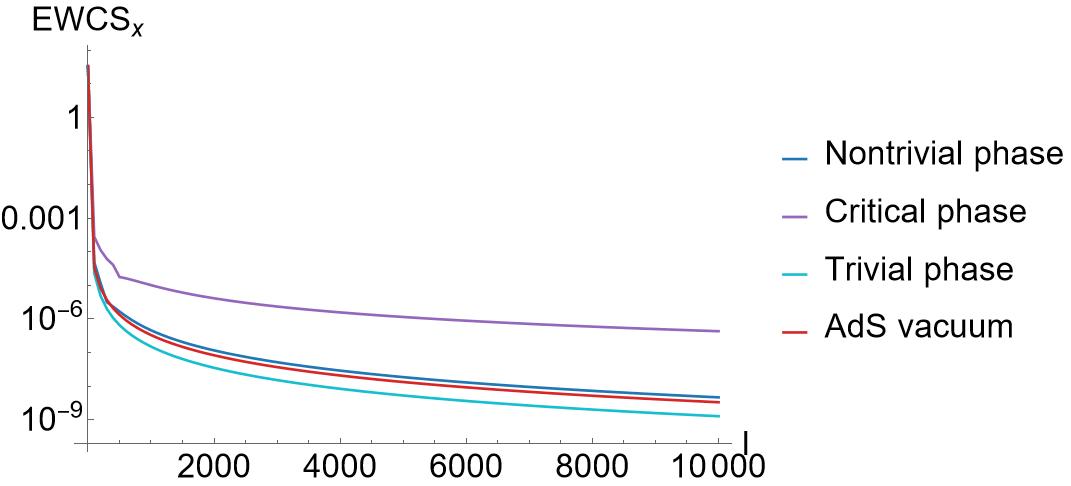}
        \small (a) EWCS\(_x(l_x)\)
    \end{minipage}
    \hfill
    \begin{minipage}[b]{0.48\textwidth}
        \centering
        \includegraphics[width=\linewidth]{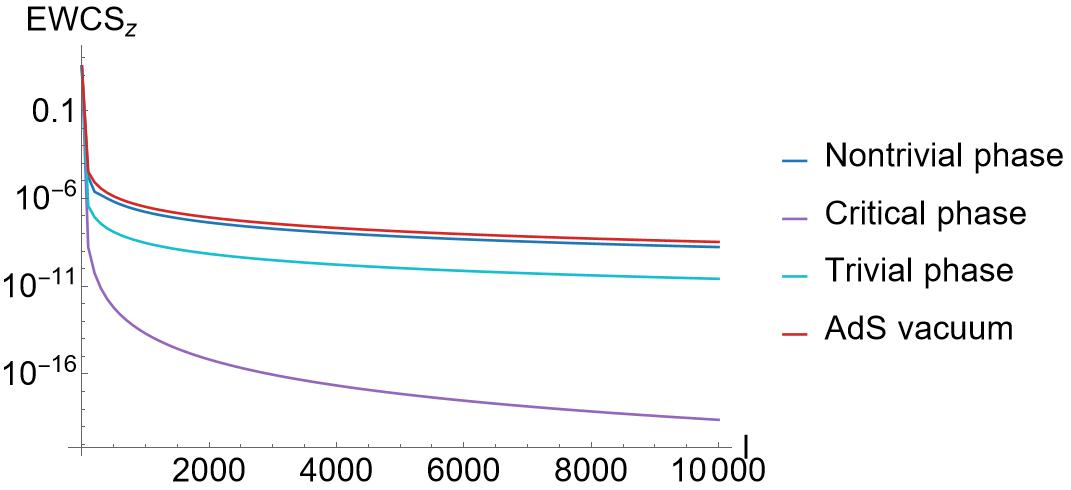}
        \small (b) EWCS\(_z(l_z)\)
    \end{minipage}

   \caption[EWCS as a function of strip width]{
The EWCS as a function of strip width for four backgrounds: the topologically nontrivial phase, the critical phase, the topologically trivial phase, and the pure AdS background. The parameters for the nontrivial and trivial phases are \(M/b=(M/b)_c\mp0.1\). Without loss of generality, an appropriate ratio of $l_{\text{strip}}$ to $l_{\text{gap}}$ has been chosen to guarantee that the entanglement wedges between different regions remain connected as the scale increases. Left: EWCS\(_x(l_x)\). Right:
    EWCS\(_z(l_z)\). 
    }
    \label{fig:fixed_l_ewcs_comparison}
\end{figure}

\begin{figure}[htbp]
    \centering

    \begin{minipage}[b]{0.48\textwidth}
        \centering
        \includegraphics[width=\linewidth]{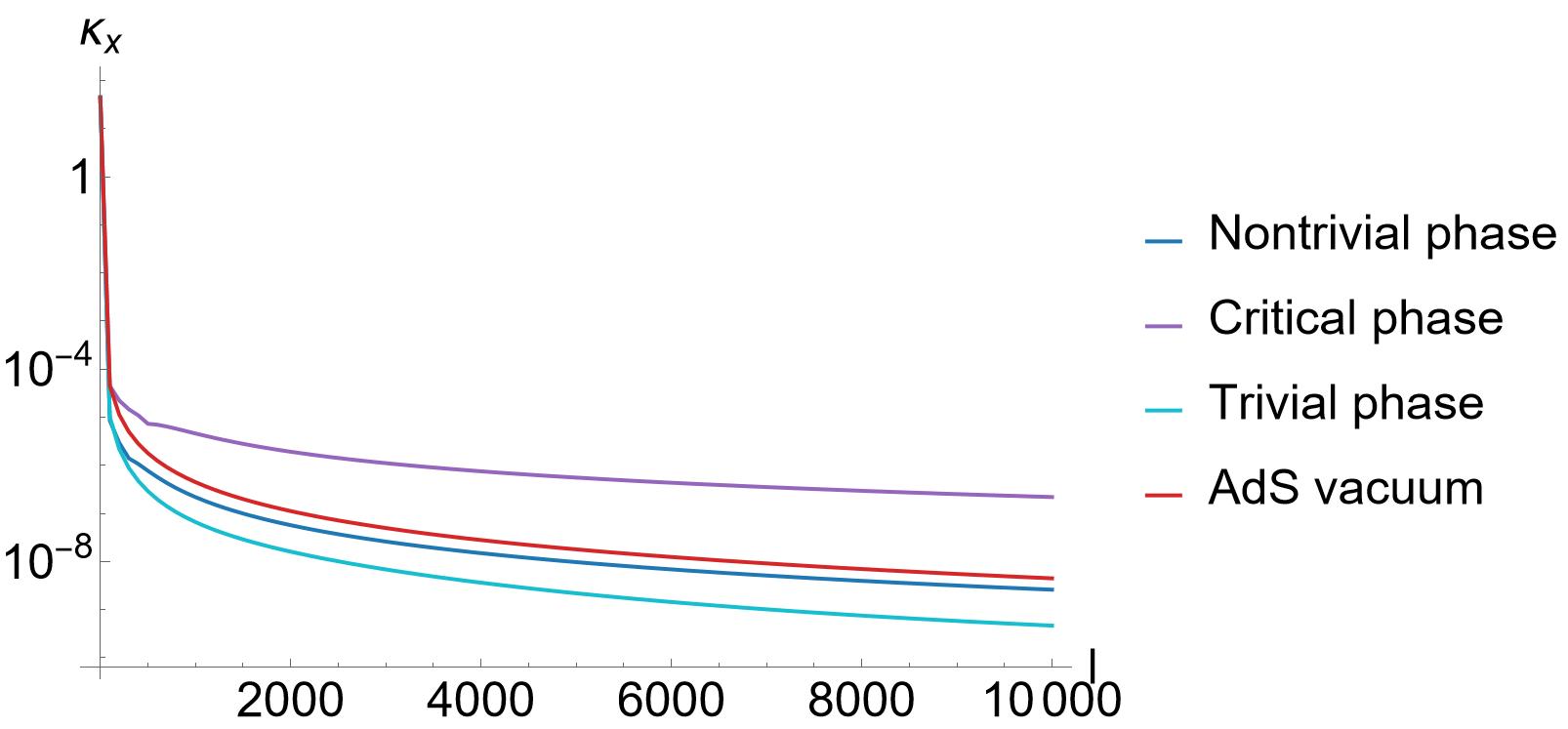}
        \small (a) \(\kappa_x(l_x)\)
    \end{minipage}
    \hfill
    \begin{minipage}[b]{0.48\textwidth}
        \centering
        \includegraphics[width=\linewidth]{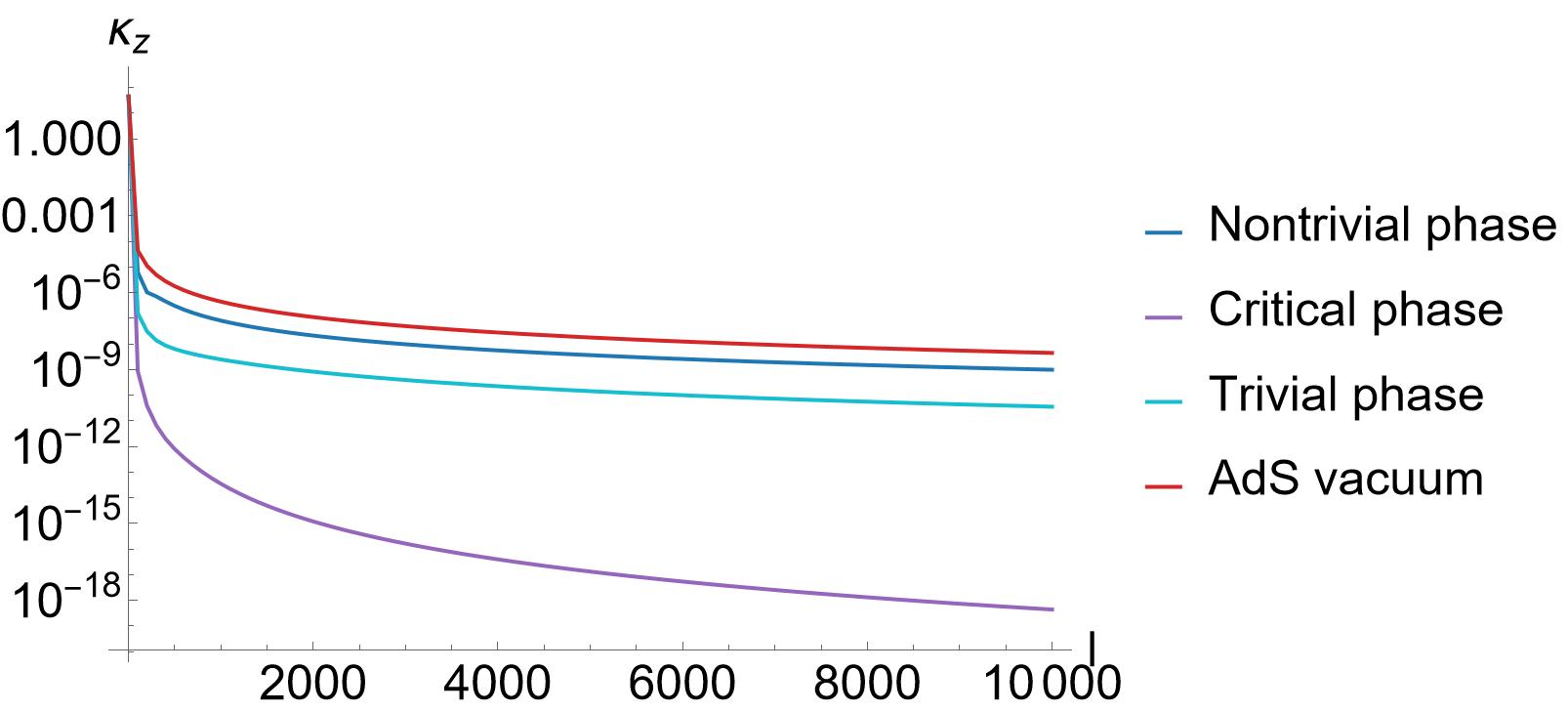}
        \small (b) \(\kappa_z(l_z)\)
    \end{minipage}

   \caption[\(\kappa\) as a function of strip width]{
The \(\kappa\) as a function of strip width for four backgrounds: the topologically nontrivial phase, the critical phase, the topologically trivial phase, and the pure AdS background. The parameters for the nontrivial and trivial phases are \(M/b=(M/b)_c\mp0.1\). Left:
    \(\kappa_x(l_x)\). Right: \(\kappa_z(l_z)\).
    }
    \label{fig:fixed_l_kappa_comparison}
\end{figure}

\begin{figure}[htbp]
    \centering

    \begin{minipage}[b]{0.48\textwidth}
        \centering
        \includegraphics[width=\linewidth]{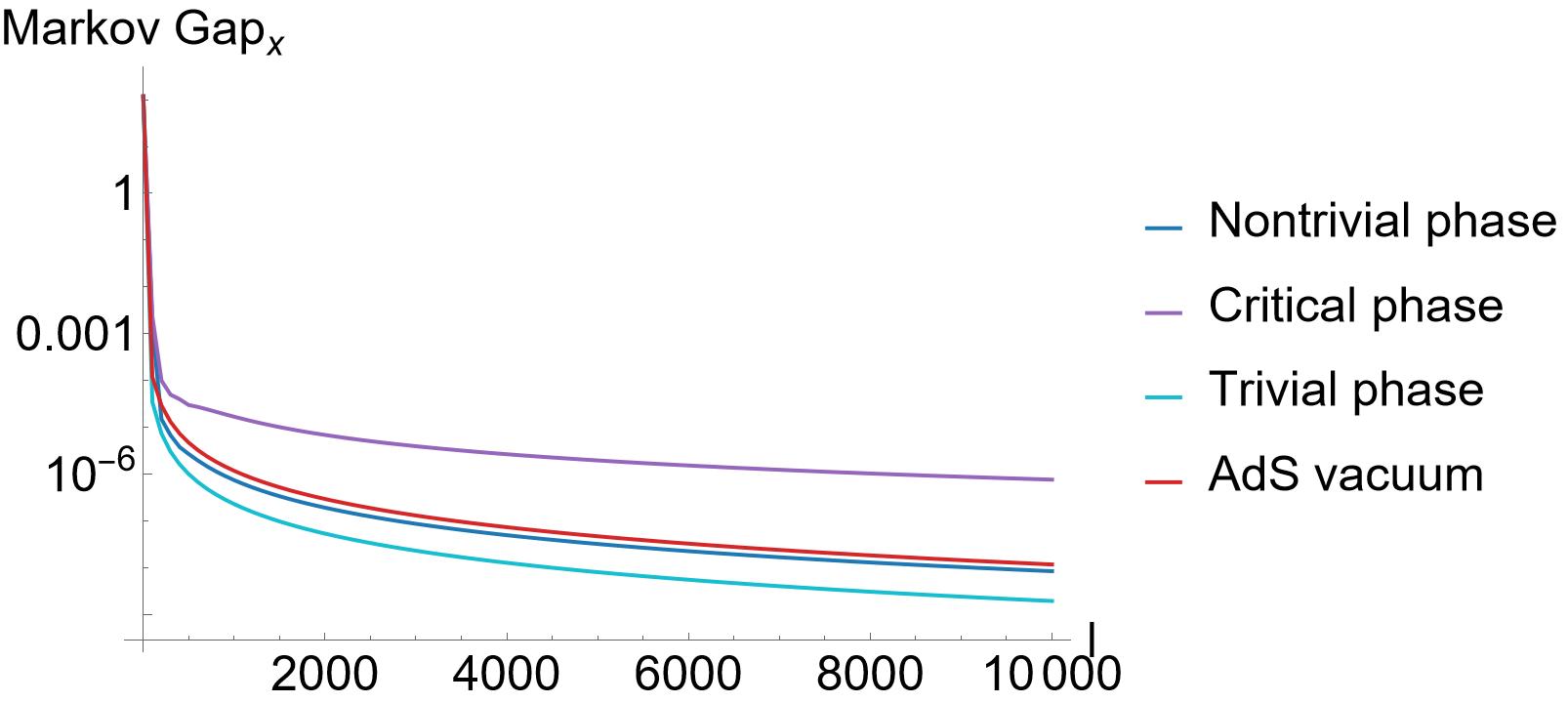}
        \small (a) Markov gap\(_x(l_x)\)
    \end{minipage}
    \hfill
    \begin{minipage}[b]{0.48\textwidth}
        \centering
        \includegraphics[width=\linewidth]{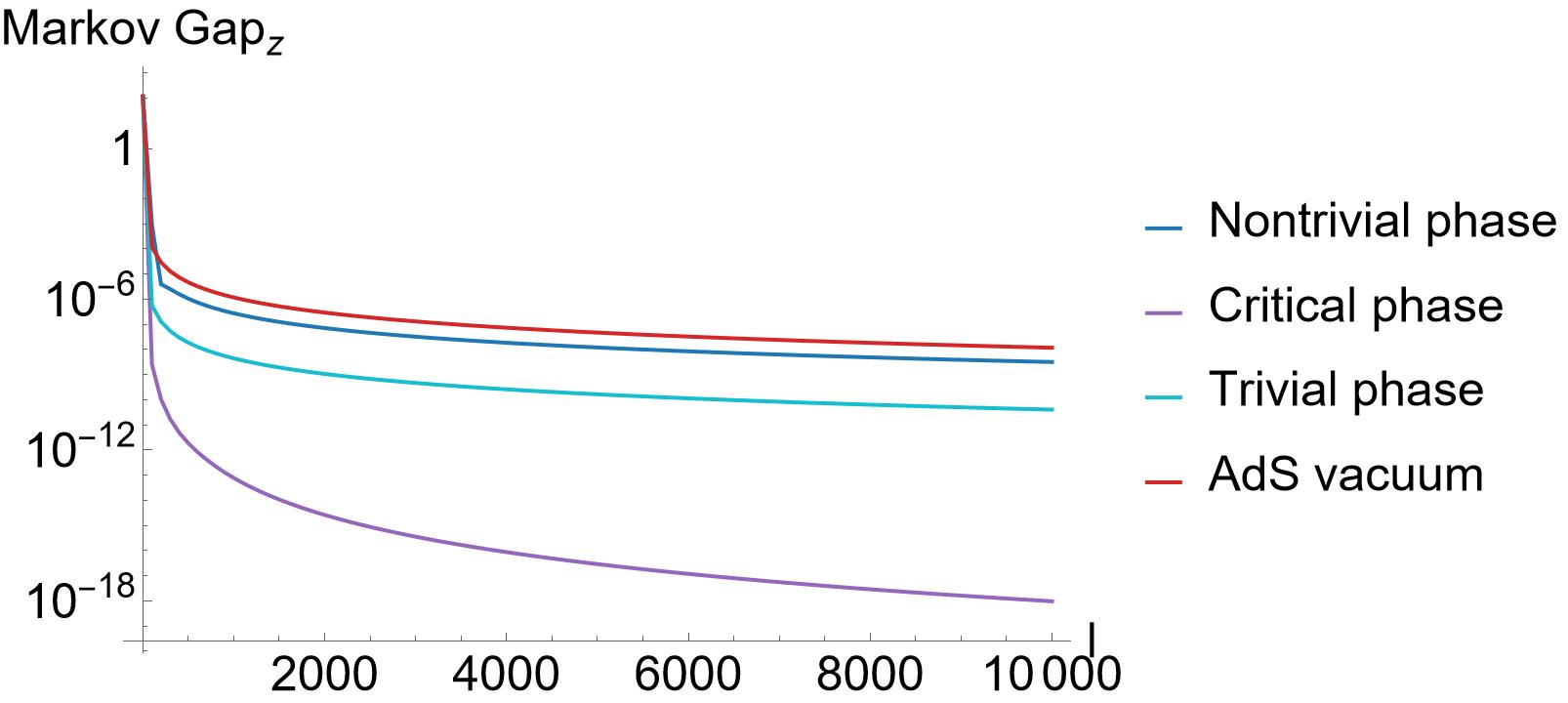}
        \small (b) Markov gap\(_z(l_z)\)
    \end{minipage}

   \caption[Markov gap as a function of strip width]{
The Markov gap as a function of strip width for four backgrounds: the topologically nontrivial phase, the critical phase, the topologically trivial phase, and the pure AdS background. The parameters for the nontrivial and trivial phases are \(M/b=(M/b)_c\mp0.1\). Without loss of generality, an appropriate ratio of $l_{\text{strip}}$ to $l_{\text{gap}}$ has been chosen to guarantee that the entanglement wedges between different regions remain connected as the scale increases. Left:
    Markov gap\(_x(l_x)\). Right: Markov gap\(_z(l_z)\). 
    }
    \label{fig:fixed_l_markovgap_comparison}
\end{figure}

\begin{figure}[htbp]
\centering
\begin{minipage}[t]{0.49\textwidth}
\centering
\includegraphics[width=\linewidth]{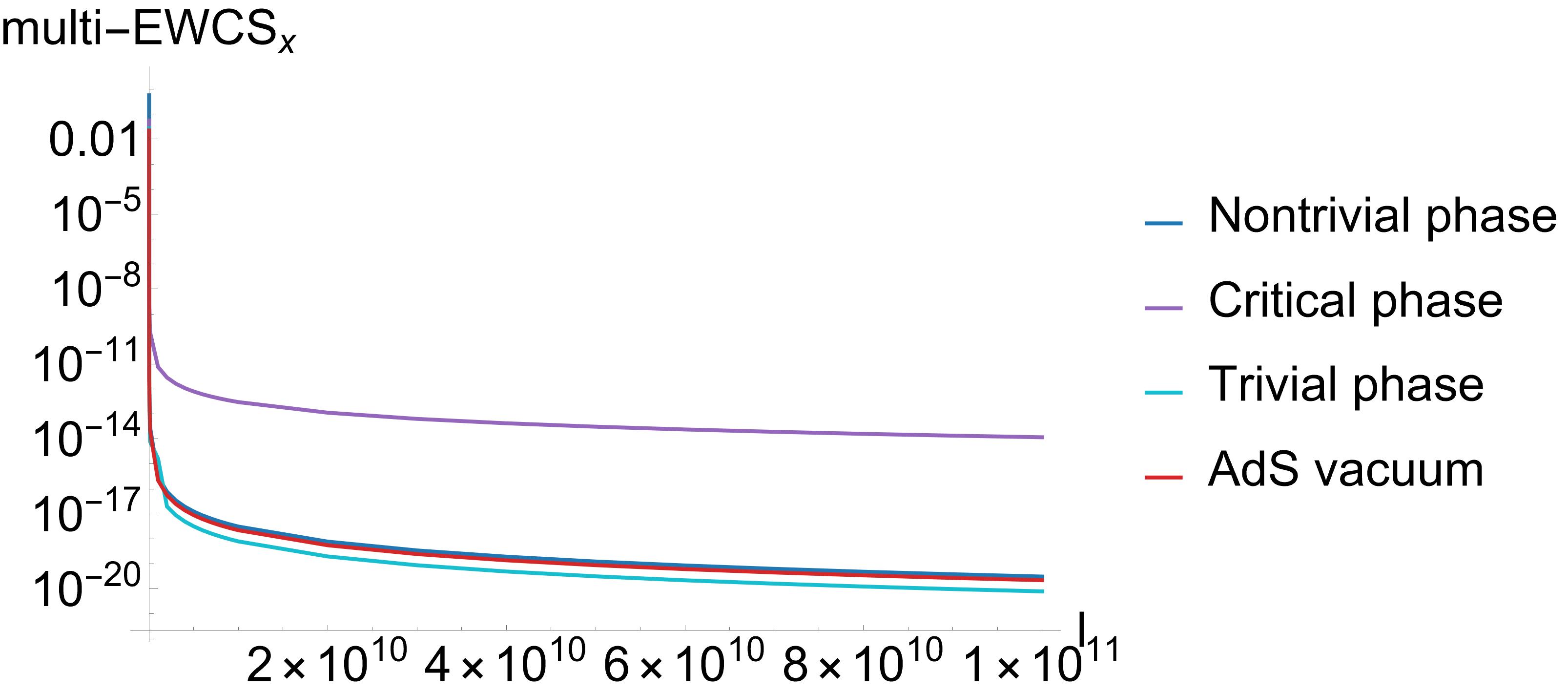}
\end{minipage}
\hfill
\begin{minipage}[t]{0.49\textwidth}
\centering
\includegraphics[width=\linewidth]{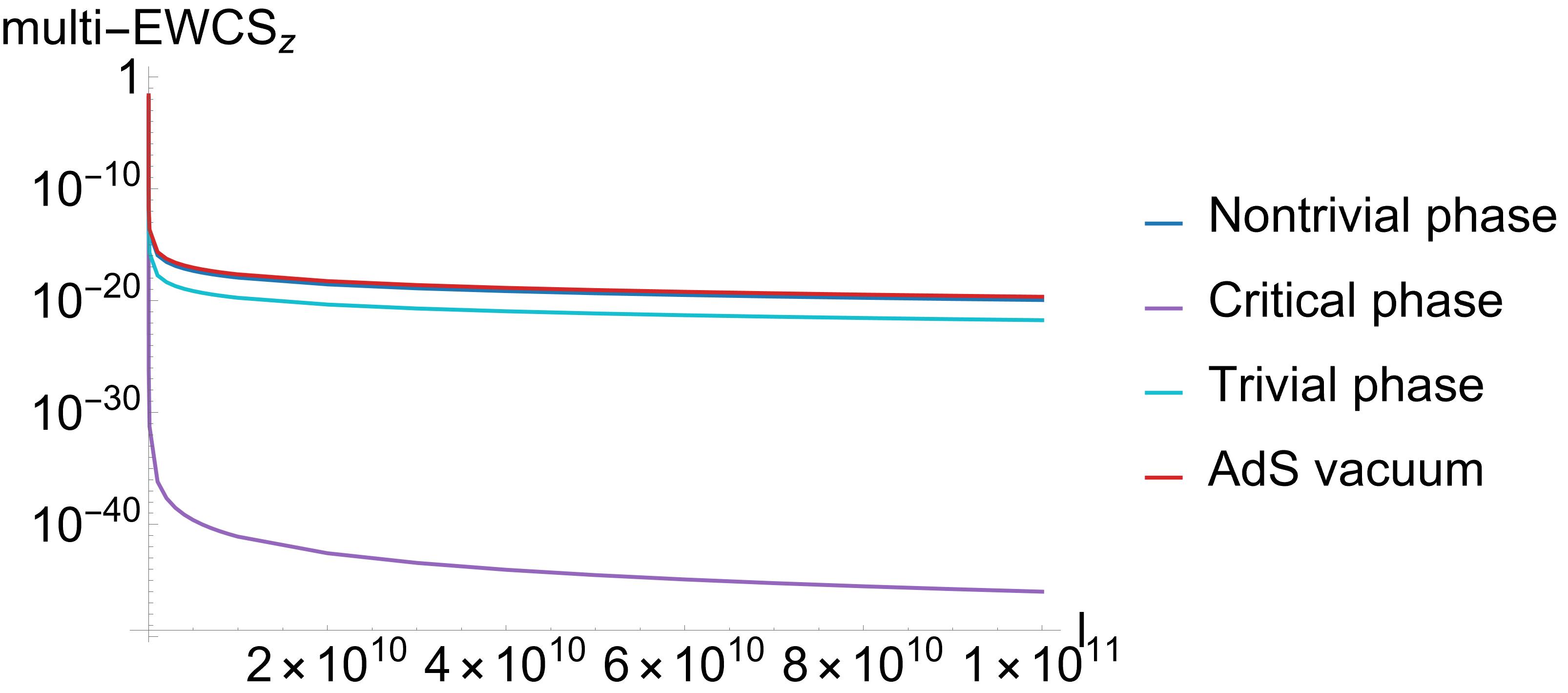}
\end{minipage}
\caption[multi-EWCS as a function of strip width]{
The multi-EWCS as a function of strip width for four backgrounds: the topologically nontrivial phase, the critical phase, the topologically trivial phase, and the pure AdS background. The parameters for the nontrivial and trivial phases are \(M/b=(M/b)_c\mp0.1\). Without loss of generality, we set $l_{\text{strip}} = 10 l_{\text{gap}}$. Left: multi-EWCS${}_x(l_x)$. Right:
multi-EWCS${}_z(l_z)$. 
}
\label{fig:b0p1_multiewcs_fixed_direction}
\end{figure}

\begin{figure}[htbp]
\centering
\begin{minipage}[t]{0.49\textwidth}
\centering
\includegraphics[width=\linewidth]{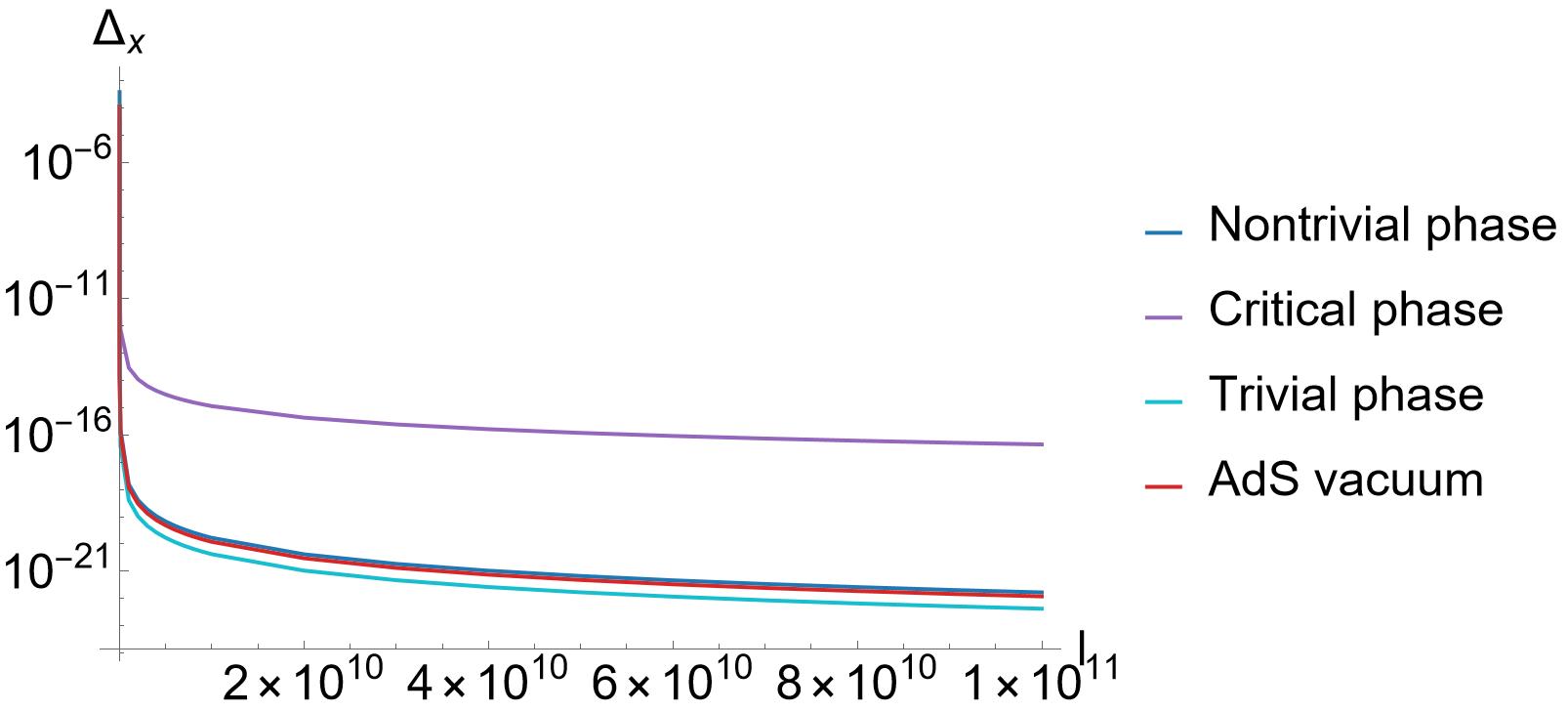}
\end{minipage}
\hfill
\begin{minipage}[t]{0.49\textwidth}
\centering
\includegraphics[width=\linewidth]{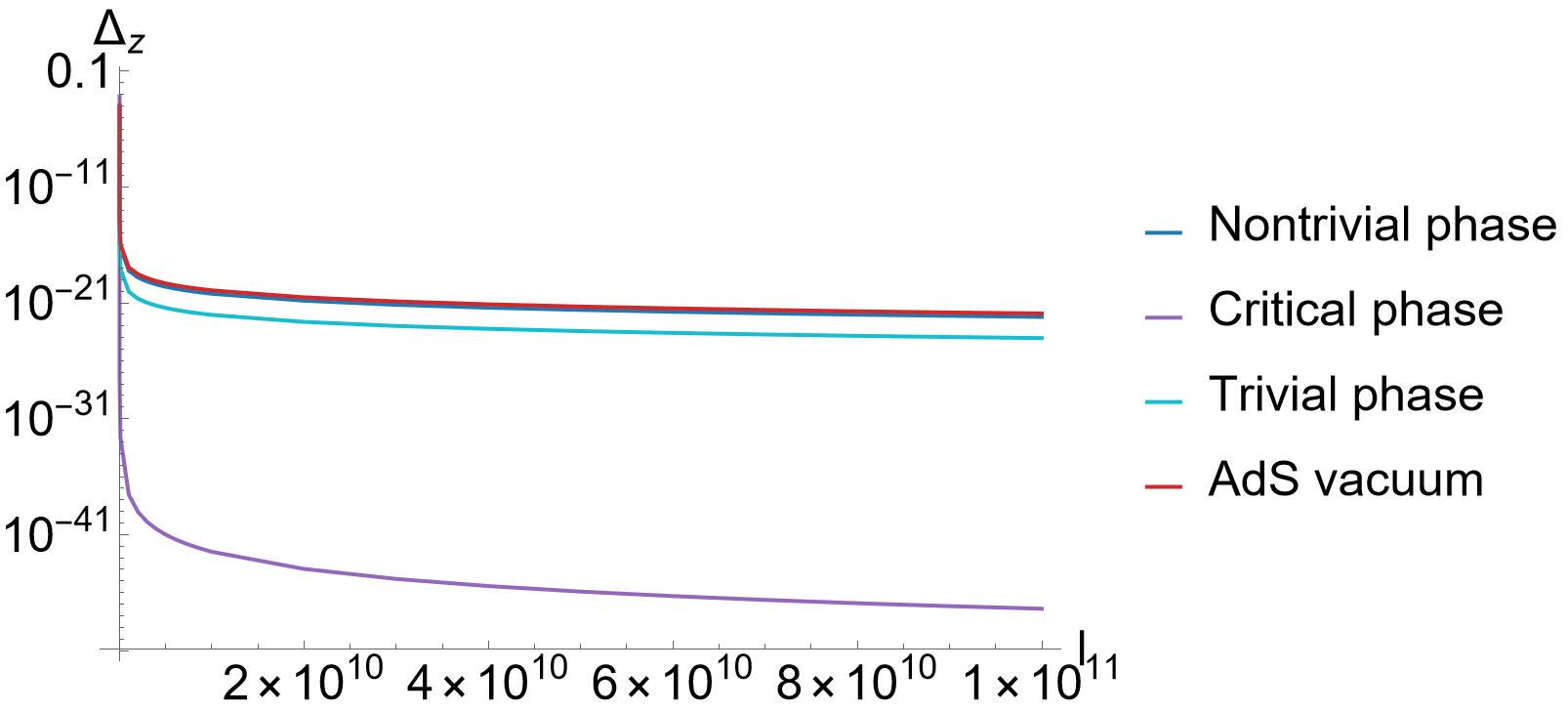}
\end{minipage}
\caption[$\Delta$ as a function of strip width]{The $\Delta$ as a function of strip width for four backgrounds: the topologically nontrivial phase, the critical phase, the topologically trivial phase, and the pure Ads background. The parameters for the nontrivial and trivial phases are \(M/b=(M/b)_c\mp0.1\). Without loss of generality, we set $l_{\text{strip}} = 10 l_{\text{gap}}$. Left: \(\Delta_x(l_x)\). Right: \(\Delta_z(l_z)\).
}
\label{fig:delta}
\end{figure}

\begin{figure}[htbp]
\centering
\begin{minipage}[t]{0.49\textwidth}
\centering
\includegraphics[width=\linewidth]{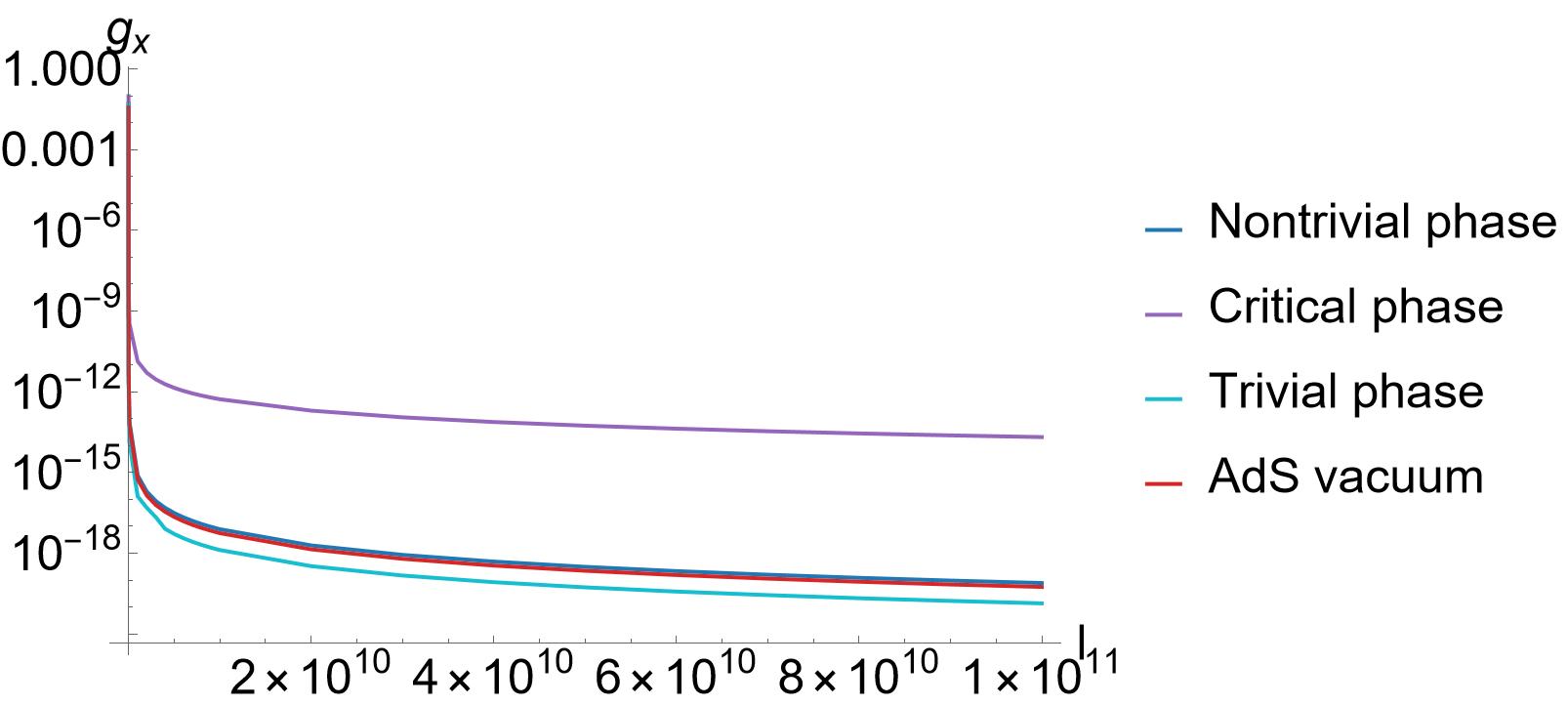}
\end{minipage}
\hfill
\begin{minipage}[t]{0.49\textwidth}
\centering
\includegraphics[width=\linewidth]{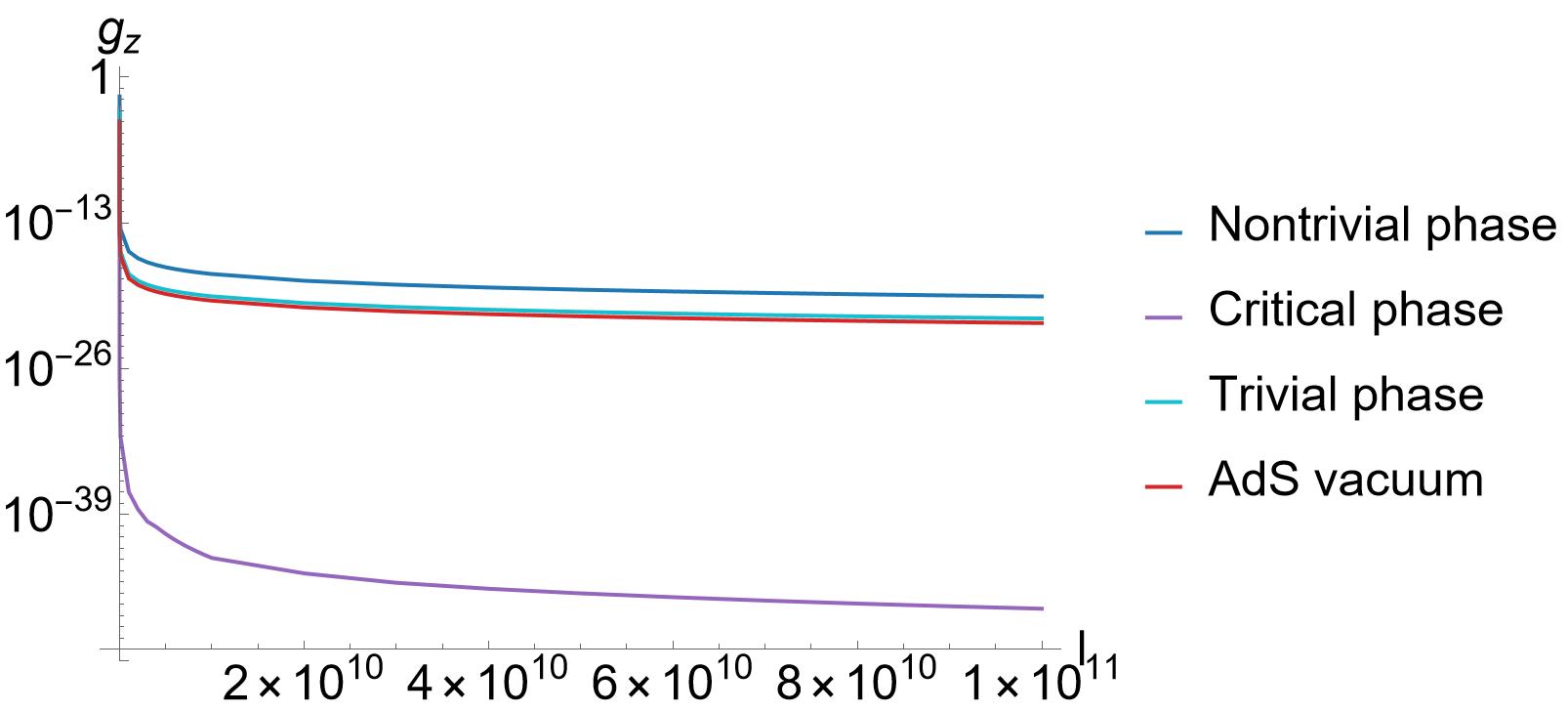}
\end{minipage}
\caption[$g$ as a function of strip width]{
The $g$ as a function of strip width for four
backgrounds: the topologically nontrivial phase, the critical phase, the
topologically trivial phase, and the pure AdS background. The parameters for the nontrivial and
trivial phases are \(M/b=(M/b)_c\mp0.1\). Without loss of generality, we set $l_{\text{strip}} = 10 l_{\text{gap}}$. Left: $g_x(l_x)$. Right:
$g_z(l_z)$. 
}
\label{fig:g}
\end{figure}

Figures~\ref{fig:fixed_l_cmi_comparison}--\ref{fig:g} show the corresponding behavior of CMI, EWCS, \(\kappa\), Markov gap, the multi-EWCS, \(\Delta\) and \(g\) in all three phases and the pure AdS background. These entanglement quantities and measures probe different types of entanglement structures of the strip subregions as introduced in Section~\ref{Section2}, and their behavior in the figures indicates that a large amount of multi-partite entanglement exists in all three phases of the holographic Weyl semimetal.  The dependence of these entanglement quantities and measures on the strip width follows similar qualitative patterns. At small strip width the curves are close to each other, showing the common UV behavior. As the strip width is increased, the values in different phases depart from each other, and the large width behavior shows distinct qualitative behavior. The detailed power-law behavior in this large width regime will be extracted in the next subsection.

Note that the quantities of CMI, EWCS, \(\kappa\) and Markov gap are tripartite quantities that capture various types of tripartite entanglement structure, as explained in Section~\ref{Section2}, while the quantities of multi-EWCS, \(\Delta\) and \(g\) are four-partite quantities that are more sensitive to four-partite quantum entanglement structures. The fact that all these quantities remain nonzero and show controlled power-law decay at large \(l\) indicates that the holographic Weyl semimetal contains long-distance multipartite correlations in several distinct geometric sectors. Their common large \(l\) behavior further shows that these sectors are governed by the same near-horizon IR physics.

\subsection{Scaling behavior at large strip width}
\label{subsec:large_width_scaling_fixed_direction}
Having displayed the strip width dependence of the entanglement quantities and measures in both the $x$ and $z$ directions, we now extract the large \(l\) behavior of them. This regime is the long-distance limit of the boundary entangling subregions. By the holographic UV/IR relation, increasing \(l_i\) moves the turning point of the corresponding RT surface deeper into the bulk~\cite{PhysRevD.59.065011}. The same is true for the multipartite wedge networks constructed from
these strip regions. Once the turning point lies in the near-horizon scaling region, the leading dependence on \(l_i\) is determined by the IR geometry of the corresponding zero temperature solution.

The purpose of this subsection is to determine these leading powers from the full numerical backgrounds. For each quantity \(X_i(l_i)\), with \(i=x,z\), we fit the data at sufficiently large \(l_i\) to obtain
\begin{equation}
  X_i(l_i)
  \sim
  A_i l_i^{b_i}.
\end{equation}
The fit is restricted to the range in which the turning point \(r_*\) is already close enough to the near-horizon region for the leading IR scaling to be visible. When a quantity approaches a finite limiting value, as happens for the holographic \(c\)-function in the nontrivial and trivial phases, we record the leading behavior as a plateau, \(l_i^0\).

The difference between the nontrivial phase, the trivial phase, and the critical solution can then be understood from the leading powers of the IR metric. In both the nontrivial phase and the trivial phase, the strict IR metric has the AdS form at leading order. Therefore these two phases give the same leading powers at large \(l_i\). Their difference is not reflected in the exponent \(b_i\), but in the fitted coefficient \(A_i\), in the scale at which the asymptotic regime becomes visible, and in the relative values of the quantities for strips along the \(x\) and \(z\) directions. At the critical point, by contrast, the near-horizon geometry is anisotropic, with \(u(r)\sim r^2\) and \(h(r)\sim r^{2\beta}\). The exponent \(\beta\) modifies the relation between \(l_i\) and \(r_*\), especially for strips along the \(z\) direction. As a result, the critical solution has different leading powers from the nontrivial and trivial phases. The resulting large \(l\) scaling behavior is summarized in Table~\ref{tab:fixed_direction_scaling}.

\begin{table}[htbp]
\centering
\small
\setlength{\tabcolsep}{5pt}
\renewcommand{\arraystretch}{1.2}
\begin{tabular}{@{}c c c c c c c@{}}
\hline
Phase & Direction & \(c_i\) & CMI & EWCS & \(\kappa\) & MG / multi-EWCS \\
\hline
nontrivial or trivial & \(x\) &
\(l_x^{0}\) &
\(l_x^{-4}\) &
\(l_x^{-2}\) &
\(l_x^{-2}\) &
\(l_x^{-2}\) \\
nontrivial or trivial & \(z\) &
\(l_z^{0}\) &
\(l_z^{-4}\) &
\(l_z^{-2}\) &
\(l_z^{-2}\) &
\(l_z^{-2}\) \\
critical & \(x\) &
\(l_x^{1-\beta}\) &
\(l_x^{-3-\beta}\) &
\(l_x^{-1-\beta}\) &
\(l_x^{-1-\beta}\) &
\(l_x^{-1-\beta}\) \\
critical & \(z\) &
\(l_z^{2-2/\beta}\) &
\(l_z^{-2-2/\beta}\) &
\(l_z^{-2/\beta}\) &
\(l_z^{-2/\beta}\) &
\(l_z^{-2/\beta}\) \\
\hline
\end{tabular}
\caption{
Scaling behavior at large strip width of the entanglement quantities and measures in both the $x$ and $z$ directions. MG denotes the Markov gap. The entries summarize the leading powers at large \(l_i\)
obtained from numerical fits organized by the IR scaling exponent. For the
noncritical \(c\)-function, \(l_i^0\) denotes a long-distance plateau. At the
critical point, \(\beta=0.407\) determined from \(h(r)\sim r^{2\beta}\). 
}
\label{tab:fixed_direction_scaling}
\end{table}

This large \(l_i\) behavior has a direct physical meaning. Since the system is gapless in the IR, the long-distance entanglement structures are not cut off by a finite correlation length. Accordingly, all the tripartite and four-partite entanglement quantities considered above vanish at large \(l_i\) through power laws rather than through exponential suppression. The corresponding powers are fixed by the near-horizon scaling. The nontrivial and trivial phases have the same leading AdS-type IR metric powers, while the critical solution has an anisotropic scaling geometry and therefore gives different powers in the \(x\) and \(z\) directions. Thus the large \(l_i\) behavior separates the critical solution from the two phases away from the critical point and shows that these entanglement quantities can characterize the topological quantum phase transition. In the next subsection we fix the strip width and vary \(M/b\), in order to see how the same quantities change across the phase diagram.

\subsection{Entanglement measures over the phase diagram}
\label{subsec:fixed_width_phase_scan}
\begin{figure}[htbp]
\centering
\begin{minipage}[t]{0.49\textwidth}
\centering
\includegraphics[width=\linewidth]{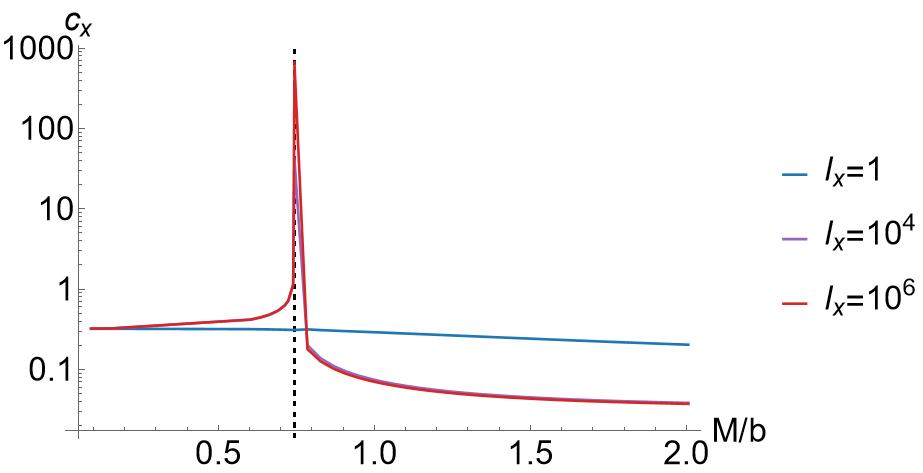}
\end{minipage}
\hfill
\begin{minipage}[t]{0.49\textwidth}
\centering
\includegraphics[width=\linewidth]{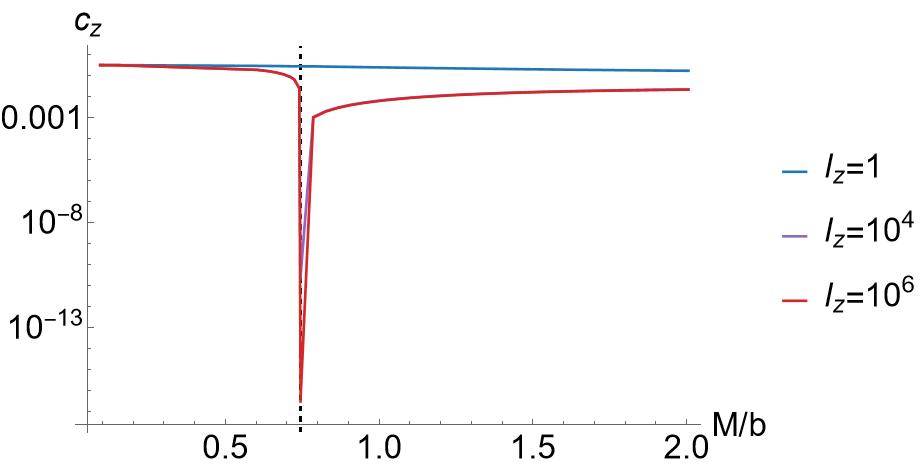}
\end{minipage}
\caption{
The holographic \(c\)-function \(c_x\) (left) and \(c_z\) (right) in the
holographic Weyl semimetal at zero temperature as functions of \(M/b\) for several fixed strip widths \(l_x\) and \(l_z\). Different colors correspond to different strip widths, and the dashed line denotes the critical value
\((M/b)_c \simeq 0.744\).
}
\label{fig:moverb_cfunction_fixed_direction}
\end{figure}

We next fix the strip width and vary the dimensionless ratio \(M/b\) across the zero temperature phase diagram. The large \(l\) analysis above identifies the range in which the geometric objects entering the entanglement quantities are controlled by the IR geometry. For small \(l_i\), with \(i=x,z\), the corresponding RT surfaces, entanglement wedges, and multipartite networks remain close to the common asymptotically AdS region, and their dependence on \(M/b\) is weak. For sufficiently large \(l_i\), they extend into the interior IR geometry. In this regime, the dependence of the entanglement quantities on \(M/b\) develops a clear feature near the critical value and can be used to identify the quantum topological transition.

\paragraph{\texorpdfstring{\(c\)-function}{c-function}}
We begin with the holographic \(c\)-function. Since it is obtained from the first derivative of the strip entanglement entropy, it gives the most direct test of how the RT surface changes as \(M/b\) is varied. Figure~\ref{fig:moverb_cfunction_fixed_direction} shows the \(c\)-function as a function of \(M/b\) for several fixed values of \(l_x\) and \(l_z\). For small \(l_x\) and \(l_z\), the curves for different values of \(M/b\) almost overlap, as expected from their common UV behavior. As the strip width is increased, the dependence on \(M/b\) becomes sharper, and a visible feature appears near \((M/b)_c\simeq0.744\). This is the regime in which the RT surfaces reach deeply enough into the bulk to distinguish the nontrivial phase, the critical solution, and the trivial phase.

The relative value of the \(c\)-function in the three phases depends on whether the strip is along the \(x\) or \(z\) directions. For strips in the \(x\) direction, the critical point has the largest value, and the nontrivial phase lies above the trivial phase in the numerical large \(l_x\) range. For strips along \(z\), the nontrivial phase is the largest, the trivial phase lies in between, and the critical point value is the smallest. The \(c\)-function therefore already shows that the phase dependence
is anisotropic.

\paragraph{CMI}

\begin{figure}[htbp]
\centering
\begin{minipage}[t]{0.49\textwidth}
\centering
\includegraphics[width=\linewidth]{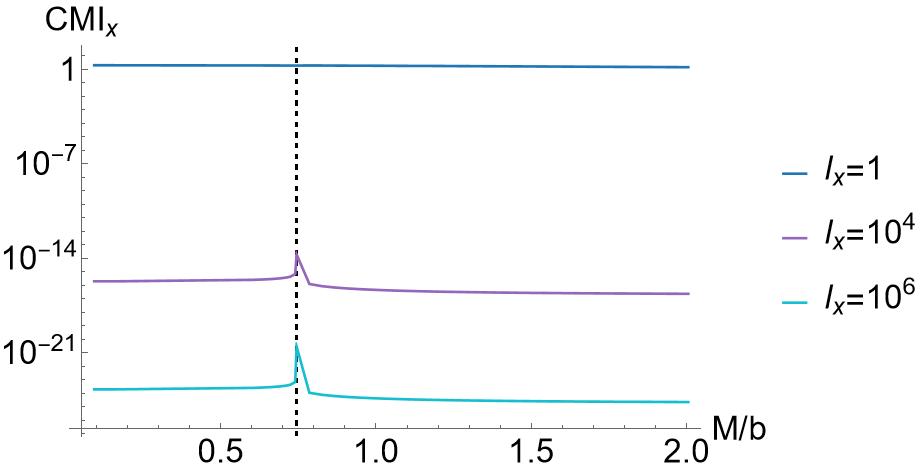}
\end{minipage}
\hfill
\begin{minipage}[t]{0.49\textwidth}
\centering
\includegraphics[width=\linewidth]{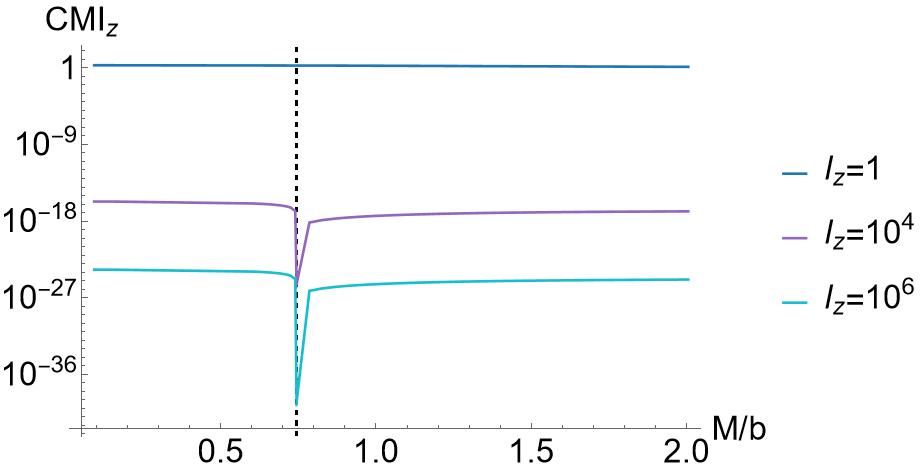}
\end{minipage}
\caption{
The CMI \(\mathrm{CMI}_x\) (left) and \(\mathrm{CMI}_z\) (right) in the holographic Weyl semimetal at zero temperature as functions of \(M/b\) for several fixed strip widths \(l_x\) and \(l_z\). Different colors correspond to different strip widths, and the dashed line denotes the critical value \((M/b)_c \simeq 0.744\).
}
\label{fig:moverb_cmi_fixed_direction}
\end{figure}

Figure~\ref{fig:moverb_cmi_fixed_direction} shows the CMI for two infinitesimal subregions conditioned on the strip between them as a function of \(M/b\) for several fixed values of \(l_x\) and \(l_z\). For small strip widths, the curves are nearly insensitive to \(M/b\), reflecting the fact that the corresponding surfaces mainly sample the common UV region. As \(l_i\) increases, the CMI develops a clear feature near \((M/b)_c\). The relative value in the nontrivial, critical, and trivial phases follows the same \(x\)- and \(z\)-dependent pattern found for the \(c\)-function.

\paragraph{EWCS}

\begin{figure}[htbp]
\centering
\begin{minipage}[t]{0.49\textwidth}
\centering
\includegraphics[width=\linewidth]{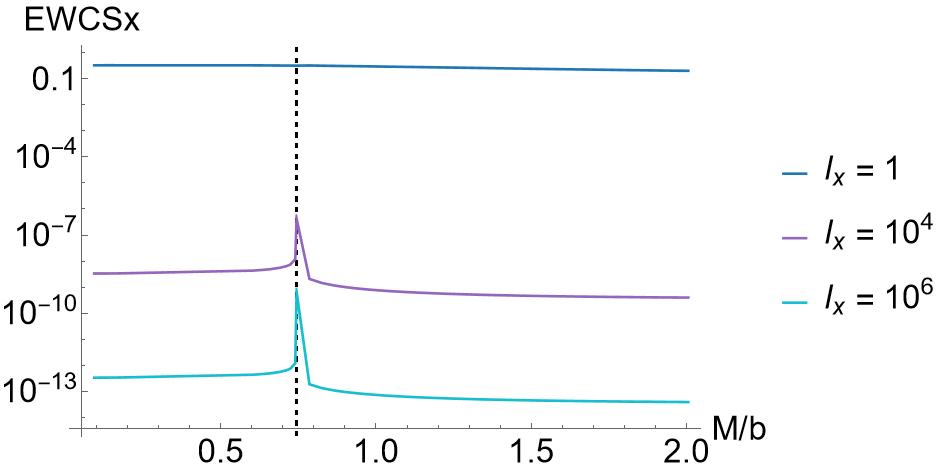}
\end{minipage}
\hfill
\begin{minipage}[t]{0.49\textwidth}
\centering
\includegraphics[width=\linewidth]{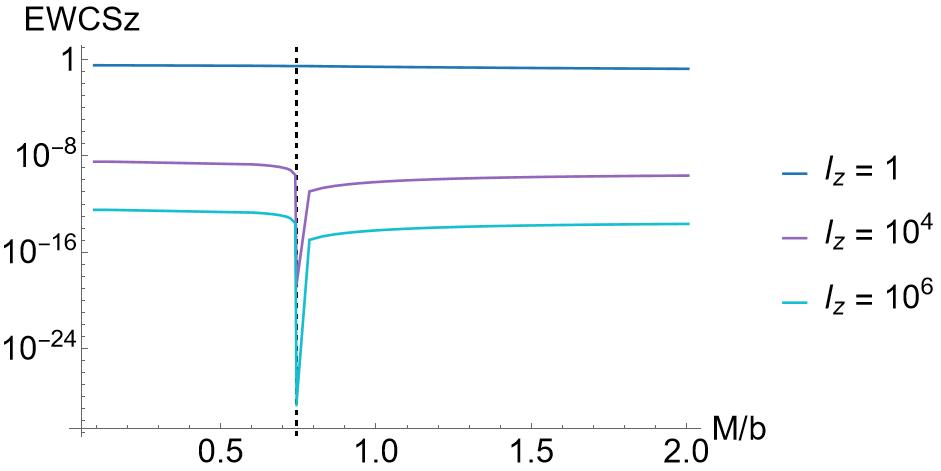}
\end{minipage}
\caption{
The EWCS \(\mathrm{EWCS}_x\) (left) and
\(\mathrm{EWCS}_z\) (right) in the holographic Weyl semimetal at zero
temperature as functions of \(M/b\) for several fixed strip widths \(l_x\)
and \(l_z\). Different colors correspond to different strip widths, and the
dashed line denotes the critical value \((M/b)_c \simeq 0.744\). Without loss of generality, an appropriate ratio of $l_{\text{strip}}$ to $l_{\text{gap}}$ has been chosen to guarantee that the entanglement wedges between different regions remain connected as the scale increases.
}
\label{fig:moverb_ewcs_fixed_direction}
\end{figure}

Figure~\ref{fig:moverb_ewcs_fixed_direction} shows the corresponding behavior of the EWCS, which is similar to the behavior of the two quantities above. The same difference between strips along the \(x\) and \(z\) directions is visible. It clearly shows that as an IR quantity, EWCS itself could serve as the nonlocal order parameter of quantum topological phase transition in the holographic Weyl semimetal. 

\paragraph{\texorpdfstring{\(\boldsymbol{\kappa}\)}{kappa}}

\begin{figure}[htbp]
\centering
\begin{minipage}[t]{0.49\textwidth}
\centering
\includegraphics[width=\linewidth]{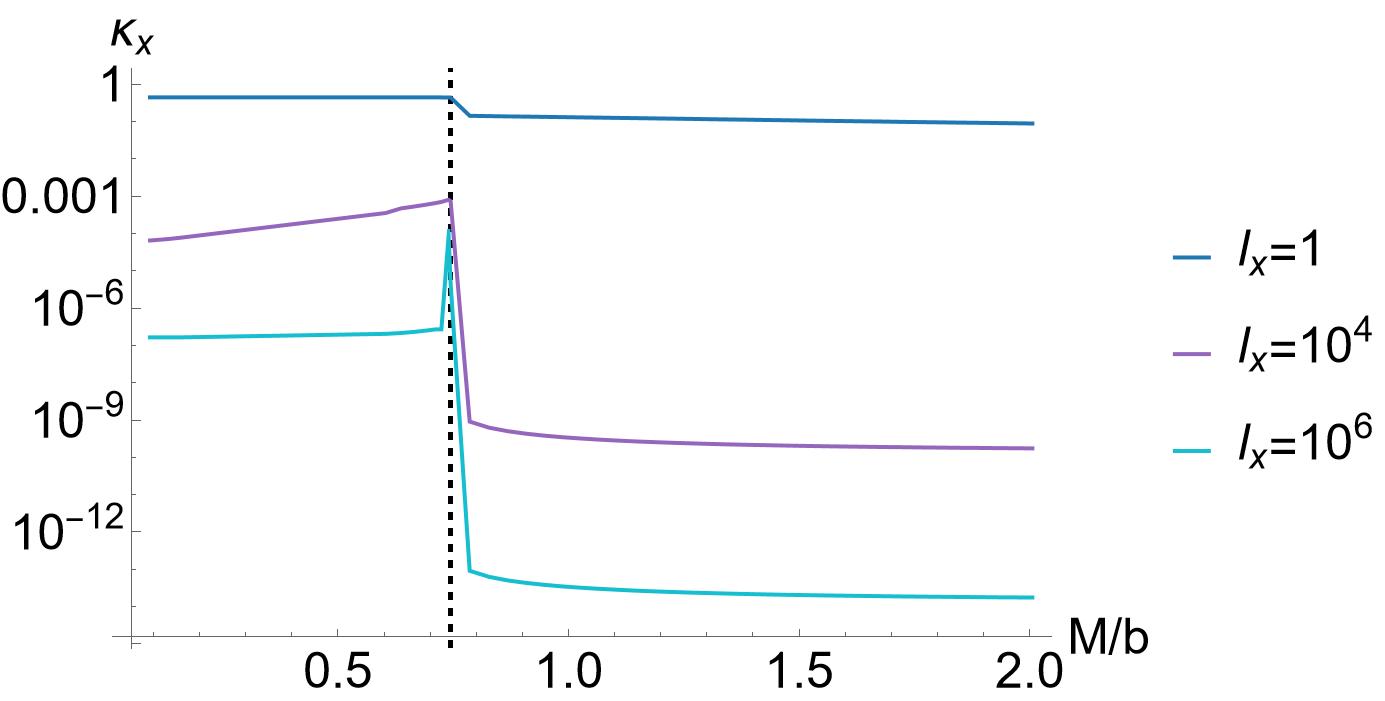}
\end{minipage}
\hfill
\begin{minipage}[t]{0.49\textwidth}
\centering
\includegraphics[width=\linewidth]{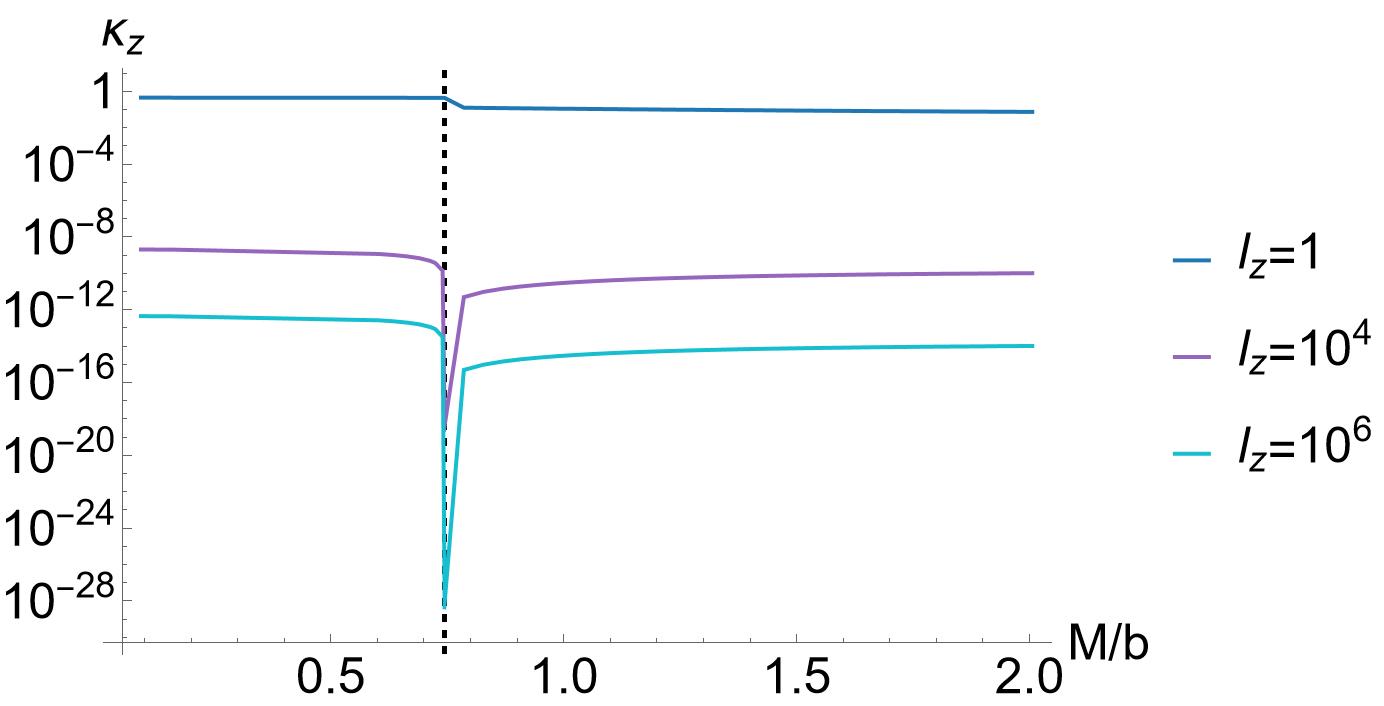}
\end{minipage}
\caption{
\(\kappa_x\) (left) and \(\kappa_z\) (right) in the holographic Weyl semimetal at zero temperature as functions of \(M/b\) for several fixed strip widths \(l_x\) and \(l_z\). Different colors correspond to different strip widths, and the dashed line denotes the critical value
\((M/b)_c \simeq 0.744\).
}
\label{fig:moverb_kappa_fixed_direction}
\end{figure}

Figure~\ref{fig:moverb_kappa_fixed_direction} shows the result for \(\kappa\), which was defined in \ref{KappaDefinition}. The behavior of \(\kappa\) again is similar to other entanglement quantities above. The relative behavior of the three phases again depends on whether the strip is finite along \(x\) or along \(z\). Clearly, the tripartite entanglement structure captured by \(\kappa\) could serve as a nonlocal order parameter for the topological phase transition in the holographic Weyl semimetal, too.

\paragraph{Markov gap}

\begin{figure}[htbp]
\centering
\begin{minipage}[t]{0.49\textwidth}
\centering
\includegraphics[width=\linewidth]{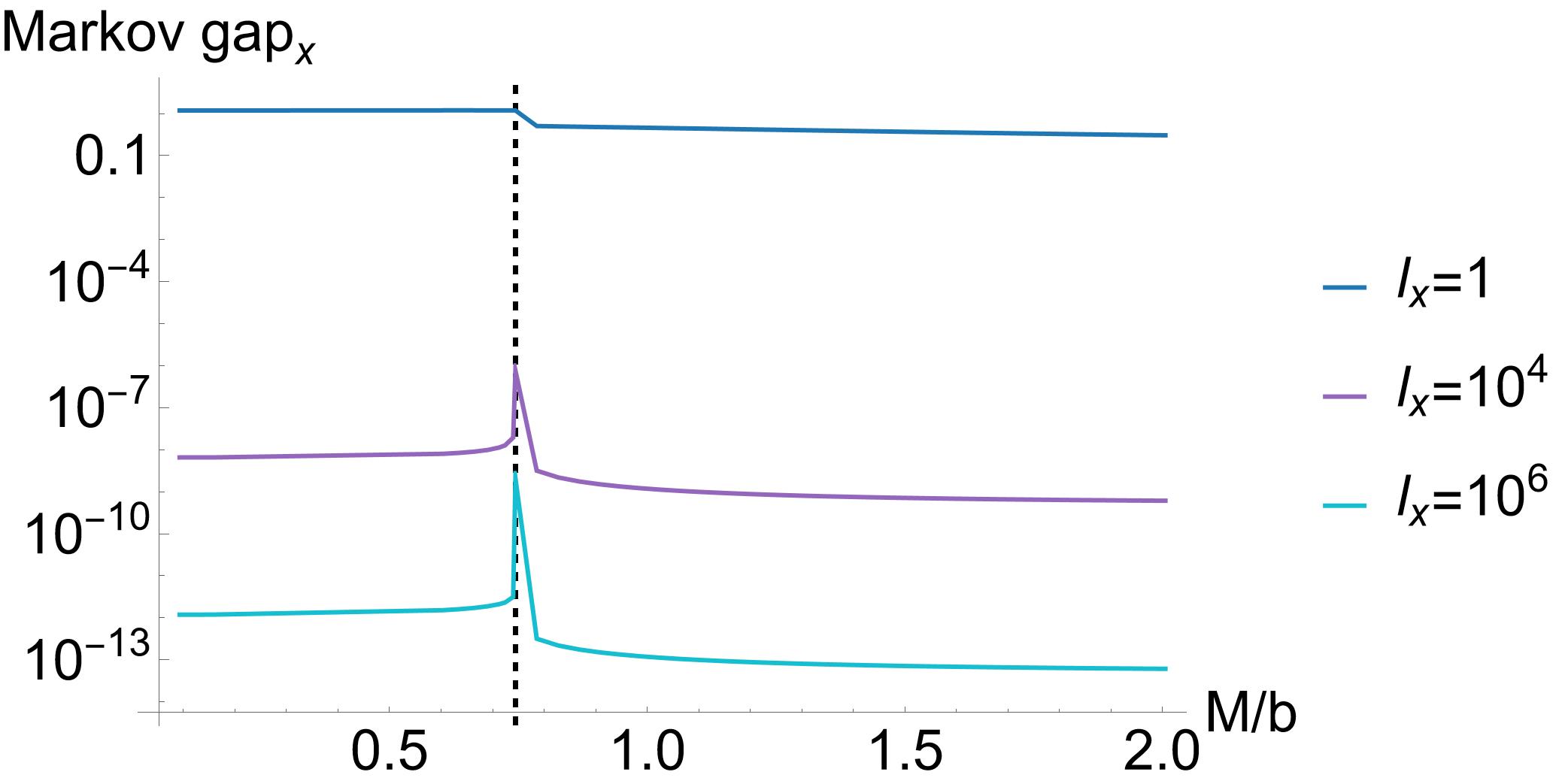}
\end{minipage}
\hfill
\begin{minipage}[t]{0.49\textwidth}
\centering
\includegraphics[width=\linewidth]{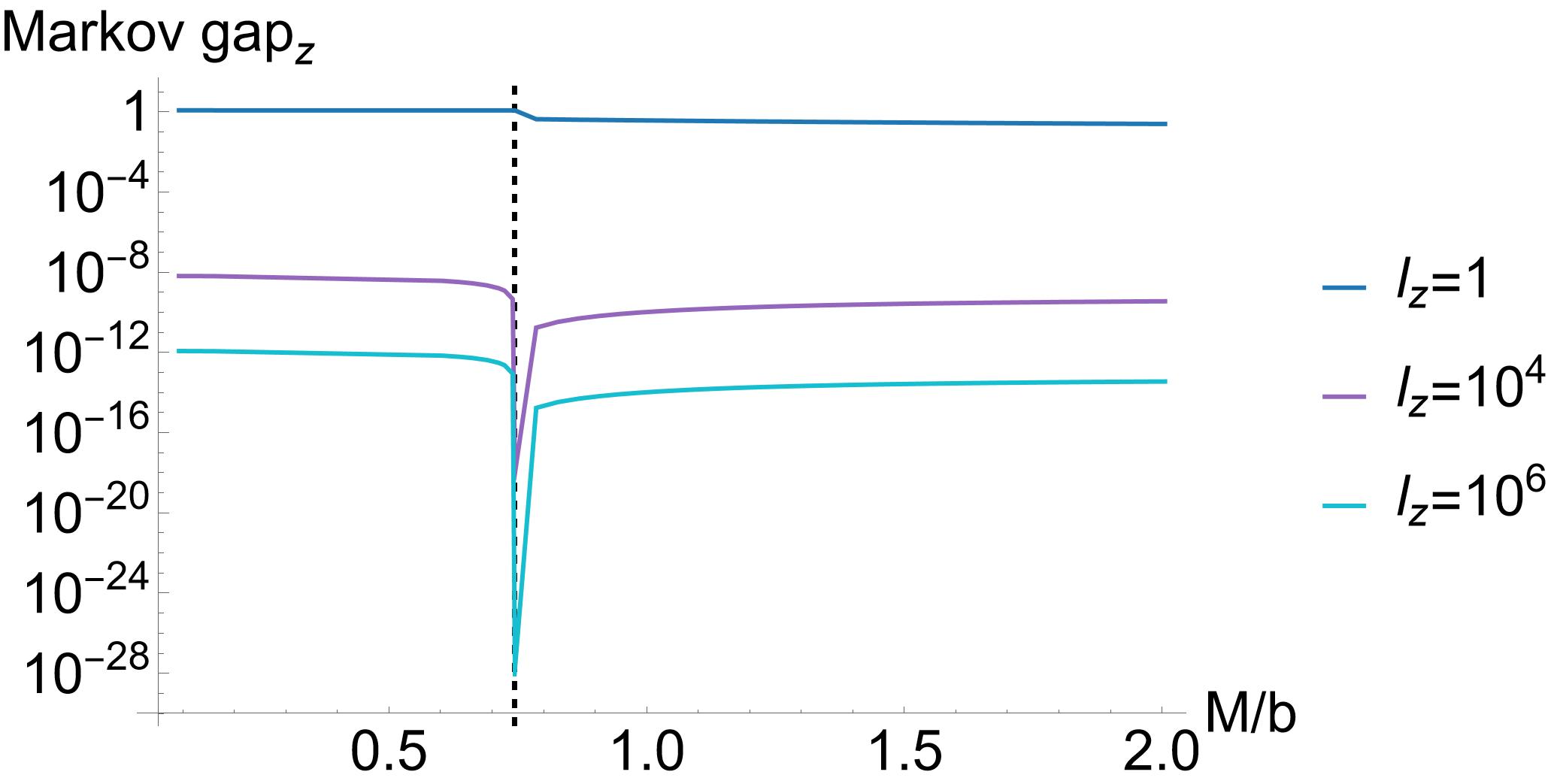}
\end{minipage}
\caption{
The Markov gap \(\mathrm{MG}_x\) (left) and \(\mathrm{MG}_z\) (right) in
the holographic Weyl semimetal at zero temperature as functions of \(M/b\)
for several fixed strip widths \(l_x\) and \(l_z\). Different colors
correspond to different strip widths, and the dashed line denotes the
critical value \((M/b)_c \simeq 0.744\). Without loss of generality, an appropriate ratio of $l_{\text{strip}}$ to $l_{\text{gap}}$ has been chosen to guarantee that the entanglement wedges between different regions remain connected as the scale increases.
}
\label{fig:moverb_markovgap_fixed_direction}
\end{figure}

Figure~\ref{fig:moverb_markovgap_fixed_direction} shows the result for the Markov gap, which was defined in Eq.~\eqref{MarkovGapDefinition}. Its behavior is again similar to that of the entanglement quantities discussed above. The relative behavior of the nontrivial phase, the critical solution, and the trivial phase depends on whether the strip is finite along \(x\) or along \(z\). Therefore, the part of the tripartite structure measured by the Markov gap also carries clear information about the topological phase transition in the holographic Weyl semimetal. 

\paragraph{multi-EWCS}

\begin{figure}[htbp]
\centering
\begin{minipage}[t]{0.49\textwidth}
\centering
\includegraphics[width=\linewidth]{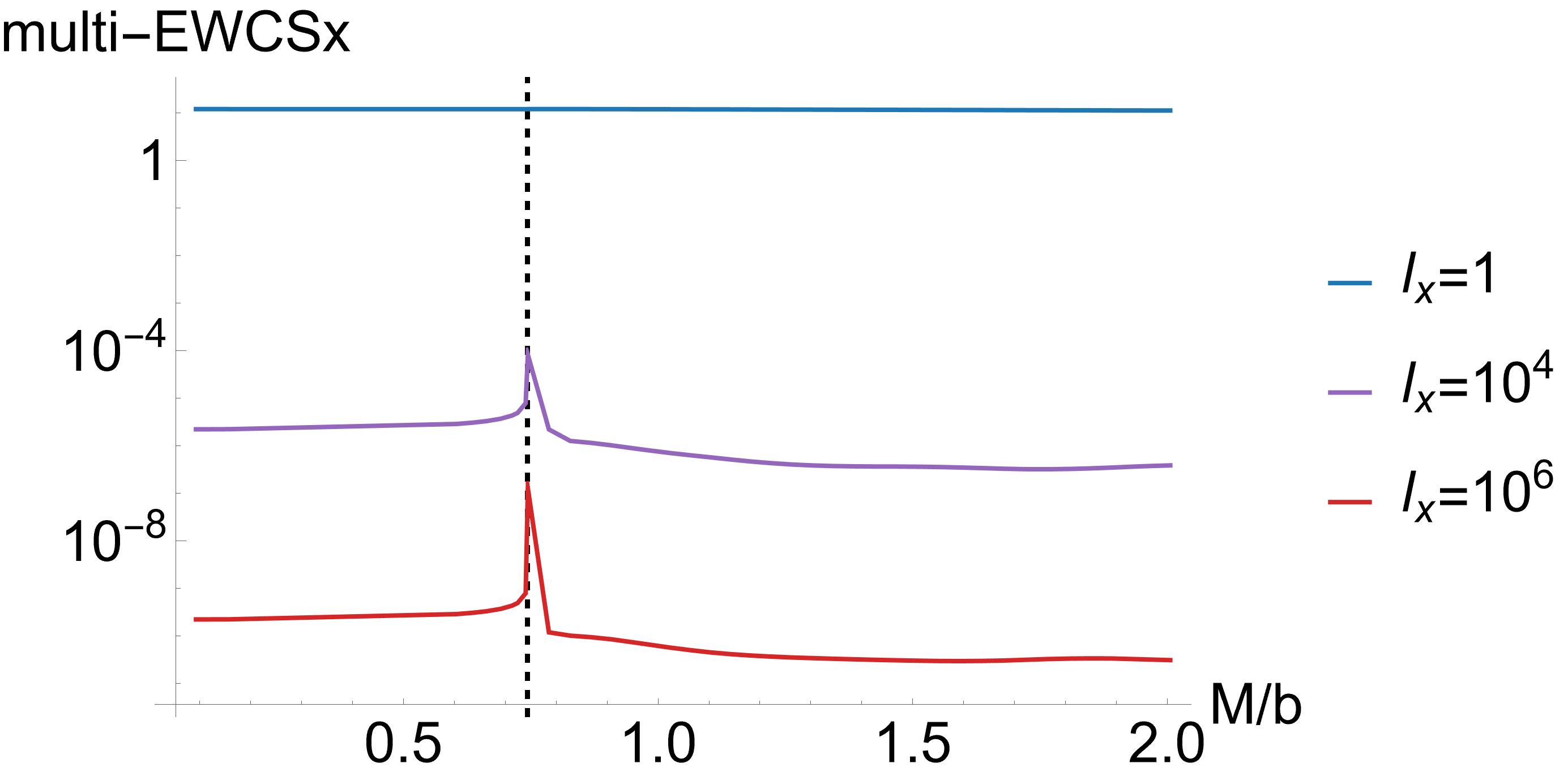}
\end{minipage}
\hfill
\begin{minipage}[t]{0.49\textwidth}
\centering
\includegraphics[width=\linewidth]{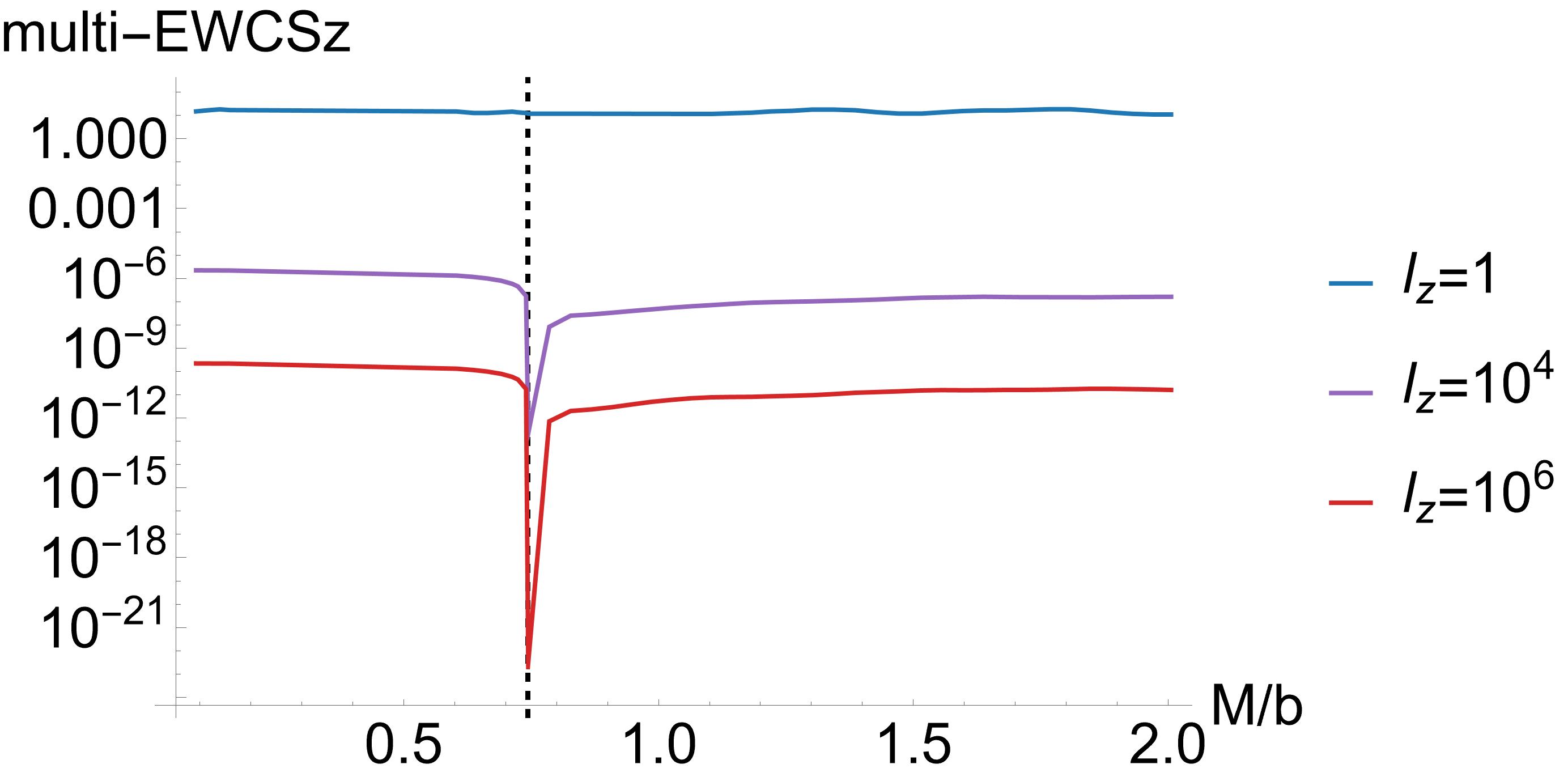}
\end{minipage}
\caption{
The multi-EWCS \(\mathrm{multi-EWCS}_x\) (left) and
\(\mathrm{multi-EWCS}_z\) (right) in the holographic Weyl semimetal at zero
temperature as functions of \(M/b\) for several fixed strip widths \(l_x\)
and \(l_z\). Different colors correspond to different strip widths, and
the dashed line denotes the critical value \((M/b)_c \simeq 0.744\). Without loss of generality, we set $l_{\text{strip}} = 10 l_{\text{gap}}$.
}
\label{fig:moverb_multiewcs_fixed_direction}
\end{figure}

Figure~\ref{fig:moverb_multiewcs_fixed_direction} shows the result for the multi-EWCS, which was defined in Eq.~\eqref{MultiEWCSDefinition}. Its behavior again follows the same pattern as the quantities considered above. The relative behavior of the nontrivial phase, the critical solution, and the trivial phase depends on whether the strip is finite along \(x\) or along \(z\). Thus the multipartite wedge geometry captured by the multi-EWCS also provides a nonlocal indicator of the topological phase transition in the holographic Weyl semimetal. This agreement shows that the anisotropic phase dependence is not specific to bipartite or tripartite quantities, but persists at the level of a higher multipartite wedge construction.

\paragraph{\texorpdfstring{\(\boldsymbol{\Delta}\)}{delta}}

\begin{figure}[htbp]
\centering
\begin{minipage}[t]{0.49\textwidth}
\centering
\includegraphics[width=\linewidth]{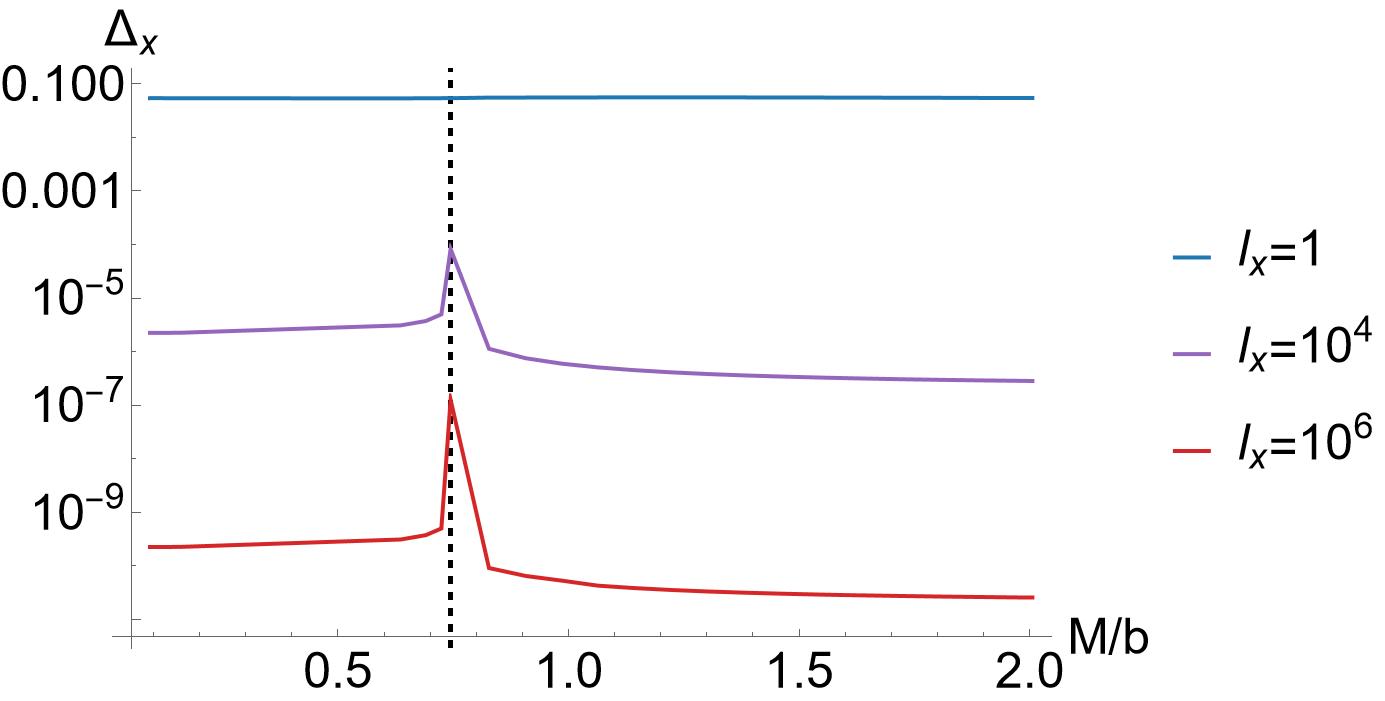}
\end{minipage}
\hfill
\begin{minipage}[t]{0.49\textwidth}
\centering
\includegraphics[width=\linewidth]{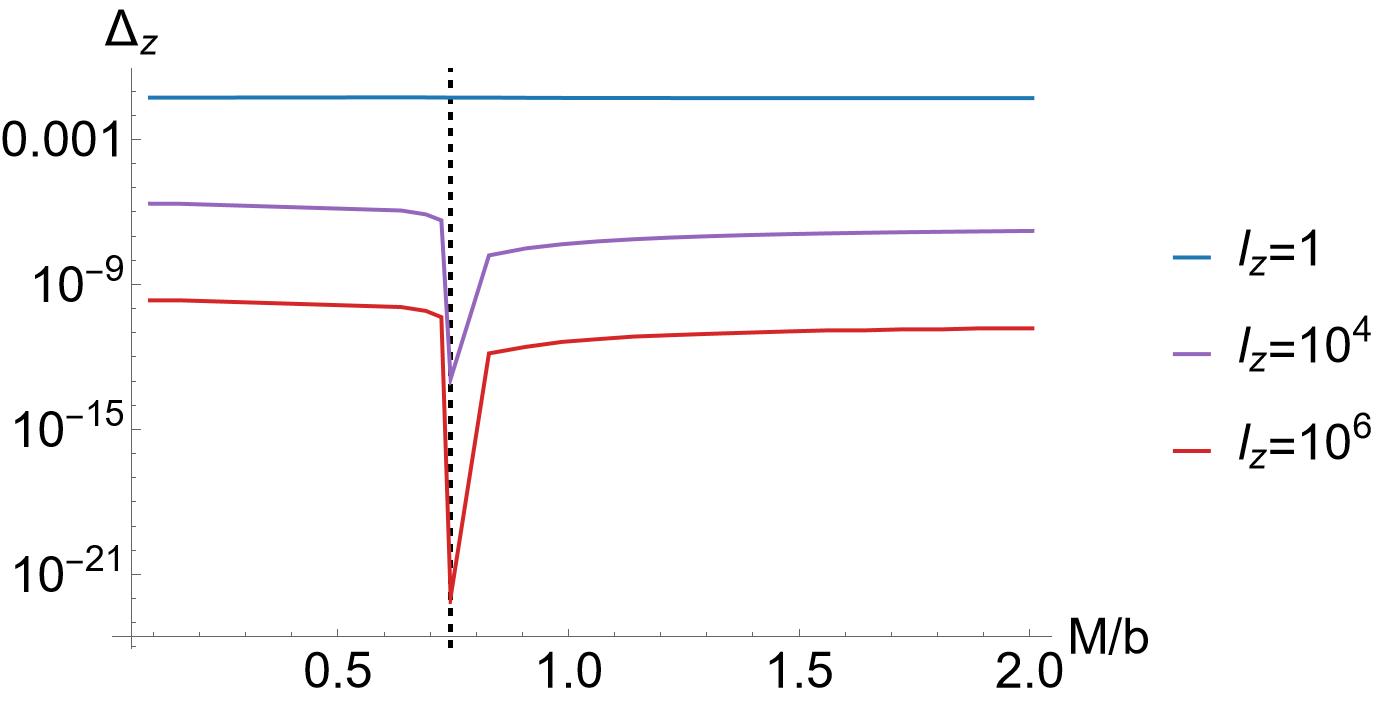}
\end{minipage}
\caption{
The four-partite signal $\Delta_W$.  \(\Delta_x\)(left) and
\(\Delta_z\) (right) in the holographic Weyl semimetal at zero
temperature as functions of \(M/b\) for several fixed strip widths \(l_x\)
and \(l_z\). Different colors correspond to different strip widths, and
the dashed line denotes the critical value \((M/b)_c \simeq 0.744\). Without loss of generality, we set $l_{\text{strip}} = 10 l_{\text{gap}}$.
}
\label{fig:moverb_delta_fixed_direction}
\end{figure}

Figure~\ref{fig:moverb_delta_fixed_direction} shows the result for the multi-EWCS based signal \(\Delta\). For small strip widths, the curves are almost insensitive to \(M/b\), as in the quantities discussed above. As the strip width is increased, \(\Delta\) develops a sharp feature near \((M/b)_c\). The form of this feature is direction dependent. In the \(x\) direction, \(\Delta_x\) is enhanced near the critical point, while in the \(z\) direction \(\Delta_z\) is strongly suppressed there. Thus the residual multipartite contribution isolated by \(\Delta\) also detects the topological transition, and it carries the same anisotropic phase dependence as the underlying wedge geometry.

\paragraph{$g$}

\begin{figure}[htbp]
\centering
\begin{minipage}[t]{0.49\textwidth}
\centering
\includegraphics[width=\linewidth]{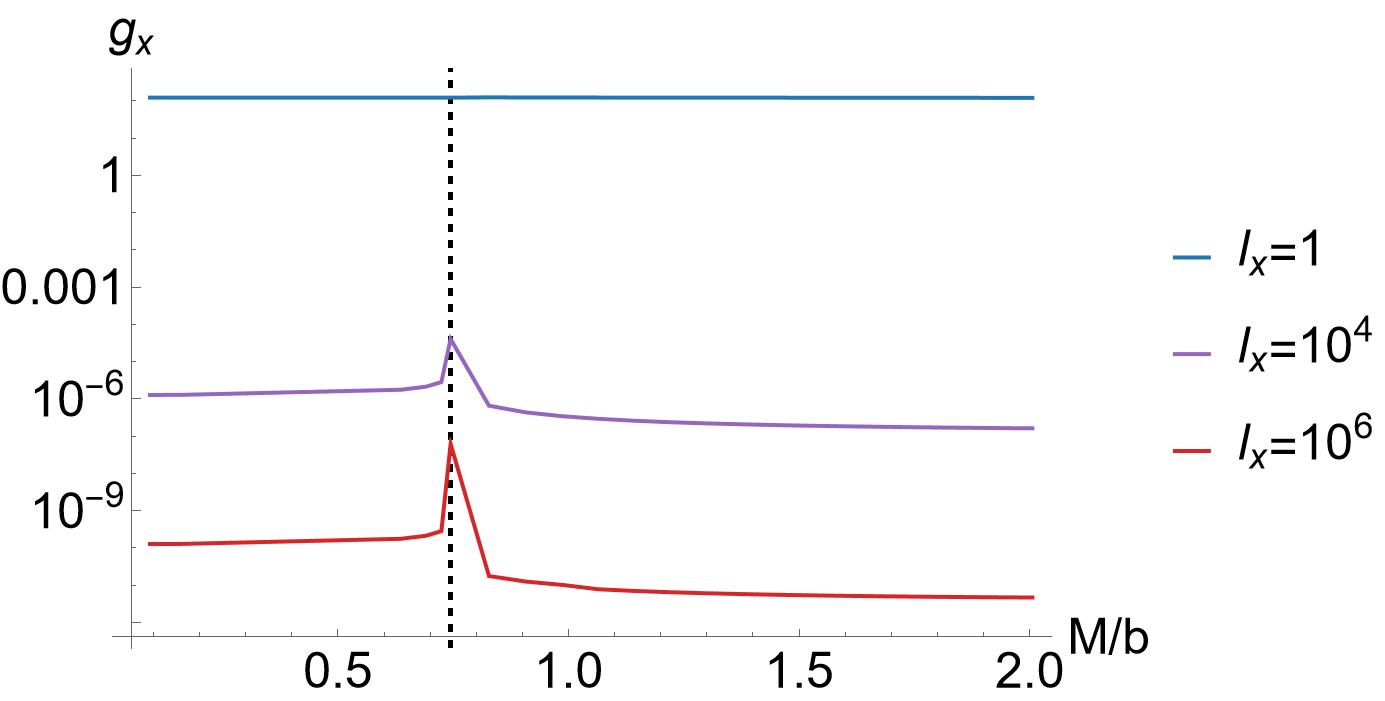}
\end{minipage}
\hfill
\begin{minipage}[t]{0.49\textwidth}
\centering
\includegraphics[width=\linewidth]{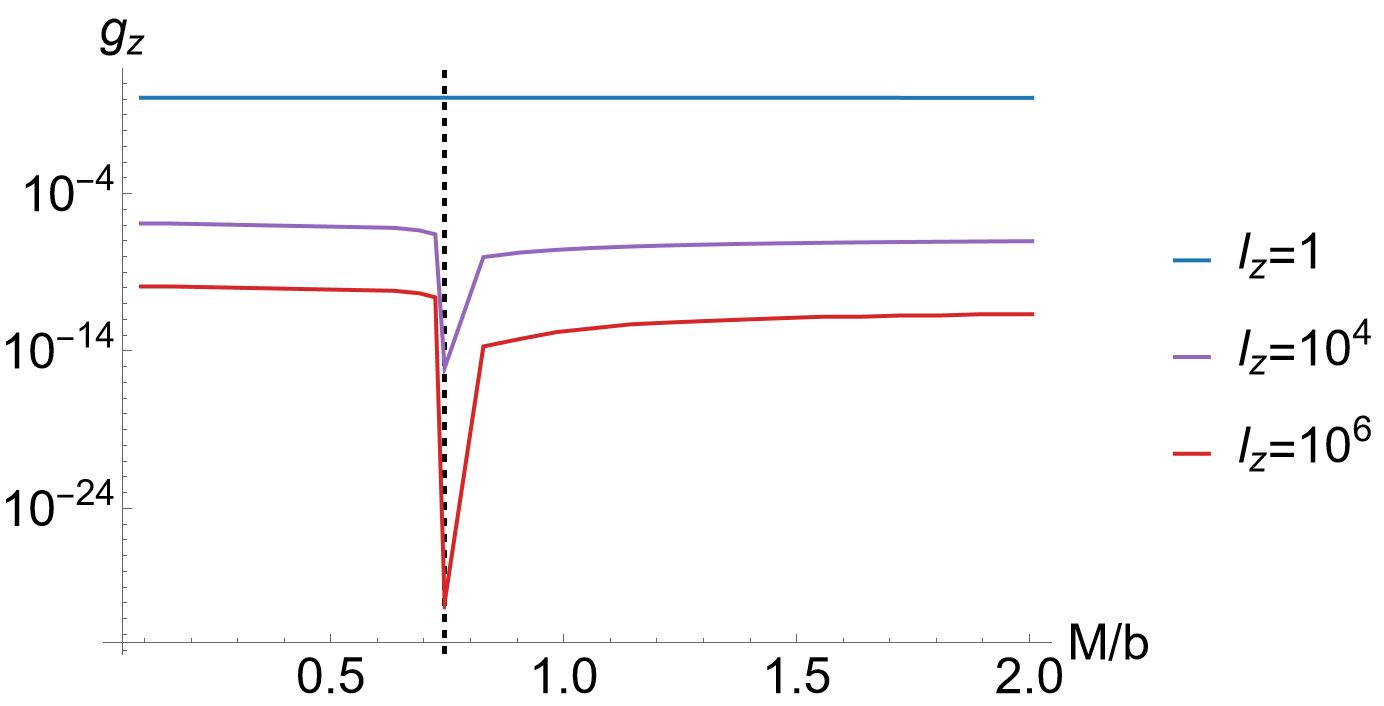}
\end{minipage}
\caption{
The four-partite signal $g$. \(g_x\) (left) and \(g_z\) (right) in
the holographic Weyl semimetal at zero temperature as functions of \(M/b\)
for several fixed strip widths \(l_x\) and \(l_z\). Different colors
correspond to different strip widths, and the dashed line denotes the
critical value \((M/b)_c \simeq 0.744\). Without loss of generality, we set $l_{\text{strip}} = 10 l_{\text{gap}}$.
}
\label{fig:moverb_g_fixed_direction}
\end{figure}

Figure~\ref{fig:moverb_g_fixed_direction} shows the result for the second multi-EWCS based signal $g$. The behavior of \(g\) is qualitatively consistent with that of \(\Delta\). For small strip widths, \(g_x\) and \(g_z\) depend only weakly on \(M/b\). For larger strip widths, both directions show a clear response near the critical value. The \(x\) direction exhibits a critical enhancement, whereas the \(z\) direction exhibits a sharp critical suppression. This shows that the phase sensitivity is not only present in the raw multi-EWCS, but also remains visible after subtracting the mutual-information contribution. The signal therefore reflects a genuine direction dependent residual multipartite structure of the entanglement wedge.

Taken together, the fixed width analyses show that the quantum topological phase transition is visible in all the entanglement quantities or measures considered here, with a common signal appearing near the critical value of \(M/b\). 

\subsection{Dependence on the distance from the critical point}
\label{subsec:symmetric_slices}
The fixed width behavior shows that the entanglement quantities develop clear features near the critical value of \(M/b\). We now ask a slightly different question: within the nontrivial phase and the trivial phase, how does the distance from the critical point affect the behavior of the entanglement quantities~\cite{ylpl-96p1}? This is important because a solution close to the critical point can remain close to the critical scaling geometry over an extended radial range before approaching the near-horizon geometry of the corresponding phase at deep IR. The strip width \(l\) then determines whether the entanglement surface mainly sees this intermediate critical regime or the large \(l\) behavior of the nontrivial or trivial phase.

To make this comparison symmetric across the transition, we define
\begin{equation}
\Delta(M/b)
=
\left|
M/b
-
(M/b)_c
\right| ,
\label{eq:delta_mb}
\end{equation}
and choose $M/b
=(M/b)_c-\Delta(M/b)$ for the nontrivial phase and $M/b(M/b)_c+\Delta(M/b)$ for the trivial phase. 

We compare two values, $\Delta(M/b)$=0.02 and 0.1. This comparison isolates the effect of proximity to the critical point. At small \(l\), all curves are dominated by the common asymptotically AdS region and depend weakly on \(\Delta(M/b)\). At larger \(l\), moving closer to the critical point enlarges the range over which the curves retain the influence of the critical scaling geometry. Moving farther away shortens this range and makes the nontrivial and trivial large \(l\) behavior visible earlier. Thus changing \(\Delta(M/b)\) mainly shifts the crossover in \(l\), rather than producing a new large \(l\) scaling pattern. In the following, we only pick two representative quantities: the $c$-function and the multi-EWCS and other quantities would exhibit similar behaviors, which could be found in Appendix~\ref{app:fixed_direction_supplement}.

\paragraph{\texorpdfstring{\(c\)-function}{c-function}}

\begin{figure}[htbp]
\centering

\begin{minipage}[b]{0.48\textwidth}
\centering
\includegraphics[width=\linewidth]{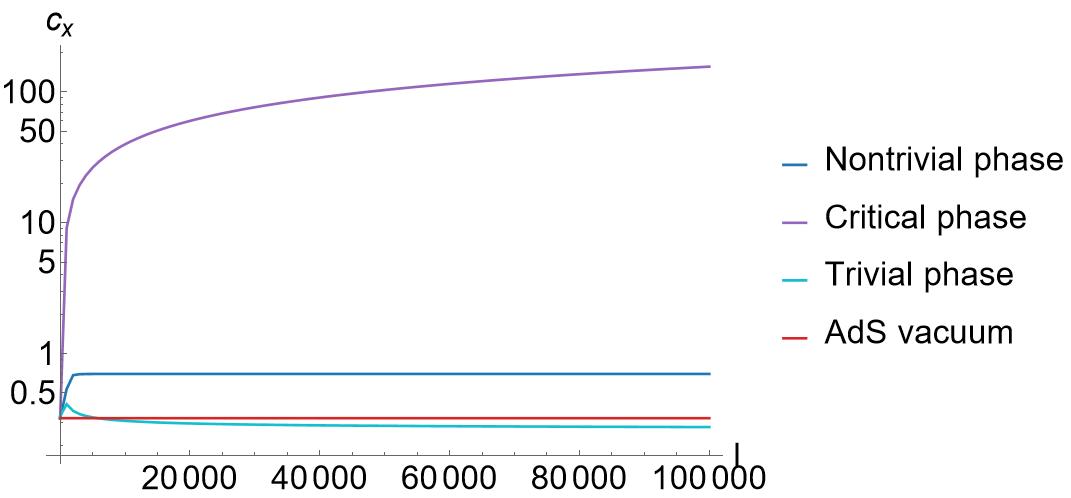}
\small (a) \(\Delta(M/b)=0.02\): \(c_x(l)\)
\end{minipage}
\hfill
\begin{minipage}[b]{0.48\textwidth}
\centering
\includegraphics[width=\linewidth]{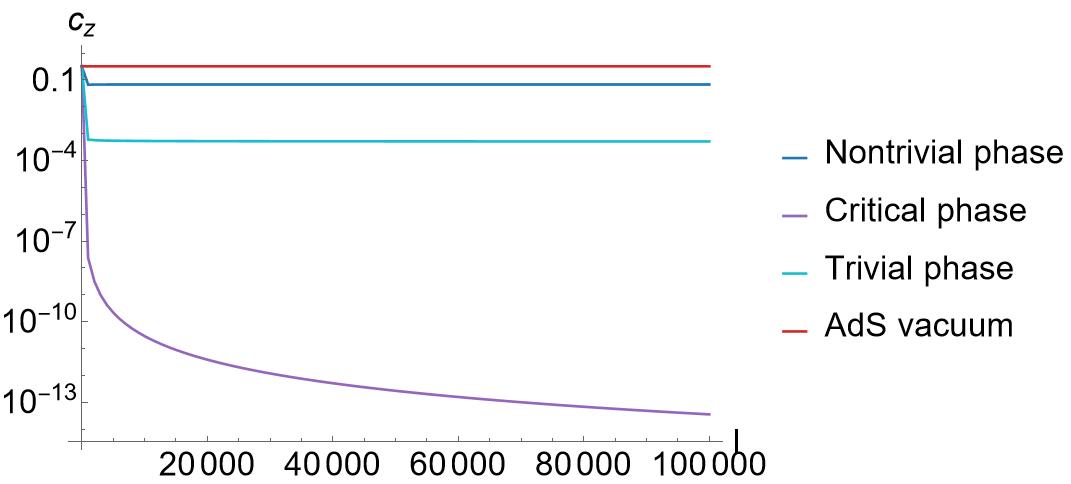}
\small (b) \(\Delta(M/b)=0.02\): \(c_z(l)\)
\end{minipage}

\vspace{0.5em}

\begin{minipage}[b]{0.48\textwidth}
\centering
\includegraphics[width=\linewidth]{Figures/DeltaMoverb=0.1-l_x-c.jpg}
\small (c) \(\Delta(M/b)=0.1\): \(c_x(l)\)
\end{minipage}
\hfill
\begin{minipage}[b]{0.48\textwidth}
\centering
\includegraphics[width=\linewidth]{Figures/DeltaMoverb=0.1-l_z-c.jpg}
\small (d) \(\Delta(M/b)=0.1\): \(c_z(l)\)
\end{minipage}

\caption{Comparison of the holographic \(c\)-function as a function of strip width \(l\) at two distances from the critical point. Panels (a,b) show the results for strips finite along \(x\) and \(z\) at \(\Delta(M/b)=0.02\), while panels (c,d) show the corresponding results at \(\Delta(M/b)=0.1\). For the smaller distance from the critical point, the nontrivial and trivial curves remain close to the critical curve over a wider range of \(l\). For the larger distance, they separate earlier and approach their own large \(l\) plateau behavior more clearly.}
\label{fig:delta_cfunction}
\end{figure}

The comparison is clearest for the holographic \(c\)-function; see Figure~\ref{fig:delta_cfunction}. For both values of \(\Delta(M/b)\), the nontrivial and trivial phases approach finite plateaus at large \(l\), while the critical solution keeps running over the accessible range. The difference between \(\Delta(M/b)=0.02\) and \(\Delta(M/b)=0.1\) lies in the crossover. When \(\Delta(M/b)=0.02\), the nontrivial and trivial curves remain close to the critical curve over a wider range of \(l\). When \(\Delta(M/b)=0.1\), they separate from the critical curve earlier and approach their own plateau behavior more clearly.

Thus tuning \(M/b\) within the same phase mainly changes the length scale at which the asymptotic behavior becomes visible. It does not change the final large \(l\) behavior of the nontrivial or trivial phase.

\paragraph{multi-EWCS}

\begin{figure}[htbp]
        \centering
        \begin{minipage}[b]{0.48\textwidth}
            \centering
            \includegraphics[width=\linewidth]{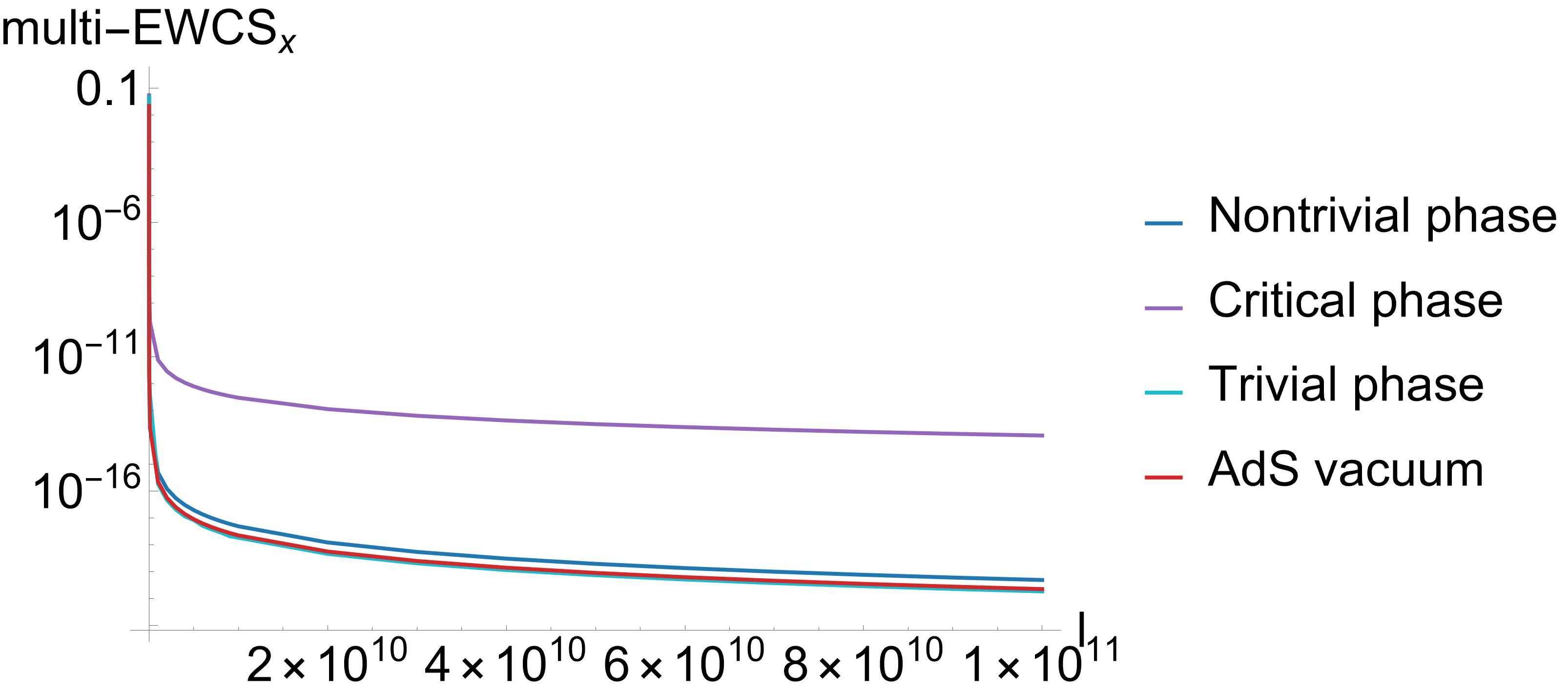}
            \small (a) \(\Delta(M/b)=0.02\): multi-EWCS\(_x(l)\)
        \end{minipage}
        \hfill
        \begin{minipage}[b]{0.48\textwidth}
            \centering
            \includegraphics[width=\linewidth]{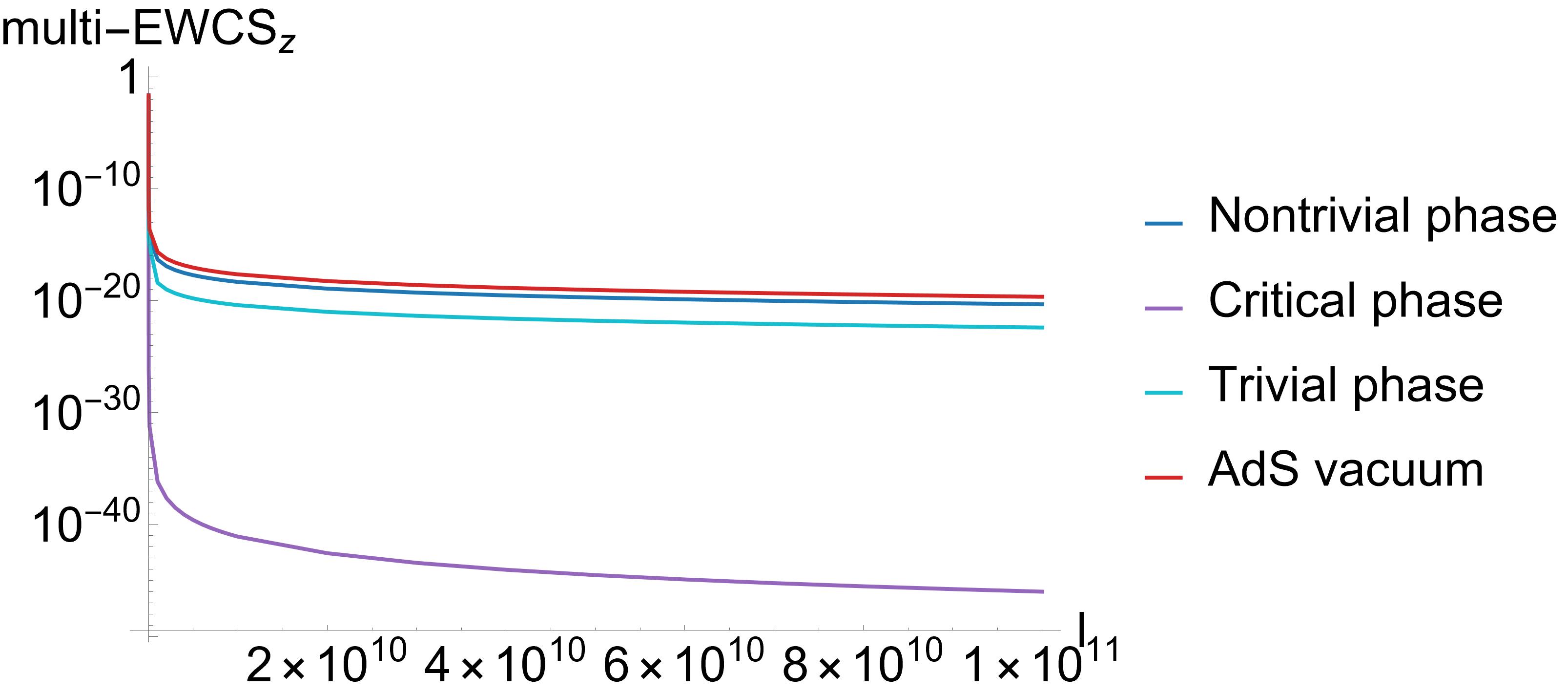}
            \small (b) \(\Delta(M/b)=0.02\): multi-EWCS\(_z(l)\)
        \end{minipage}
        \vspace{0.5em}
        \begin{minipage}[b]{0.48\textwidth}
            \centering
            \includegraphics[width=\linewidth]{Figures/DeltaMoverb=0.1-l_x-MultiEWCS.jpg}
            \small (c) \(\Delta(M/b)=0.1\): multi-EWCS\(_x(l)\)
        \end{minipage}
        \hfill
        \begin{minipage}[b]{0.48\textwidth}
            \centering
            \includegraphics[width=\linewidth]{Figures/DeltaMoverb=0.1-l_z-MultiEWCS.jpg}
            \small (d) \(\Delta(M/b)=0.1\): multi-EWCS\(_z(l)\)
        \end{minipage}

        \caption{ Comparison of the multi-EWCS as a function of strip width \(l\) at two distances from the critical point. Panels (a,b) show the results for strips finite along \(x\) and \(z\) at \(\Delta(M/b)=0.02\), while panels (c,d) show the corresponding results at \(\Delta(M/b)=0.1\). For the smaller distance from the critical point, the nontrivial and trivial curves retain the influence of the critical scaling region over a wider range of \(l\). For the larger distance, the large \(l\) decay of the nontrivial and trivial phases becomes visible earlier. Without loss of generality, we set $l_{\text{strip}} = 10 l_{\text{gap}}$.}
        \label{fig:delta_multiewcs}
    \end{figure}

The result for multi-EWCS is in Figure~\ref{fig:delta_multiewcs}. Unlike the noncritical \(c\)-function, the multi-EWCS decays at large \(l\), but the role of \(\Delta(M/b)\) is the same. For \(\Delta(M/b)=0.02\), the curves keep a visible influence of the critical scaling region over a longer range of \(l\). For \(\Delta(M/b)=0.1\), the nontrivial and trivial large \(l\) behavior is reached earlier.

Therefore the multi-EWCS confirms that changing the distance from the critical point does not produce a new large \(l\) scaling pattern. It changes the crossover range between the critical scaling regime and the asymptotic behavior of the nontrivial or trivial phase. This shows that the same finite scale effect appears not only in the RT surface quantity \(c(l)\), but also in the
multipartite wedge cross section geometric quantity.

        The corresponding plots for the CMI, the EWCS, \(\kappa\), and the Markov gap as functions of \(l\) are collected in Appendix~\ref{app:benchmark_comparisons}. These figures complete the set of data not shown in the main text and support the same conclusion. Increasing \(\Delta(M/b)\) from \(0.02\) to \(0.1\) does not introduce a new large \(l\) pattern in the numerical regime. It mainly shortens the finite scale window in which the nearby critical geometry remains visible and makes the asymptotic behavior of the noncritical phases more transparent. This comparison therefore shows how changing the same dimensionless control parameter within a given phase affects the entanglement measure: it changes amplitudes and crossover scales, while leaving the IR character of the corresponding phase intact.
\section{Angular dependence of entanglement quantities for  strips at different directions}
\label{Section4}
The analysis in the previous section showed that the entanglement quantities have different large \(l\) behavior for strips along the transverse \(x\) direction and for strips along the \(z\) direction selected by the axial source. This difference suggests that the same quantities should also depend on the angle of the strip when it is pointing to an arbitrary direction in the \(x\)-\(z\) plane. In this section, we therefore study strip regions whose finite direction makes an angle \(\theta\) with the \(x\) direction.

The motivation for the study of this angular dependence is the anisotropy of the holographic Weyl semimetal. Previous studies in anisotropic systems have used strips with arbitrary orientations to examine how anisotropy affects holographic entanglement entropy and mutual information~\cite{LIU2019155}, and have shown that the entropic \(c\)-function must be treated carefully when Lorentz invariance and spatial rotational symmetry are broken. In the present model, the axial source fixes the \(z\) direction, while \(x\) is a transverse direction. Varying \(\theta\) gives a continuous comparison between strips close to the \(x\) direction and strips close to the \(z\) direction, and therefore tests how the long-distance entanglement behavior encodes the anisotropic IR geometry. Here we present the angular dependence of the entanglement quantities as functions of \(l\); the extraction and comparison of the angular coefficients are given in Section~\ref{Section5}.

\subsection{Basic setups for strips with an angle in the \texorpdfstring{\(x\)-\(z\)}{x-z} plane}
\label{subsec:rotated_strip_geometry}
We now extend the construction of subsection~\ref{Subsection2.2} to strips that have finite width along a generic direction in the \(x\)-\(z\) plane. Starting from the background metric in \eqref{ZeroTemperatureSolutionAnsatz}, we restrict to a constant time slice. The spatial line element relevant for the RT surface is
\begin{equation}
ds^2_{\Sigma}=u(r) dx^2+u(r)dy^2+h(r)dz^2+\frac{dr^2}{u(r)} .
\end{equation}
The directional dependence of the strip calculation comes from the inequivalence between the metric coefficients in the transverse \(x\) direction and the \(z\) direction.

For a strip at an arbitrary direction with angle \(\theta\) from the transverse \(x\) direction, we choose worldvolume coordinates \((\sigma,y,\zeta)\) 
\begin{equation}
x=\sigma\cos\theta-\zeta\sin\theta,
\qquad
z=\sigma\sin\theta+\zeta\cos\theta.
\qquad
\end{equation}
At the boundary, \(\sigma\) is the coordinate along the finite width direction of the strip and \(\zeta\) is the coordinate along the extended direction in the \(x\)-\(z\) plane. The strip region is taken to be
\begin{equation}
\qquad
-\frac{l}{2}\leq\sigma\leq\frac{l}{2},
\qquad
-\frac{L_y}{2}\leq y\leq\frac{L_y}{2},
\qquad
-\frac{L_\perp}{2}\leq\zeta\leq\frac{L_\perp}{2},
\end{equation}
with \(L_y,L_\perp\to\infty\) at the end. Since the background is independent of \(x\), \(y\), and \(z\), and since the strip is homogeneous along \(y\) and \(\zeta\), the corresponding RT surface can be parametrized by a single profile \(r=r(\sigma)\). The angle \(\theta\) is measured from the transverse \(x\) direction. Thus \(\theta=0\) gives the strip finite along \(x\), while \(\theta=\pi/2\) gives the strip finite along the distinguished \(z\) direction. This parametrization is not a symmetry transformation of the anisotropic bulk geometry. It only uses the translational symmetries of the background to describe a boundary strip with a fixed angle.

With this parametrization, the relevant components of the induced metric are
\begin{equation}
\begin{aligned}
\gamma_{\sigma\sigma}
&=u(r)\cos^2\theta+h(r)\sin^2\theta+\frac{r'(\sigma)^2}{u(r)},
\\
\gamma_{\zeta\zeta}
&=u(r)\sin^2\theta+h(r)\cos^2\theta,
\\
\gamma_{\sigma\zeta}
&=\left(h(r)-u(r)\right)\sin\theta\cos\theta,
\\
\gamma_{yy}
&=u(r).
\end{aligned}
\end{equation}

The area density is therefore
\begin{equation}
\sqrt{\det\gamma}
=
\sqrt{
u(r)
\left[
\gamma_{\sigma\sigma}\gamma_{\zeta\zeta}
-
\gamma_{\sigma\zeta}^{\,2}
\right]
}.
\end{equation}

Using
\begin{equation}
\left(u\cos^2\theta+h\sin^2\theta
\right)\left(u\sin^2\theta+h\cos^2\theta
\right)-\left(h-u\right)^2\sin^2\theta\cos^2\theta
=u h,
\end{equation}
the area functional can be written as
\begin{equation}
\mathcal A_\theta=L_y L_\zeta\int_{-l/2}^{l/2}
d\sigma\,\sqrt{u(r)\,\mathcal G_\perp(r,\theta)\left[\frac{r'(\sigma)^2}{u(r)}+\mathcal G_{\ell}^{\rm eff}(r,\theta)\right]
},
\end{equation}
where
\begin{equation}
\mathcal G_\perp(r,\theta)=u(r)\sin^2\theta+h(r)\cos^2\theta
\end{equation}
is the metric factor along the extended direction \(\zeta\), and
\begin{equation}
\mathcal G_{\ell}^{\rm eff}(r,\theta)
=\frac{u(r)h(r)}
{u(r)\sin^2\theta+h(r)\cos^2\theta}
\end{equation}
is the effective metric factor along the finite direction of the strip. This expression reduces to the usual strip finite along \(x\) at \(\theta=0\), and to the usual strip finite along \(z\) at
\(\theta=\pi/2\).
\subsection{Angular dependence of the entanglement quantities and measures}
\label{subsec:raw_angular_response_01}
Having specified the basic setups, we now evaluate the same set of entanglement quantities and measures for these strips. For each phase, we display these quantities as functions of the strip width \(l\) at several angles \(\theta\), in order to see how the curves change as the angle changes. We evaluate the entanglement quantities and measures at \(\theta=0^\circ,\;5^\circ,\;15^\circ,\;30^\circ,\;45^\circ,\;60^\circ,\;90^\circ\). This set evolves from the transverse $x$-direction to the axial $z$-direction, and the inclusion of \(5^\circ\) helps test the sensitivity to small angles near the transverse direction. For the representative noncritical backgrounds, we take \(M/b=(M/b)_c\mp0.1\), corresponding to the topologically nontrivial and trivial phases, respectively. The extraction of angular coefficients and the associated fits are postponed to Section~\ref{Section5}. 

In the angular analysis below we show the holographic \(c\)-function, CMI, EWCS, \(\kappa\), the Markov gap, and the multi-EWCS. We do not include separate angular analysis for the two four-partite entanglement signals \(\Delta\) and \(g\). This is because these signals are defined from the multi-EWCS together with fewer-partite quantities so that their large \(l\) powers therefore follow from the same set of ingredients and do not give additional independent scaling behavior. For this reason, and to avoid redundant discussion, we do not analyze these two signals separately in the angular study.

\begin{figure}[htbp]
    \centering

    \begin{minipage}[b]{0.48\textwidth}
        \centering
        \includegraphics[width=\linewidth]{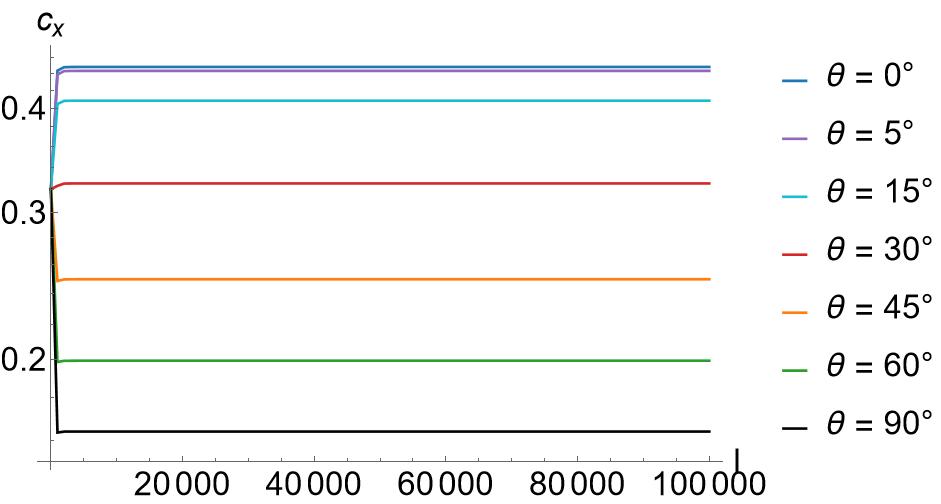}
        \small (a) Nontrivial phase: \(c_x(l)\)
    \end{minipage}
    \hfill
    \begin{minipage}[b]{0.48\textwidth}
        \centering
        \includegraphics[width=\linewidth]{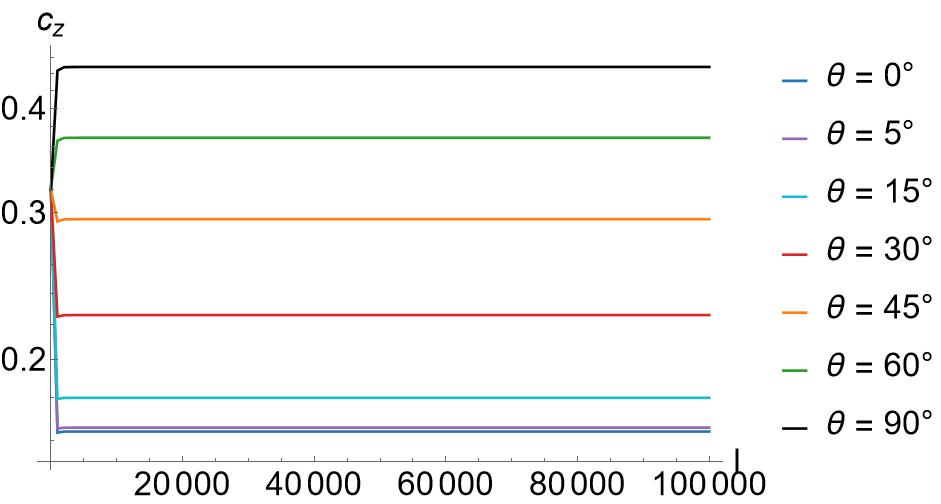}
        \small (b) Nontrivial phase: \(c_z(l)\)
    \end{minipage}

    \vspace{0.5em}

    \begin{minipage}[b]{0.48\textwidth}
        \centering
        \includegraphics[width=\linewidth]{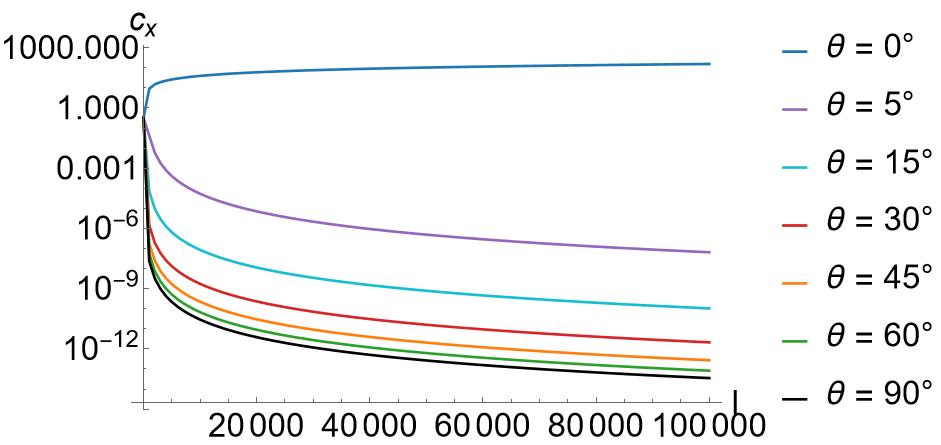}
        \small (c) Critical phase: \(c_x(l)\)
    \end{minipage}
    \hfill
    \begin{minipage}[b]{0.48\textwidth}
        \centering
        \includegraphics[width=\linewidth]{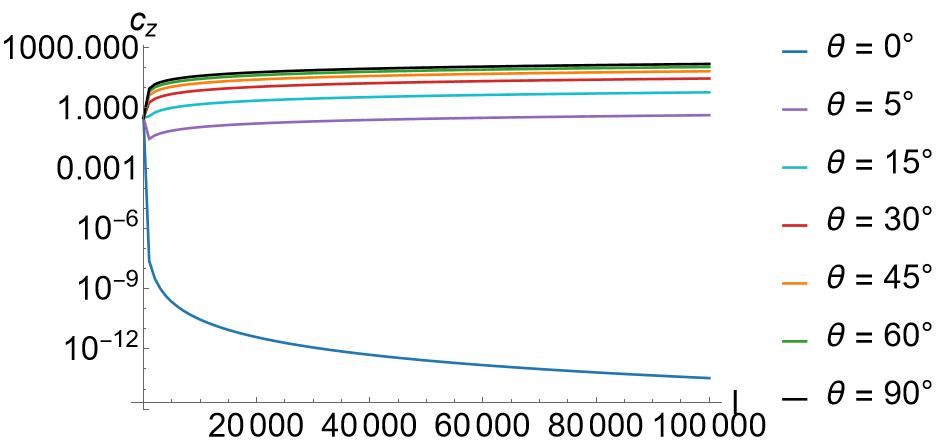}
        \small (d) Critical phase: \(c_z(l)\)
    \end{minipage}

    \vspace{0.5em}

    \begin{minipage}[b]{0.48\textwidth}
        \centering
        \includegraphics[width=\linewidth]{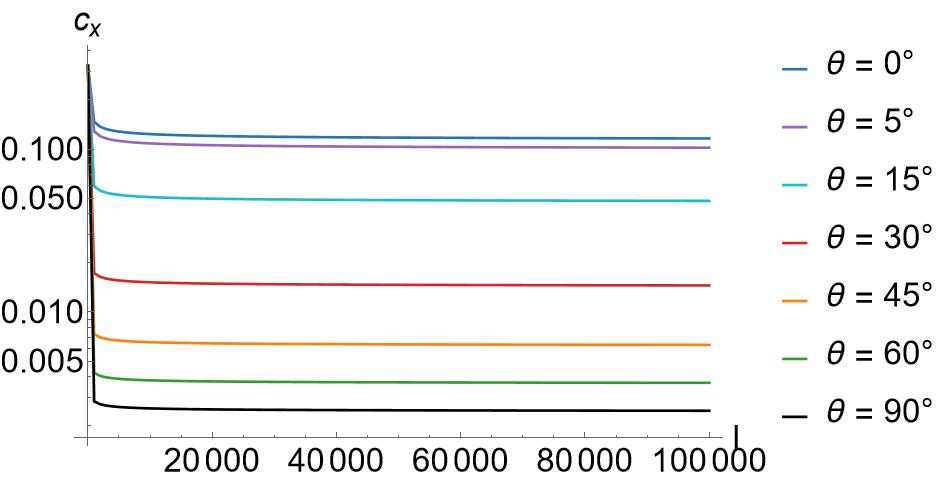}
        \small (e) Trivial phase: \(c_x(l)\)
    \end{minipage}
    \hfill
    \begin{minipage}[b]{0.48\textwidth}
        \centering
        \includegraphics[width=\linewidth]{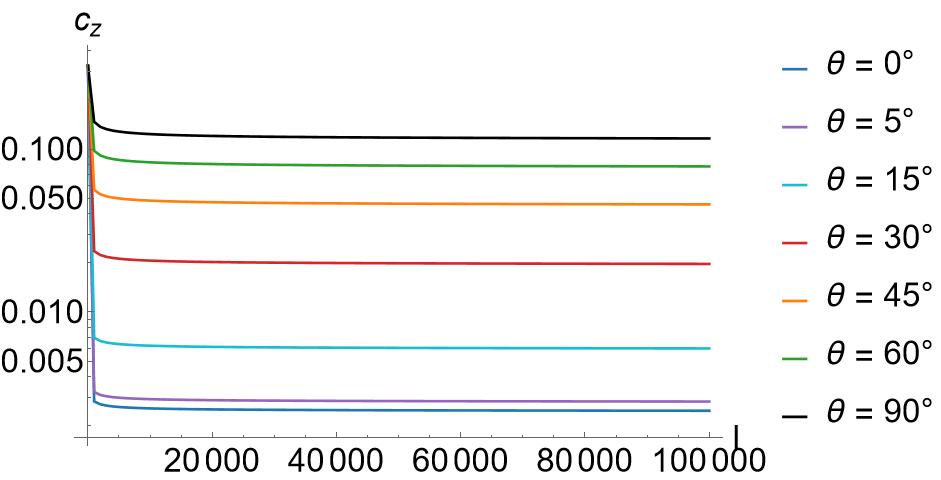}
        \small (f) Trivial phase: \(c_z(l)\)
    \end{minipage}

    \caption{
    Angular dependence of the holographic \(c\)-function for strips with an
    angle \(\theta\) from the \(x\) axis. The nontrivial, critical, and
    trivial phases are evaluated at \(M/b=(M/b)_c-0.1\),
    \(M/b=(M/b)_c\), and \(M/b=(M/b)_c+0.1\), respectively. Panels
    (a,c,e) show \(c_x(l)\), and panels (b,d,f) show \(c_z(l)\). The curves
    correspond to
    \(\theta=0^\circ,5^\circ,15^\circ,30^\circ,45^\circ,60^\circ,90^\circ\).
    }
    \label{fig:cfunction_dMoverb_01_all}
\end{figure}

We first examine the angular dependence of the holographic \(c\)-function as the strip direction is tuned from the \(x\) direction to the \(z\) direction; see Figure~\ref{fig:cfunction_dMoverb_01_all}. In the topologically nontrivial and trivial phases away from the critical point, the curves approach finite values at large \(l\). The angle \(\theta\) changes these limiting values, but does not change the fact that the leading behavior is a plateau. At the critical point, by contrast, the curves remain scale dependent over the accessible range of \(l\), and changing \(\theta\) affects the large \(l\) scaling behavior itself, reflecting the anisotropic critical geometry.

\begin{figure}[htbp]
    \centering

    \begin{minipage}[b]{0.48\textwidth}
        \centering
        \includegraphics[width=\linewidth]{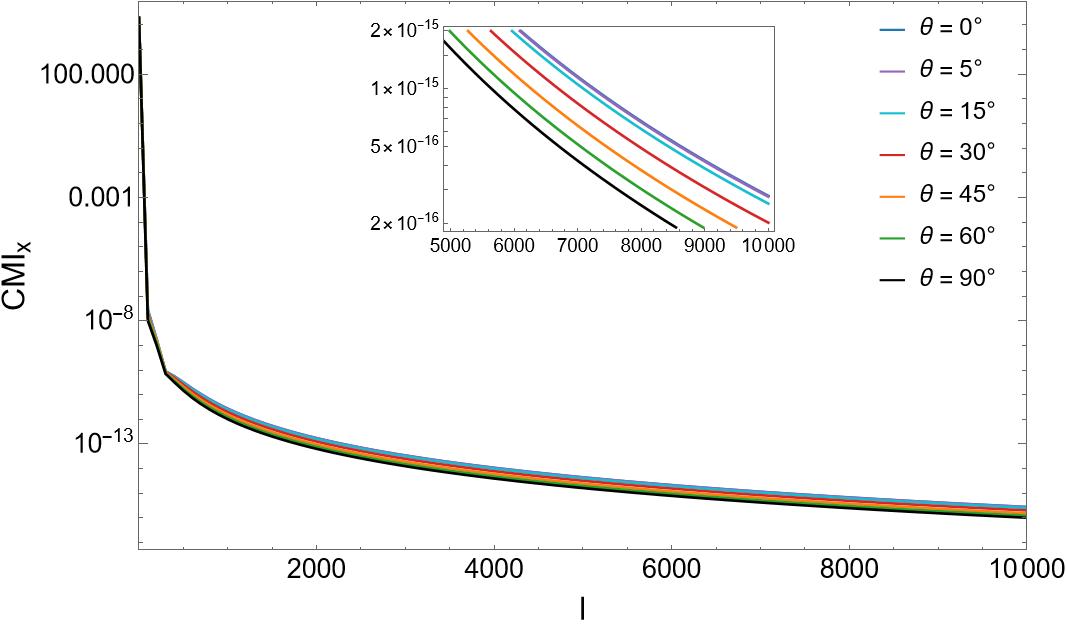}
        \small (a) Nontrivial phase: CMI\(_x(l)\)
    \end{minipage}
    \hfill
    \begin{minipage}[b]{0.48\textwidth}
        \centering
        \includegraphics[width=\linewidth]{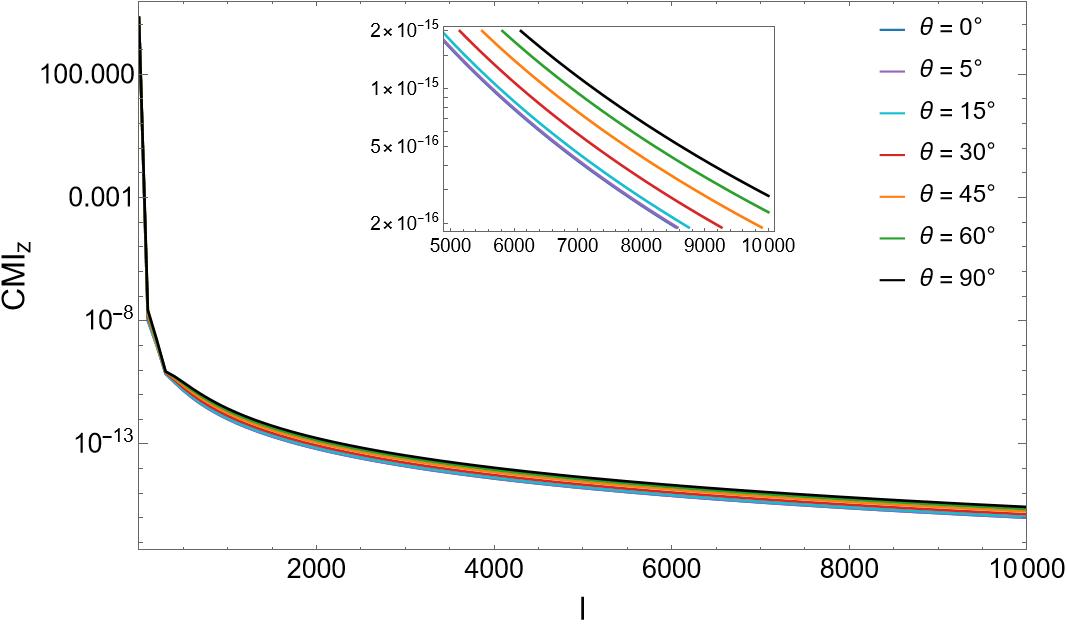}
        \small (b) Nontrivial phase: CMI\(_z(l)\)
    \end{minipage}

    \vspace{0.5em}

    \begin{minipage}[b]{0.48\textwidth}
        \centering
        \includegraphics[width=\linewidth]{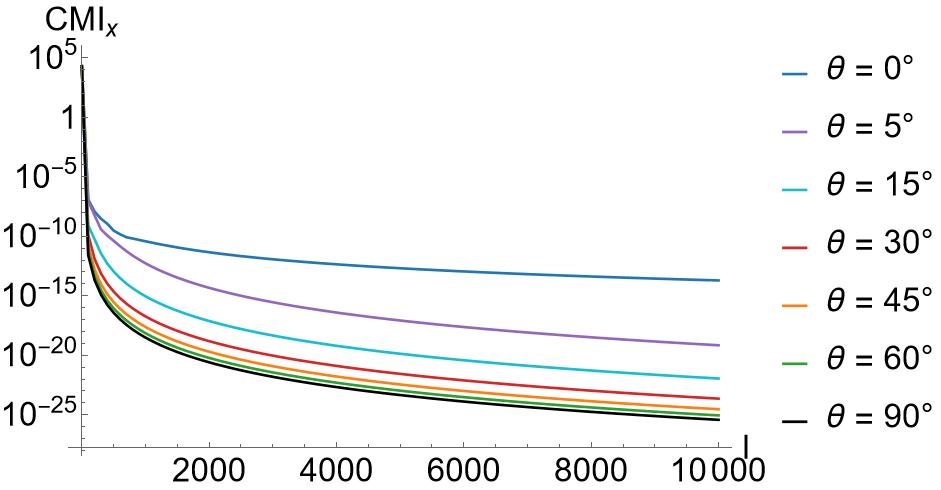}
        \small (c) Critical phase: CMI\(_x(l)\)
    \end{minipage}
    \hfill
    \begin{minipage}[b]{0.48\textwidth}
        \centering
        \includegraphics[width=\linewidth]{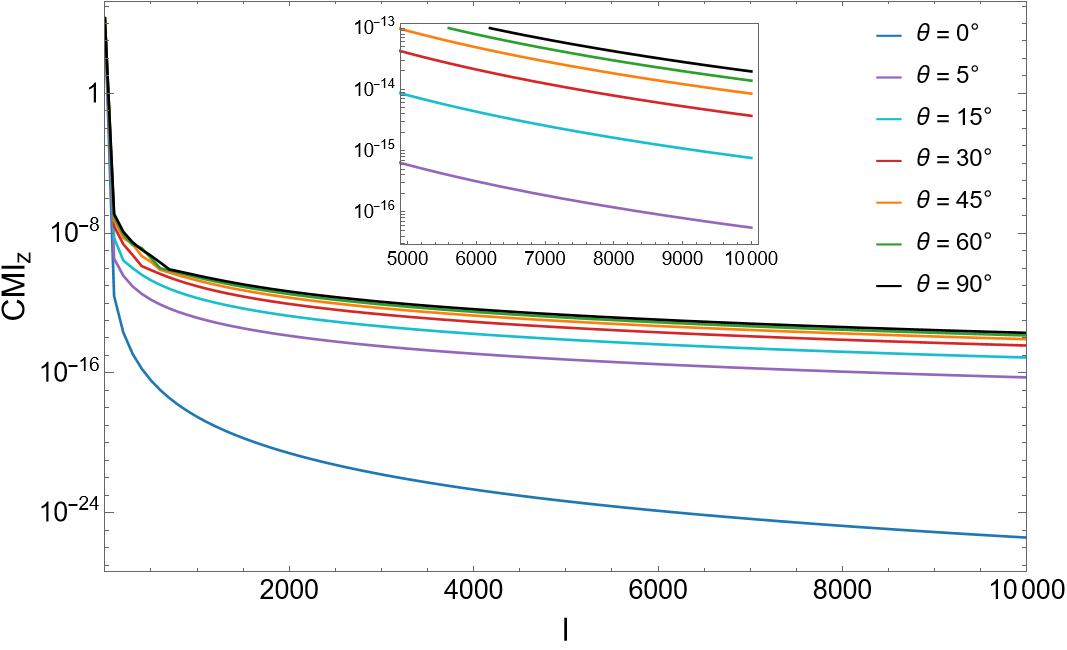}
        \small (d) Critical phase: CMI\(_z(l)\)
    \end{minipage}

    \vspace{0.5em}

    \begin{minipage}[b]{0.48\textwidth}
        \centering
        \includegraphics[width=\linewidth]{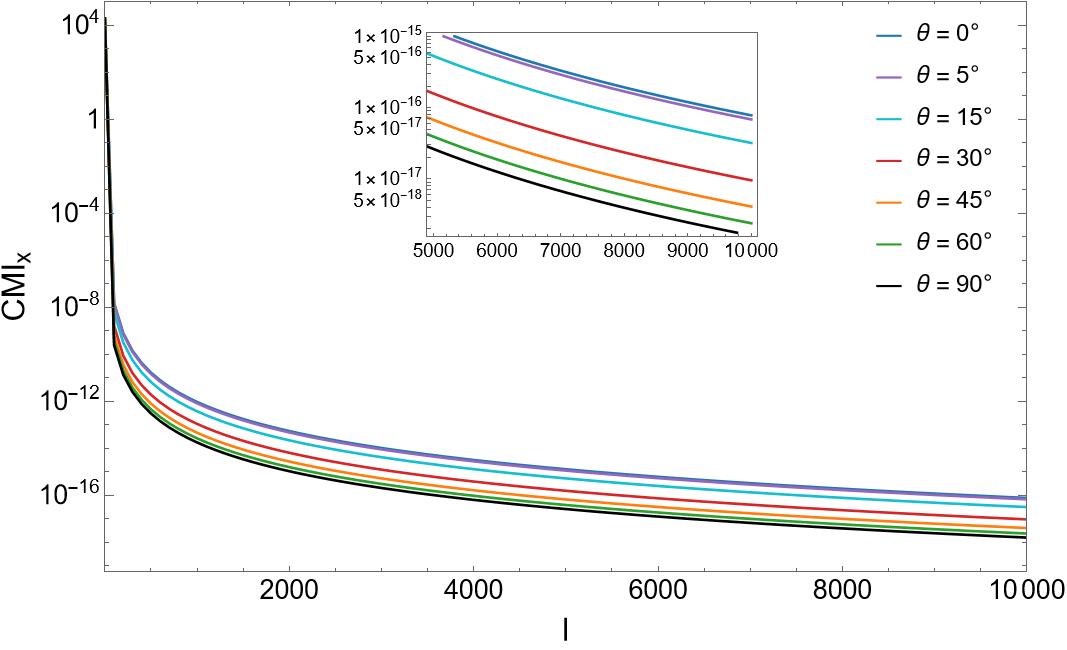}
        \small (e) Trivial phase: CMI\(_x(l)\)
    \end{minipage}
    \hfill
    \begin{minipage}[b]{0.48\textwidth}
        \centering
        \includegraphics[width=\linewidth]{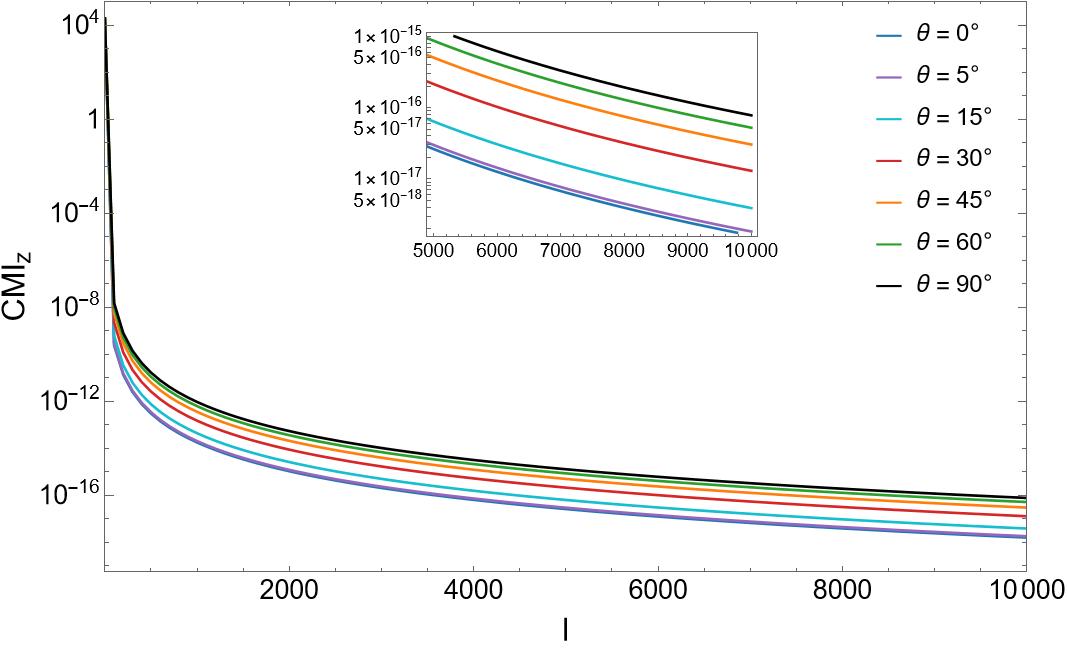}
        \small (f) Trivial phase: CMI\(_z(l)\)
    \end{minipage}

    \caption{
    Angular dependence of the CMI for strips with an
    angle \(\theta\) from the \(x\) axis. The nontrivial, critical,
    and trivial phases are evaluated at \(M/b=(M/b)_c-0.1\),
    \(M/b=(M/b)_c\), and \(M/b=(M/b)_c+0.1\), respectively. Panels
    (a,c,e) show \(\mathrm{CMI}_x(l)\), and panels (b,d,f) show
    \(\mathrm{CMI}_z(l)\). The curves correspond to
    \(\theta=0^\circ,5^\circ,15^\circ,30^\circ,45^\circ,60^\circ,90^\circ\).
    }
    \label{fig:CMI_dMoverb_01_all}
\end{figure}

\begin{figure}[htbp]
    \centering

    \begin{minipage}[b]{0.48\textwidth}
        \centering
        \includegraphics[width=\linewidth]{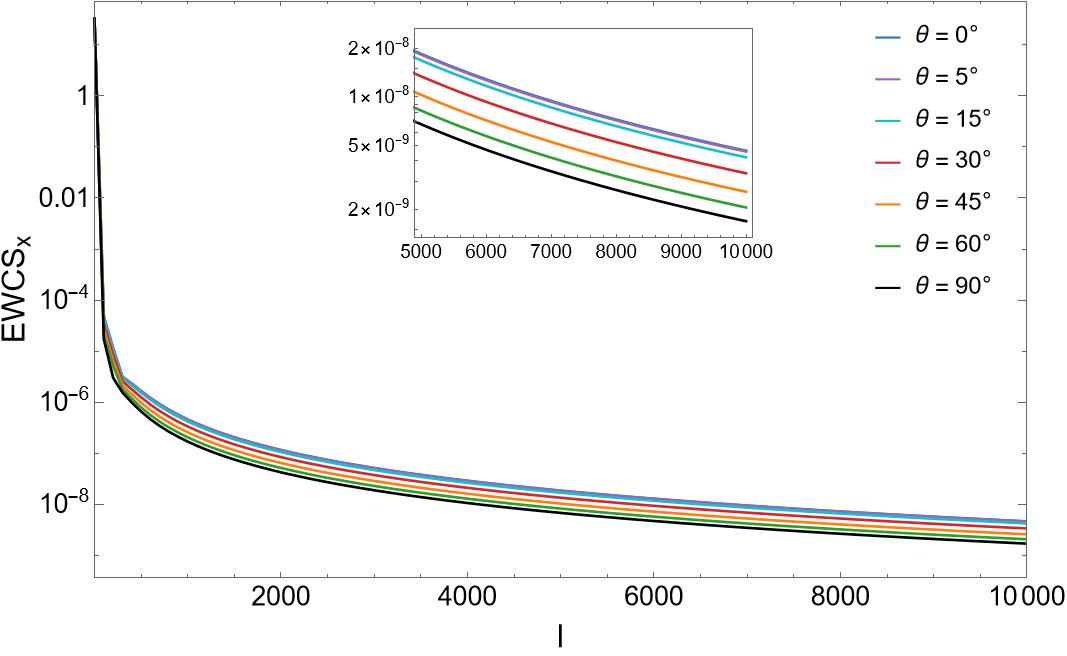}
        \small (a) Nontrivial phase: EWCS\(_x(l)\)
    \end{minipage}
    \hfill
    \begin{minipage}[b]{0.48\textwidth}
        \centering
        \includegraphics[width=\linewidth]{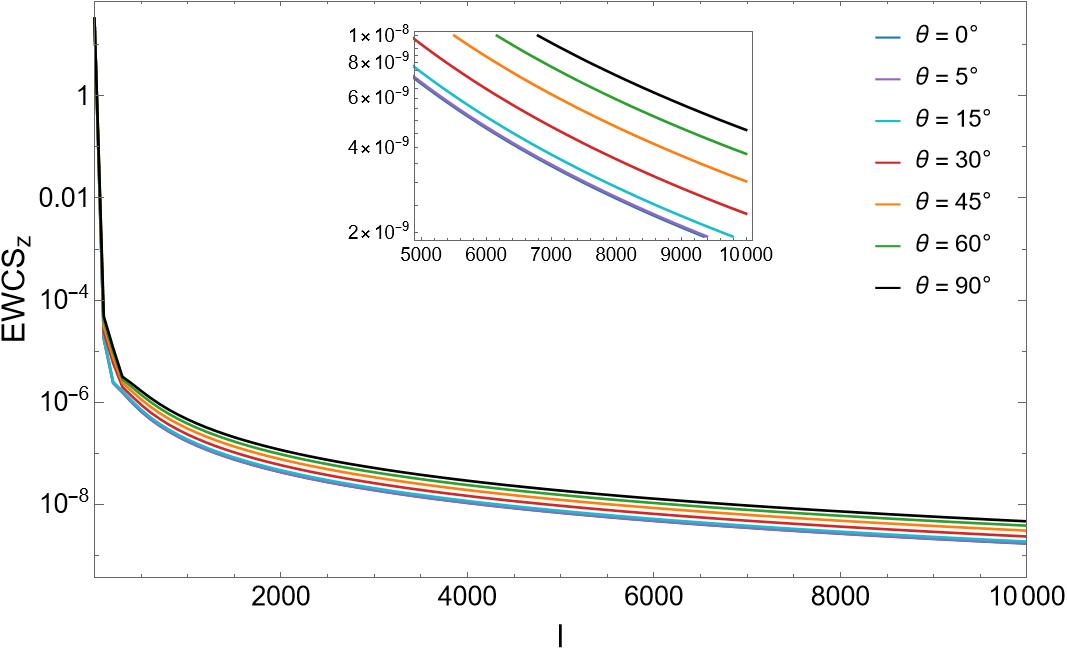}
        \small (b) Nontrivial phase: EWCS\(_z(l)\)
    \end{minipage}

    \vspace{0.5em}

    \begin{minipage}[b]{0.48\textwidth}
        \centering
        \includegraphics[width=\linewidth]{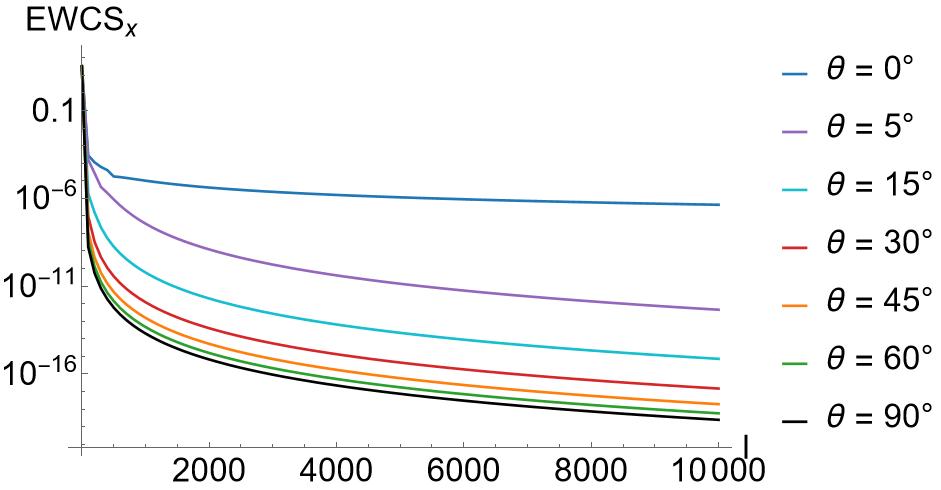}
        \small (c) Critical phase: EWCS\(_x(l)\)
    \end{minipage}
    \hfill
    \begin{minipage}[b]{0.48\textwidth}
        \centering
        \includegraphics[width=\linewidth]{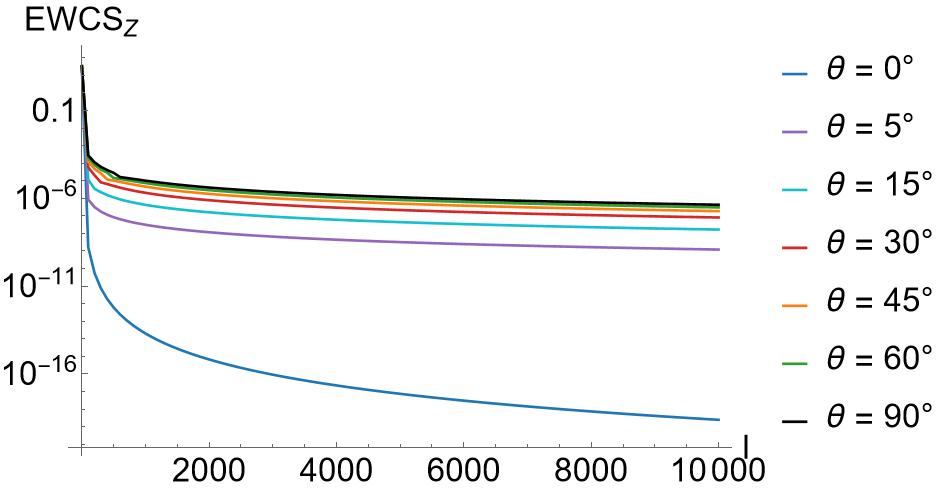}
        \small (d) Critical phase: EWCS\(_z(l)\)
    \end{minipage}

    \vspace{0.5em}

    \begin{minipage}[b]{0.48\textwidth}
        \centering
        \includegraphics[width=\linewidth]{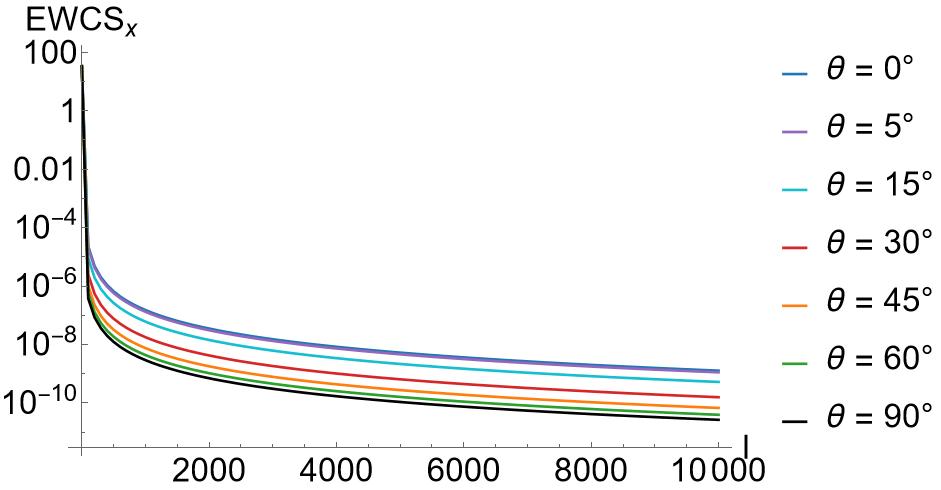}
        \small (e) Trivial phase: EWCS\(_x(l)\)
    \end{minipage}
    \hfill
    \begin{minipage}[b]{0.48\textwidth}
        \centering
        \includegraphics[width=\linewidth]{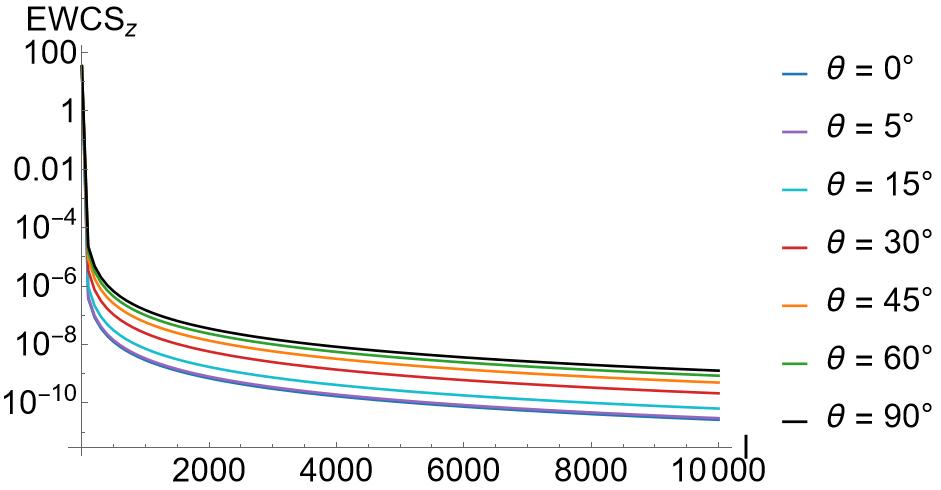}
        \small (f) Trivial phase: EWCS\(_z(l)\)
    \end{minipage}

    \caption{
   Angular dependence of the EWCS for strips with an angle \(\theta\) from the \(x\) axis. The nontrivial,
    critical, and trivial phases are evaluated at \(M/b=(M/b)_c-0.1\),
    \(M/b=(M/b)_c\), and \(M/b=(M/b)_c+0.1\), respectively. Panels
    (a,c,e) show \(\mathrm{EWCS}_x(l)\), and panels (b,d,f) show
    \(\mathrm{EWCS}_z(l)\). The curves correspond to
    \(\theta=0^\circ,5^\circ,15^\circ,30^\circ,45^\circ,60^\circ,90^\circ\). Without loss of generality, an appropriate ratio of $l_{\text{strip}}$ to $l_{\text{gap}}$ has been chosen to guarantee that the entanglement wedges between different regions remain connected as the scale increases.
    }
    \label{fig:EWCS_dMoverb_01_all}
\end{figure}

\begin{figure}[htbp]
    \centering

    \begin{minipage}[b]{0.48\textwidth}
        \centering
        \includegraphics[width=\linewidth]{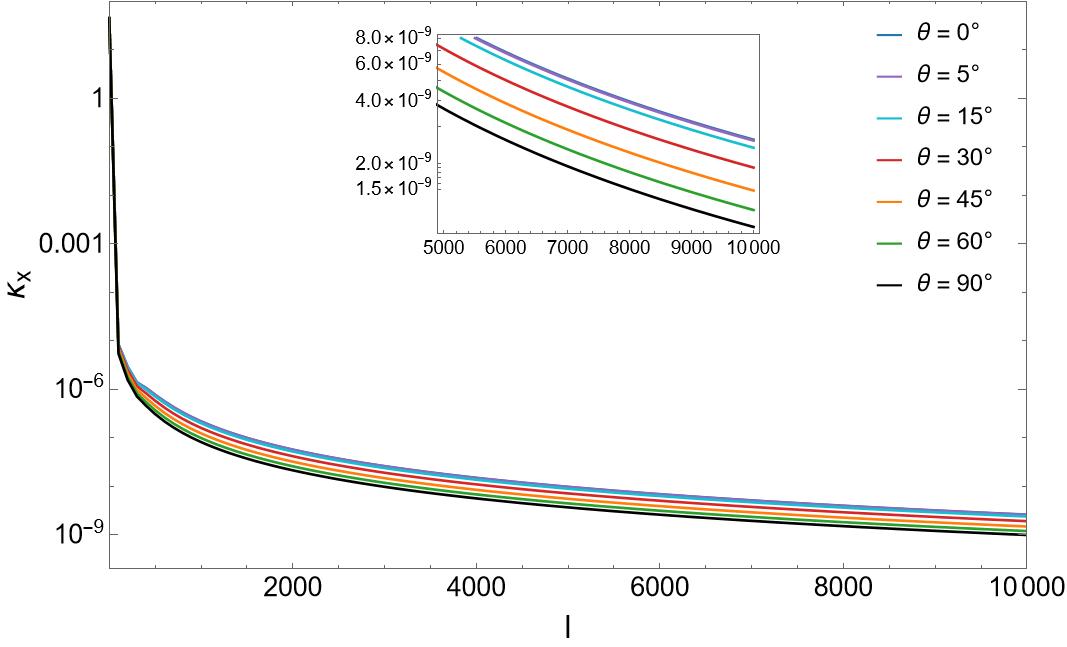}
        \small (a) Nontrivial phase: \(\kappa_x(l)\)
    \end{minipage}
    \hfill
    \begin{minipage}[b]{0.48\textwidth}
        \centering
        \includegraphics[width=\linewidth]{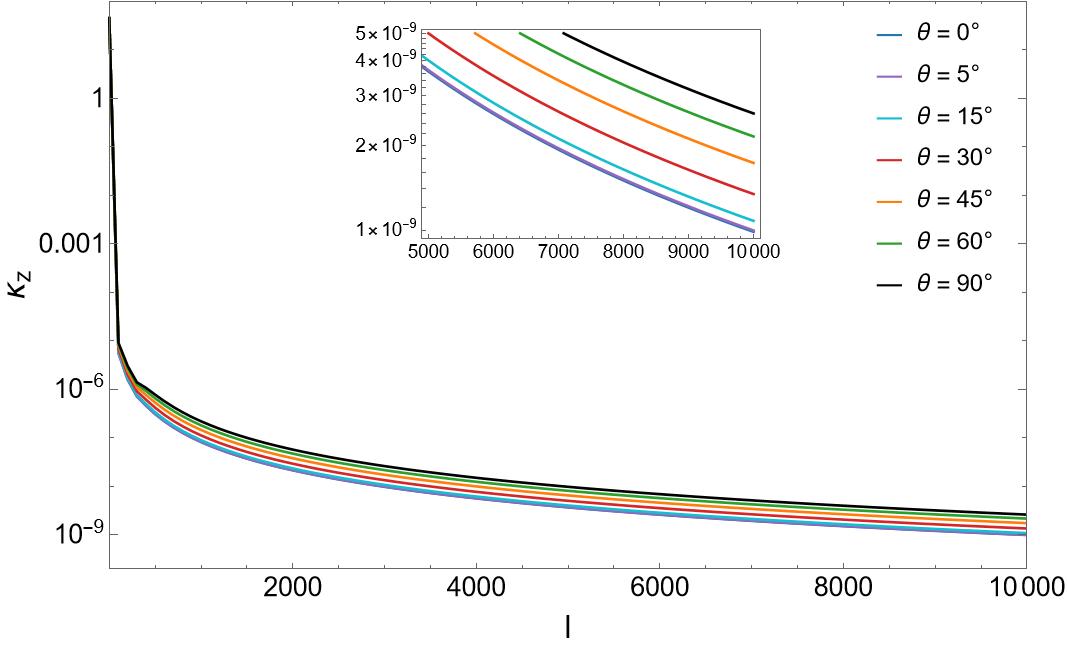}
        \small (b) Nontrivial phase: \(\kappa_z(l)\)
    \end{minipage}

    \vspace{0.5em}

    \begin{minipage}[b]{0.48\textwidth}
        \centering
        \includegraphics[width=\linewidth]{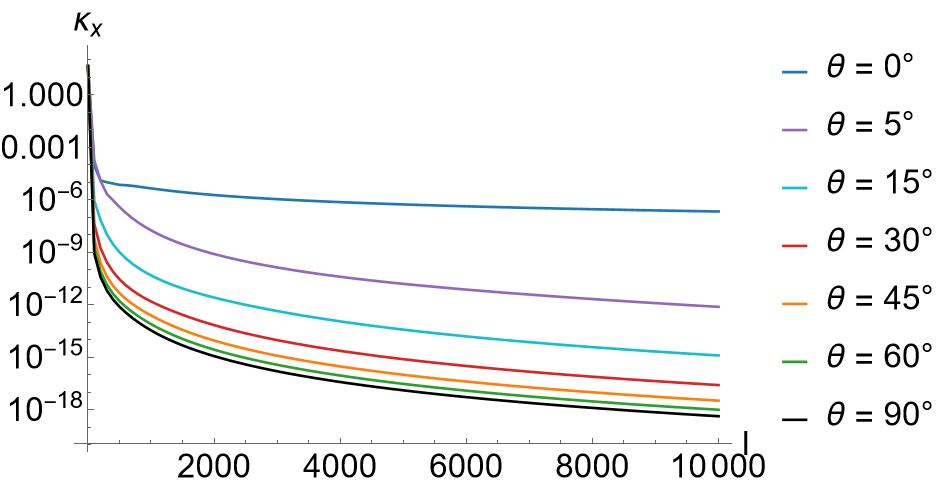}
        \small (c) Critical phase: \(\kappa_x(l)\)
    \end{minipage}
    \hfill
    \begin{minipage}[b]{0.48\textwidth}
        \centering
        \includegraphics[width=\linewidth]{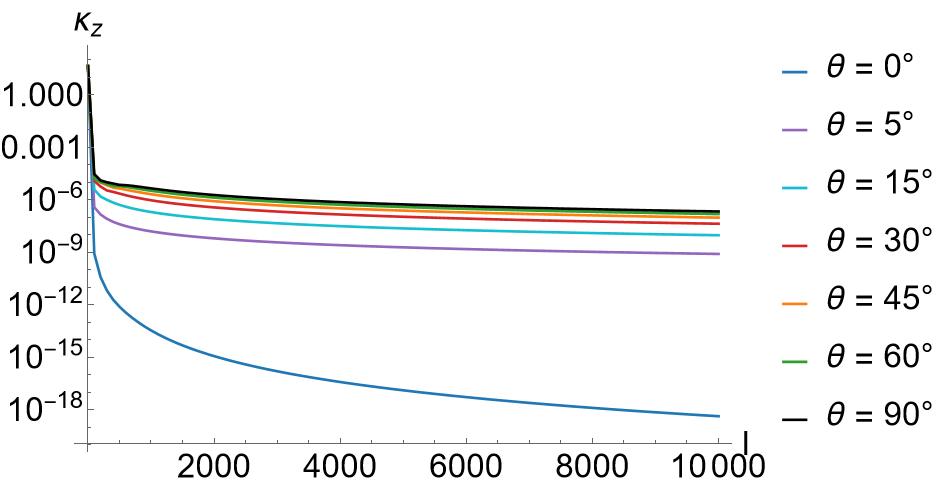}
        \small (d) Critical phase: \(\kappa_z(l)\)
    \end{minipage}

    \vspace{0.5em}

    \begin{minipage}[b]{0.48\textwidth}
        \centering
        \includegraphics[width=\linewidth]{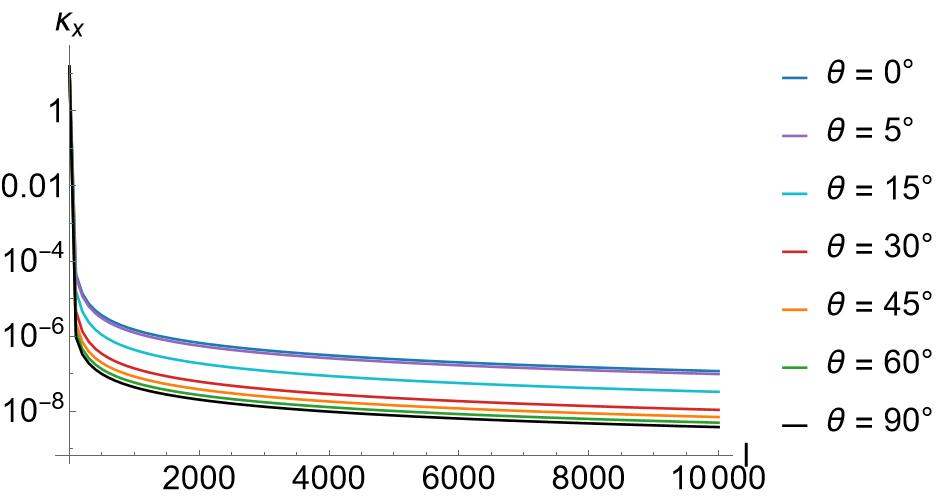}
        \small (e) Trivial phase: \(\kappa_x(l)\)
    \end{minipage}
    \hfill
    \begin{minipage}[b]{0.48\textwidth}
        \centering
        \includegraphics[width=\linewidth]{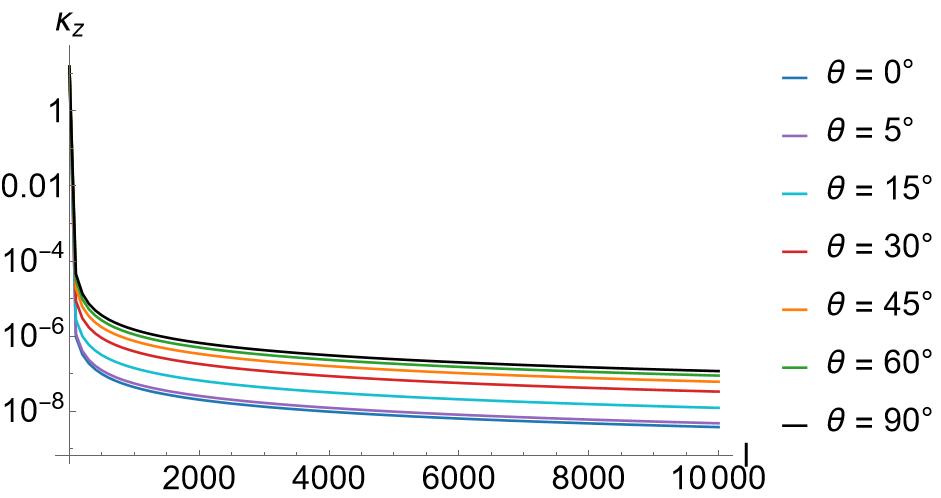}
        \small (f) Trivial phase: \(\kappa_z(l)\)
    \end{minipage}

    \caption{
    Angular dependence of the \(\kappa\) for strips with an
    angle \(\theta\) from the \(x\) axis. The nontrivial, critical, and trivial phases are evaluated at
    \(M/b=(M/b)_c-0.1\), \(M/b=(M/b)_c\), and \(M/b=(M/b)_c+0.1\),
    respectively. Panels (a,c,e) show \(\kappa_x(l)\), and panels (b,d,f)
    show \(\kappa_z(l)\). The curves correspond to
    \(\theta=0^\circ,5^\circ,15^\circ,30^\circ,45^\circ,60^\circ,90^\circ\).
    }
    \label{fig:kappa_dMoverb_01_all}
\end{figure}

\begin{figure}[htbp]
    \centering

    \begin{minipage}[b]{0.48\textwidth}
        \centering
        \includegraphics[width=\linewidth]{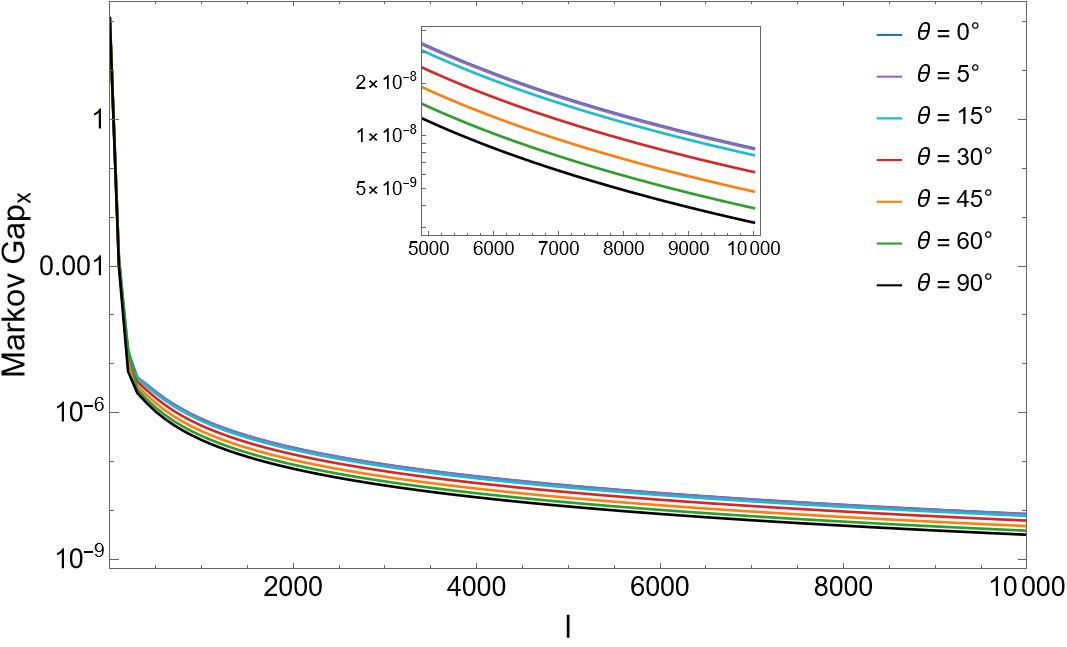}
        \small (a) Nontrivial phase: Markov gap\(_x(l)\)
    \end{minipage}
    \hfill
    \begin{minipage}[b]{0.48\textwidth}
        \centering
        \includegraphics[width=\linewidth]{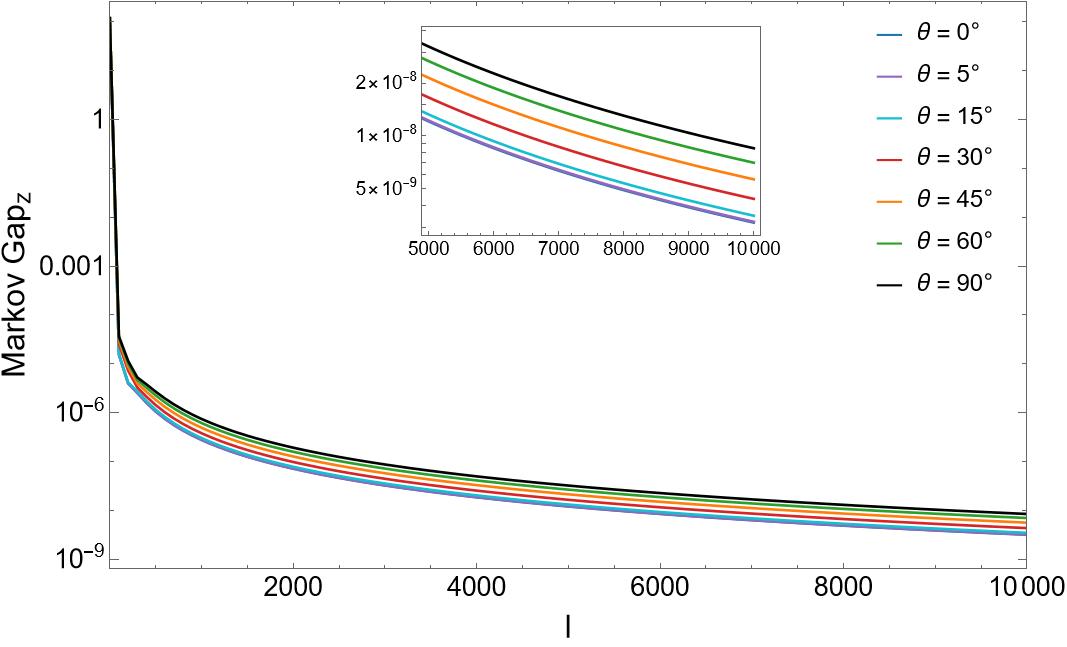}
        \small (b) Nontrivial phase: Markov gap\(_z(l)\)
    \end{minipage}

    \vspace{0.5em}

    \begin{minipage}[b]{0.48\textwidth}
        \centering
        \includegraphics[width=\linewidth]{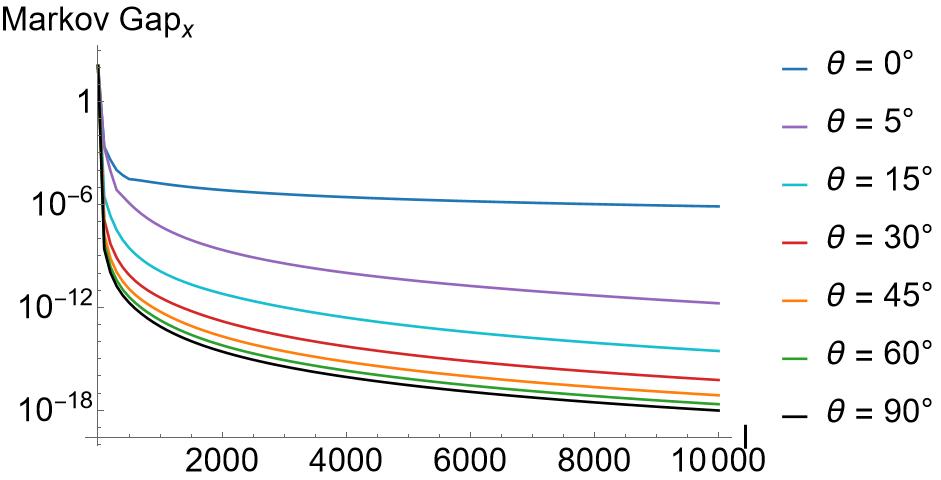}
        \small (c) Critical phase: Markov gap\(_x(l)\)
    \end{minipage}
    \hfill
    \begin{minipage}[b]{0.48\textwidth}
        \centering
        \includegraphics[width=\linewidth]{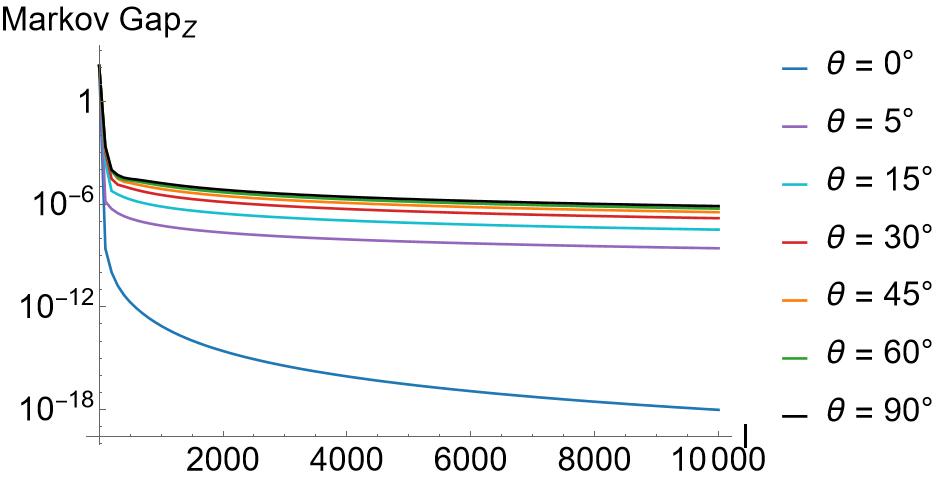}
        \small (d) Critical phase: Markov gap\(_z(l)\)
    \end{minipage}

    \vspace{0.5em}

    \begin{minipage}[b]{0.48\textwidth}
        \centering
        \includegraphics[width=\linewidth]{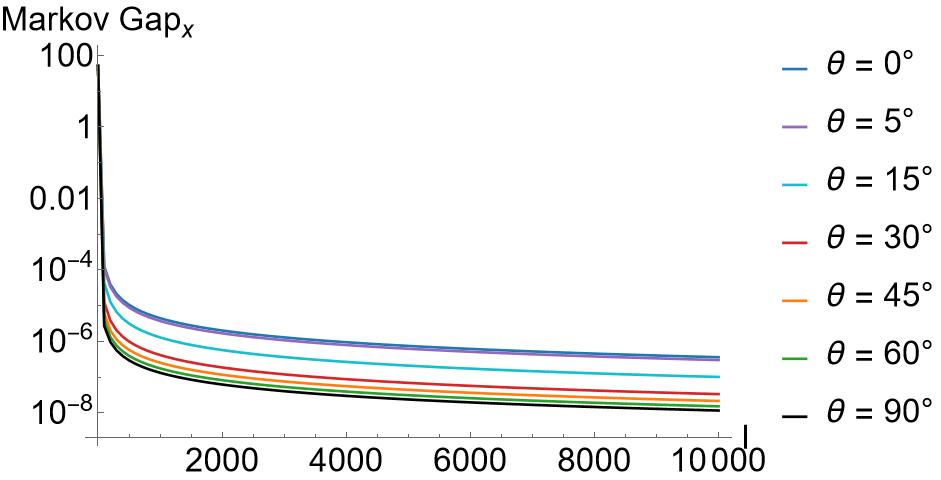}
        \small (e) Trivial phase: Markov gap\(_x(l)\)
    \end{minipage}
    \hfill
    \begin{minipage}[b]{0.48\textwidth}
        \centering
        \includegraphics[width=\linewidth]{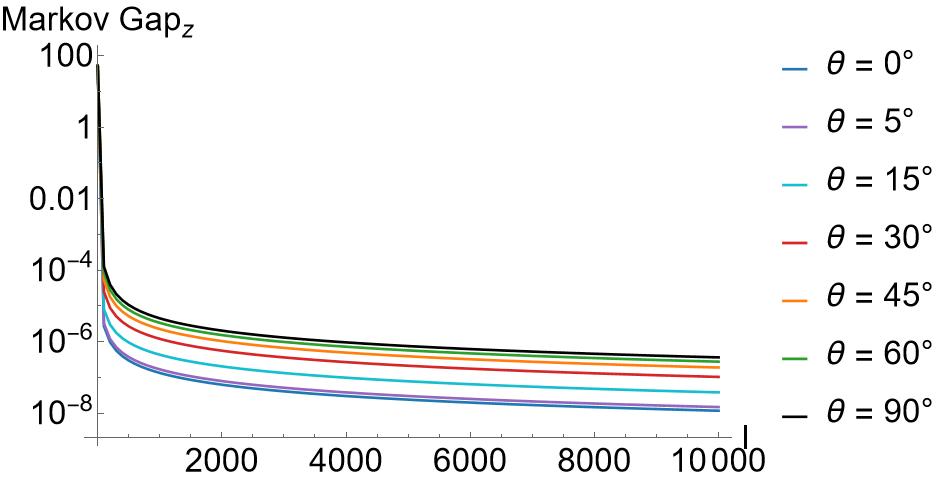}
        \small (f) Trivial phase: Markov gap\(_z(l)\)
    \end{minipage}

    \caption{
    Angular dependence of the Markov gap for strips with an angle \(\theta\) from the \(x\) axis. The nontrivial,
    critical, and trivial phases are evaluated at \(M/b=(M/b)_c-0.1\),
    \(M/b=(M/b)_c\), and \(M/b=(M/b)_c+0.1\), respectively. Panels
    (a,c,e) show \(\mathrm{Markov\ gap}_x(l)\), and panels (b,d,f) show
    \(\mathrm{Markov\ gap}_z(l)\). The curves correspond to
    \(\theta=0^\circ,5^\circ,15^\circ,30^\circ,45^\circ,60^\circ,90^\circ\). Without loss of generality, an appropriate ratio of $l_{\text{strip}}$ to $l_{\text{gap}}$ has been chosen to guarantee that the entanglement wedges between different regions remain connected as the scale increases.
    }
    \label{fig:Markov_gap_dMoverb_01_all}
\end{figure}

\begin{figure}[htbp]
    \centering

    \begin{minipage}[b]{0.48\textwidth}
        \centering
        \includegraphics[width=\linewidth]{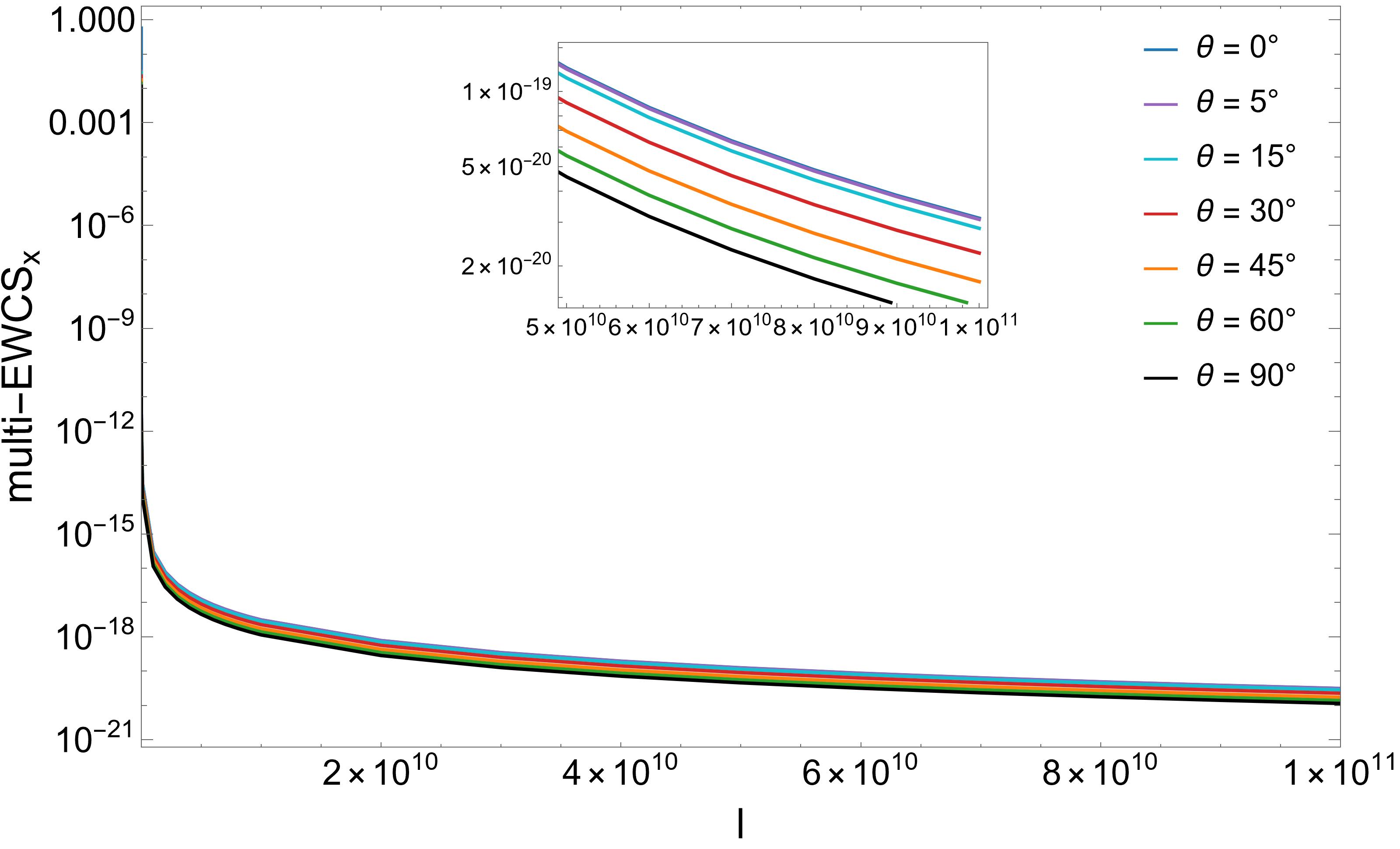}
        \small (a) Nontrivial phase: multi-EWCS\(_x(l)\)
    \end{minipage}
    \hfill
    \begin{minipage}[b]{0.48\textwidth}
        \centering
        \includegraphics[width=\linewidth]{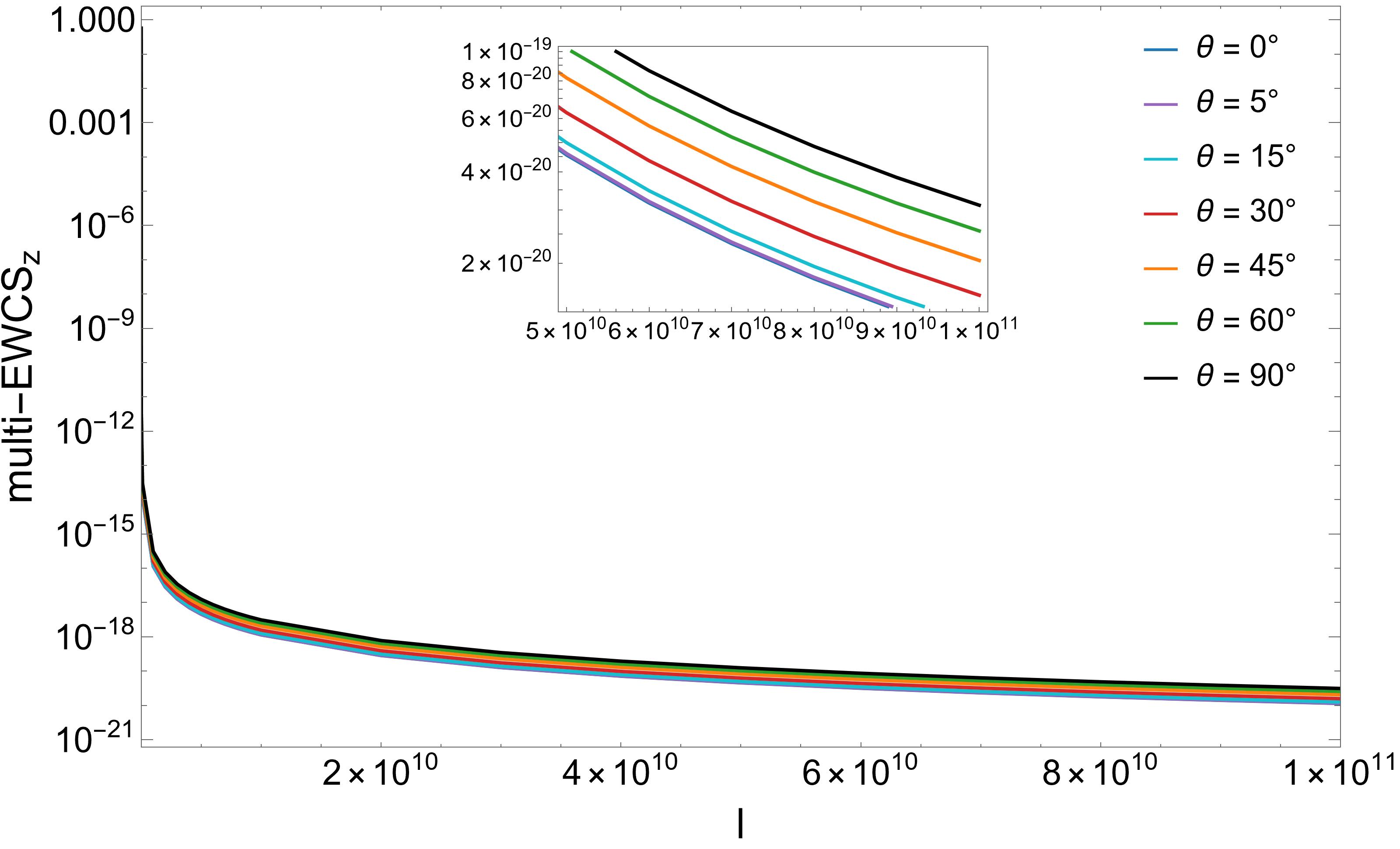}
        \small (b) Nontrivial phase: multi-EWCS\(_z(l)\)
    \end{minipage}

    \vspace{0.5em}

    \begin{minipage}[b]{0.48\textwidth}
        \centering
        \includegraphics[width=\linewidth]{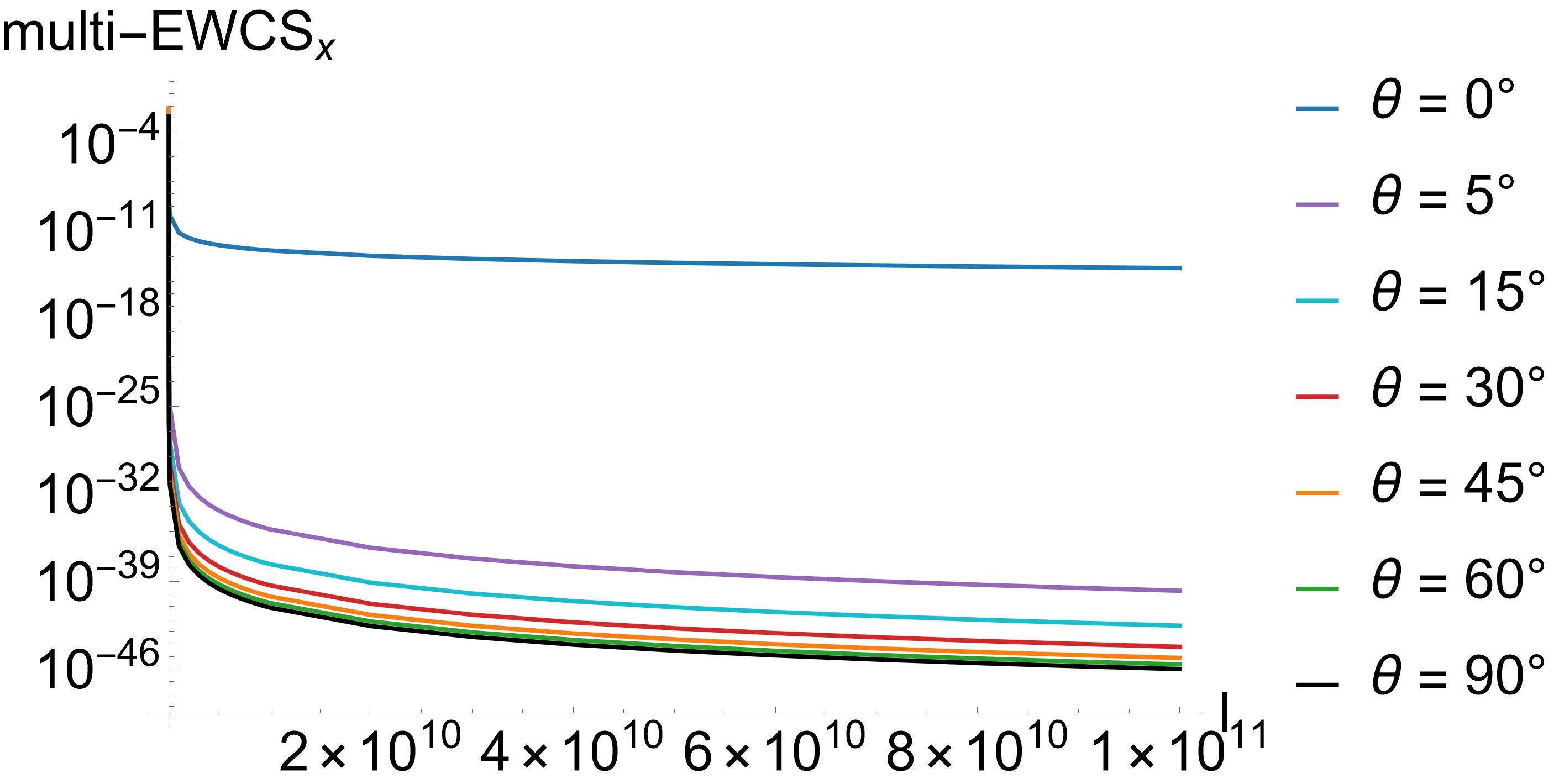}
        \small (c) Critical phase: multi-EWCS\(_x(l)\)
    \end{minipage}
    \hfill
    \begin{minipage}[b]{0.48\textwidth}
        \centering
        \includegraphics[width=\linewidth]{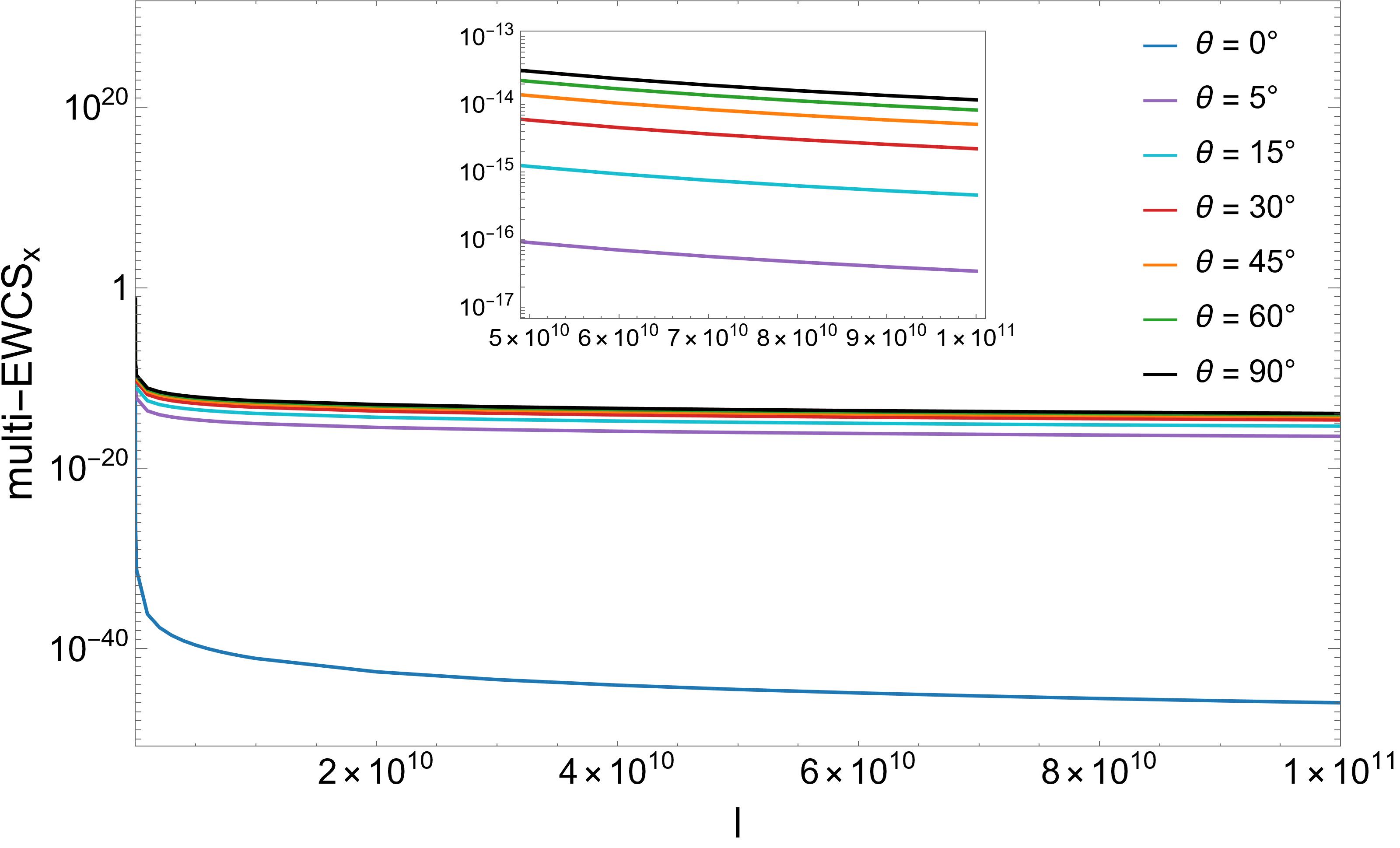}
        \small (d) Critical phase: multi-EWCS\(_z(l)\)
    \end{minipage}

    \vspace{0.5em}

    \begin{minipage}[b]{0.48\textwidth}
        \centering
        \includegraphics[width=\linewidth]{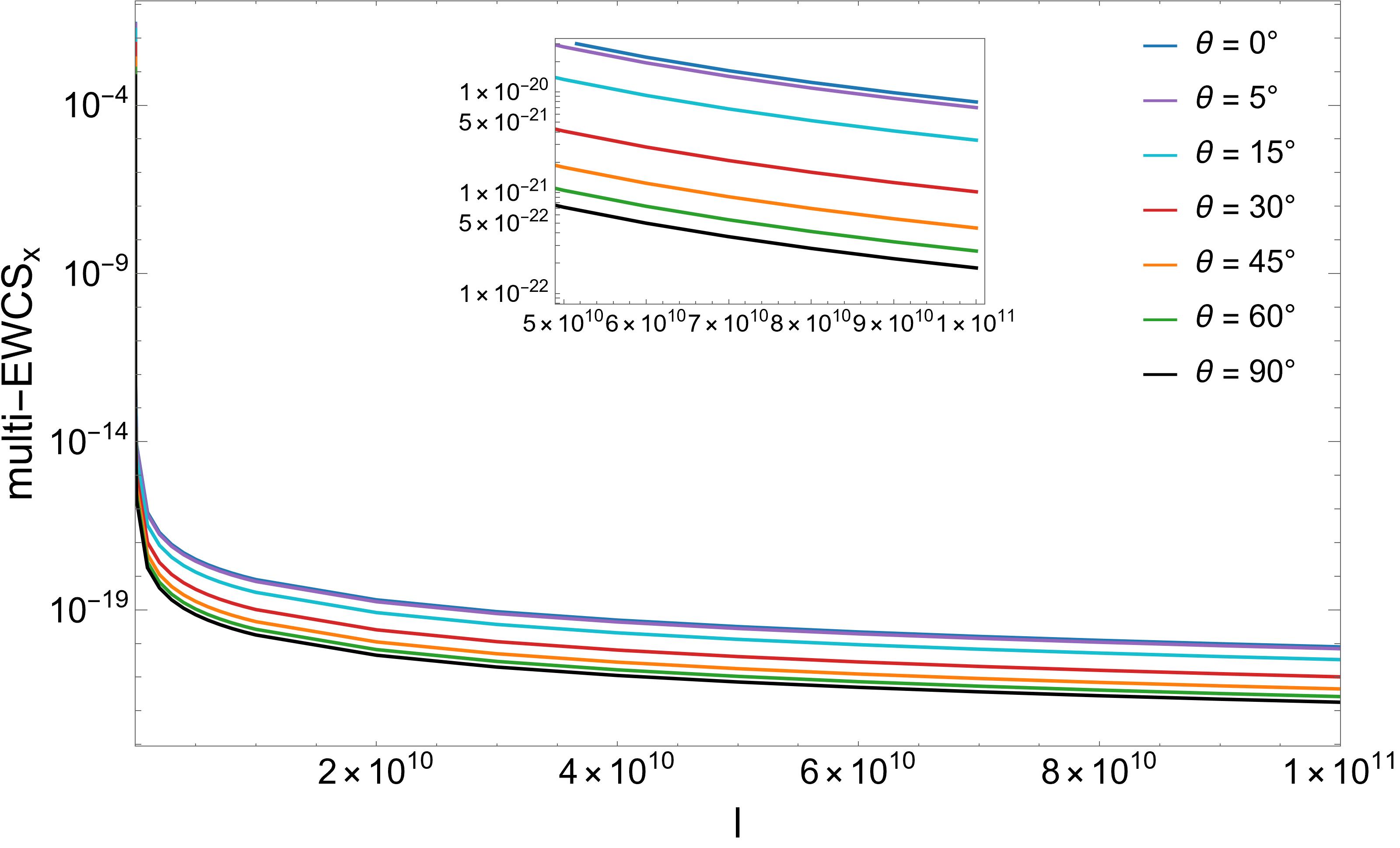}
        \small (e) Trivial phase: multi-EWCS\(_x(l)\)
    \end{minipage}
    \hfill
    \begin{minipage}[b]{0.48\textwidth}
        \centering
        \includegraphics[width=\linewidth]{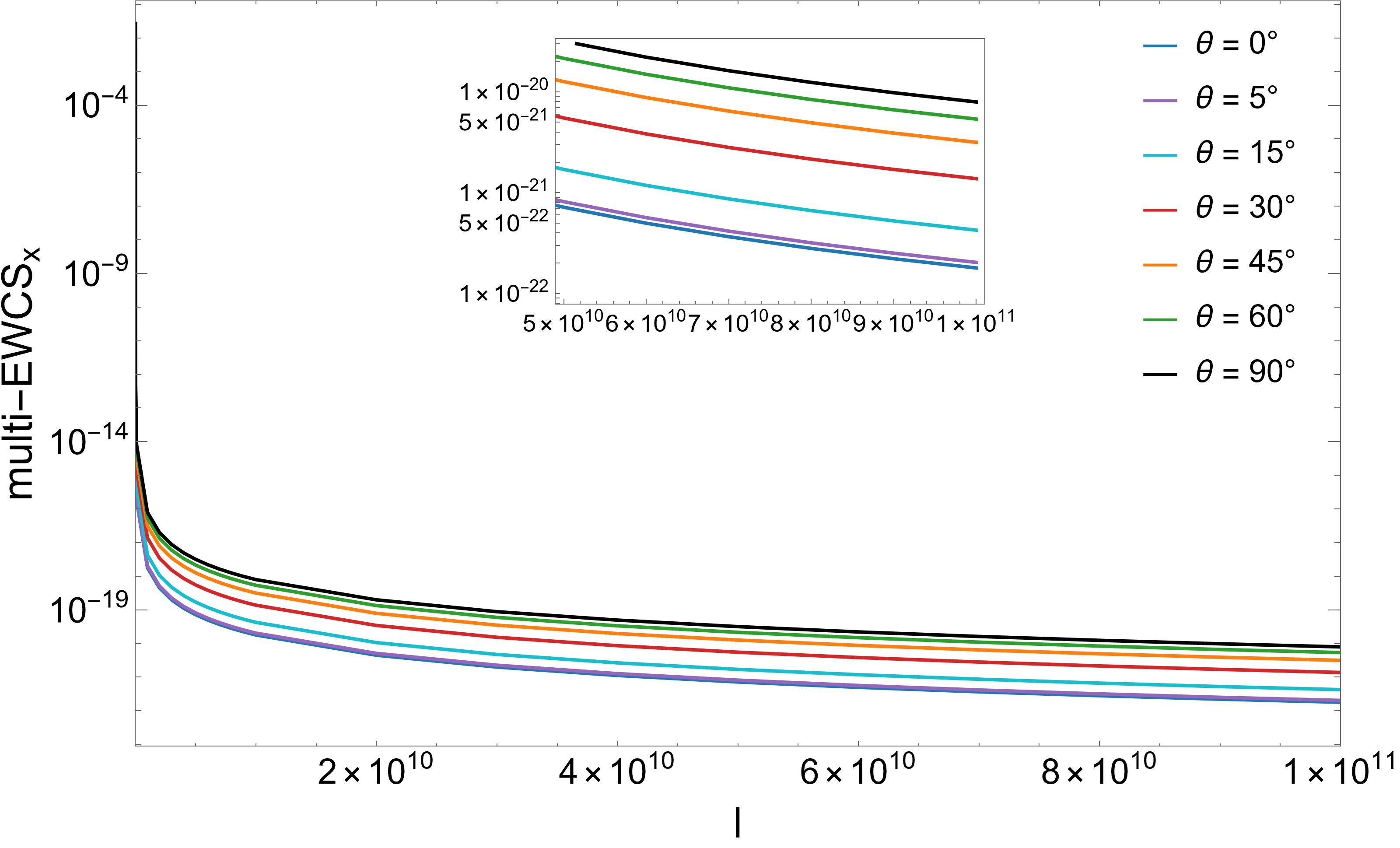}
        \small (f) Trivial phase: multi-EWCS\(_z(l)\)
    \end{minipage}

    \caption{
   Angular dependence of the multi-EWCS for strips with an
    angle \(\theta\) from the \(x\) axis. The nontrivial,
    critical, and trivial phases are evaluated at \(M/b=(M/b)_c-0.1\),
    \(M/b=(M/b)_c\), and \(M/b=(M/b)_c+0.1\), respectively. Panels
    (a,c,e) show \(\mathrm{multi-EWCS}_x(l)\), and panels (b,d,f) show
    \(\mathrm{multi-EWCS}_z(l)\). The curves correspond to
    \(\theta=0^\circ,5^\circ,15^\circ,30^\circ,45^\circ,60^\circ,90^\circ\). Without loss of generality, we set $l_{\text{strip}} = 10 l_{\text{gap}}$.
    }
    \label{fig:MultiEWCS_dMoverb_01_all}
\end{figure}

The angular behavior for the remaining quantities, CMI, EWCS, \(\kappa\), the Markov gap, and multi-EWCS, is shown in Figures~\ref{fig:CMI_dMoverb_01_all}--\ref{fig:MultiEWCS_dMoverb_01_all}. Unlike the noncritical \(c\)-function, these quantities decay at large \(l\). In the nontrivial and trivial phases, varying \(\theta\) mainly changes the  coefficient of this decay behavior, while the scaling power remains the same as in the fixed direction analysis. Thus the effect of the angle in these two phases is most cleanly captured by the coefficient \(a(\theta)\) of the power-law behavior. At the critical point, the angular dependence is stronger, because the scaling power can also vary with \(\theta\). These results show that the finite-angle strips interpolate continuously between the \(x\) and \(z\) results in the nontrivial and trivial phases, while the critical solution displays a more pronounced angular dependence. 

The angular plots above show that the finite angle of the strip affects the large \(l\) behavior through two pieces of data: the fitted power and the coefficient. For quantities with stable large \(l\) behavior, we write
\begin{equation}
X(l,\theta)
\sim
a(\theta)\,l^{\,b(\theta)} .
\label{eq:angular_large_l_general}
\end{equation}
Here \(X(l,\theta)\) denotes the density obtained after factoring out the regulator volume parallel to the strip, with all boundary lengths measured in units of \(b^{-1}\). Restoring the regulator volume only changes the overall normalization of \(a(\theta)\), and does not affect the extracted power.

In the nontrivial phase and in the trivial phase, the IR metric has the same scaling properties as the pure AdS$_5$ geometry at leading order. Therefore, the fitted power $b(\theta)$ is independent of \(\theta\) for each quantity. In these two phases one finds
\begin{equation}
b_c=0,
\qquad
b_{\rm CMI}=-4,
\qquad
b_{\rm EWCS}
=b_{\kappa}=b_{\rm MG}=b_{\rm multi-EWCS}=-2 .
\end{equation}
The \(c\)-function is the zeroth power case and appears as a plateau, whereas CMI, EWCS, \(\kappa\), the Markov gap, and multi-EWCS decay at large \(l\). For these decaying quantities, the angular dependence of the leading term is therefore carried by the coefficient \(a(\theta)\).

The critical solution is different. Its near-horizon geometry satisfies \(u(r)\sim r^2\) and \(h(r)\sim r^{2\beta}\), so the \(x\) and \(z\) directions have different IR scaling weights. As a result, the fitted power depends on \(\theta\). At the two limiting directions, the powers reduce to the fixed-direction results, summarized as:
\setlength{\tabcolsep}{0.25cm} 
\begin{table}[htbp]
\centering
\begin{tabular}{c c c c c c c}
\toprule
$\theta$ & $b_c$ & $b_{\mathrm{CMI}}$ & $b_{\mathrm{EWCS}}$ & $b_{\kappa}$ & $b_{\mathrm{MG}}$ & $b_{\mathrm{multi-EWCS}}$ \\
\midrule
$0$ & $1-\beta$ & $-3-\beta$ & $-1-\beta$ & $-1-\beta$ & $-1-\beta$ & $-1-\beta$ \\
$\frac{\pi}{2}$ & $2-\frac{2}{\beta}$ & $-2-\frac{2}{\beta}$ & $-\frac{2}{\beta}$ & $-\frac{2}{\beta}$ & $-\frac{2}{\beta}$ & $-\frac{2}{\beta}$ \\
\bottomrule
\end{tabular}
\end{table}

For intermediate angles, the critical powers are extracted numerically rather than described by a single angle independent exponent.
\section{Angular coefficients across phases and entanglement measures}
\label{Section5}
The angular analysis in Section~\ref{Section4} showed that, for strips that have an angle \(\theta\) in the \(x\)-\(z\) plane, the large \(l\) behavior can be written as
\begin{equation}
X(l,\theta)
\sim
a(\theta)\,l^{\,b(\theta)} .
\label{eq:angular_generic_scaling_form}
\end{equation}
In the nontrivial phase and in the trivial phase, the leading IR metric has the same AdS powers as the difference between these two phases mainly lies in the axial sector. Hence, for each entanglement quantity, the exponent \(b(\theta)\) is independent of \(\theta\) and is the same in the two phases. The leading power therefore does not distinguish the nontrivial phase from the trivial phase and the natural quantity to compare is the angular coefficient \(a(\theta)\). It measures how the leading large \(l\) term changes as the direction of the strip is varied between \(x\) and \(z\). Similar considerations have appeared in other anisotropic holographic settings, where the orientation of strip subregions plays an essential role in the entanglement response~\cite{Chu:2019uoh}.

In this section we extract and compare \(a(\theta)\) for CMI, EWCS, \(\kappa\), the Markov gap, and multi-EWCS. We use the representative parameters $M/b=(M/b)_c\mp0.1,$ with the minus sign for the nontrivial phase and the plus sign for the trivial phase. The critical solution is treated separately, because there the exponent \(b(\theta)\) also varies with \(\theta\).

\subsection{Fitting strategy and physical interpretation of the angular coefficient}
\label{subsec:fitting_protocol_physical_meaning}
We now parameterize the extracted angular coefficients in the nontrivial and trivial phases. Since the finite direction of the strip has projection \(\cos\theta\) onto \(x\) and projection \(\sin\theta\) onto \(z\), and since the strip does not distinguish the two opposite signs of this direction, the natural angular variables are \(\cos^2\theta\) and \(\sin^2\theta\). We fit the coefficients as
\begin{equation}
a_x(\theta)=\alpha_x^{(0)}+\alpha_x^{(1)}\left[\beta_x^{(1)}(\cos^2\theta)^{\gamma_x^{(1)}}+\bigl(1-\beta_x^{(1)}\bigr)(\cos^2\theta)^{\gamma_x^{(2)}}\right],
\label{eq:ax_general_fit}
\end{equation}
and
\begin{equation}
a_z(\theta)=\alpha_z^{(0)}+\alpha_z^{(1)}\left[\beta_z^{(1)}(\sin^2\theta)^{\gamma_z^{(1)}}+\bigl(1-\beta_z^{(1)}\bigr)(\sin^2\theta)^{\gamma_z^{(2)}}\right].
\label{eq:az_general_fit}
\end{equation}

For \(i=x,z\), the parameter \(\alpha_i^{(0)}\) gives the residual value of \(a_i(\theta)\) when the projection onto the corresponding direction vanishes. Thus \(\alpha_x^{(0)}\) is the value of \(a_x\) at \(\theta=\pi/2\), while \(\alpha_z^{(0)}\) is the value of \(a_z\) at \(\theta=0\). A larger \(\alpha_i^{(0)}\) means that the large \(l\) coefficient remains sizable even away from the corresponding limiting direction. The parameter \(\alpha_i^{(1)}\) measures the size of the angular variation.
We have
\begin{equation}
a_x(0)-a_x(\pi/2)=\alpha_x^{(1)},
\qquad
a_z(\pi/2)-a_z(0)=\alpha_z^{(1)} .
\end{equation}
It therefore quantifies the contrast between the limiting value and the residual value.

The parameter \(\beta_i^{(1)}\) fixes the relative weight of the two angular components in the fit. The exponents \(\gamma_i^{(1)}\) and \(\gamma_i^{(2)}\) control how rapidly the coefficient changes as the finite direction of the strip moves away from the corresponding limiting direction. Smaller exponents give broader angular profiles, while larger exponents give profiles more concentrated near the \(x\) or \(z\) limit.
\begin{table}[t]
\centering
\small
\setlength{\tabcolsep}{2.8pt}

\resizebox{\textwidth}{!}{
\begin{tabular}{lcccccccc}
\toprule
Measure
& \(\alpha^{(0)}_{N}\)
& \(\alpha^{(0)}_{T}\)
& \(\alpha^{(1)}_{x,N}\)
& \(\alpha^{(1)}_{z,N}\)
& \(\alpha^{(1)}_{x,T}\)
& \(\alpha^{(1)}_{z,T}\)
& \(\gamma^{(2)}_{x,N}\)
& \(\gamma^{(2)}_{x,T}\) \\
\midrule
CMI
& \(1.016\)
& \(0.231\times10^{-1}\)
& \(1.754\)
& \(1.754\)
& \(1.453\)
& \(1.453\)
& \(3.202\)
& \(25.443\) \\
EWCS
& \(0.169\)
& \(0.357\times10^{-2}\)
& \(0.293\)
& \(0.293\)
& \(0.213\)
& \(0.213\)
& \(4.115\)
& \(26.956\) \\
\(\kappa\)
& \(0.041\)
& \(0.557\times10^{-4}\)
& \(0.785\times10^{-1}\)
& \(0.785\times10^{-1}\)
& \(0.220\times10^{-2}\)
& \(0.220\times10^{-2}\)
& \(4.081\)
& \(27.110\) \\
Markov gap
& \(0.168\)
& \(0.167\times10^{-3}\)
& \(0.315\)
& \(0.315\)
& \(0.637\times10^{-2}\)
& \(0.637\times10^{-2}\)
& \(3.689\)
& \(20.000\) \\
multi-EWCS
& \(113.861\)
& \(1.792\)
& \(197.234\)
& \(197.234\)
& \(77.583\)
& \(77.583\)
& \(4.045\)
& \(19.827\) \\
\bottomrule
\end{tabular}}

\caption{Selected parameters of the angular fits in the nontrivial and trivial phases for \(M/b=(M/b)_c\mp0.1\). Here \(N\) and \(T\) denote the topologically nontrivial and trivial phases, respectively. The table lists the residual coefficient \(\alpha_{x,z}^{(0)}\), the total angular variation \(\alpha_{x,z}^{(1)}\), and the sharpness exponents \(\gamma_{x,N}^{(2)}\) and \(\gamma_{x,T}^{(2)}\) for the \(x\)-direction profile. These parameters characterize, respectively, the value that remains when the projection onto the corresponding direction vanishes, the contrast between the two limiting directions, and the degree to which the angular profile is concentrated near the \(x\) direction. The complete five-parameter fits are collected in Appendix~\ref{app:angular_fit_parameters}.
}
\label{tab:angular_fit_diagnostics}
\end{table}

The fit parameters therefore summarize three pieces of information: the residual coefficient away from the limiting direction, the total angular variation, and the sharpness of the angular profile. These are the quantities used below to compare the nontrivial and trivial phases. The most relevant parameters are listed in Table~\ref{tab:angular_fit_diagnostics}, while the complete five-parameter fits and the critical angular profiles are collected in Appendix~\ref{app:angular_fit_parameters}.
\subsection{Angular coefficients of the fitted entanglement quantities}
\label{subsec:fitted_observable_angular_coefficients}
\begin{figure}[t]
    \centering
    \includegraphics[width=0.48\textwidth]{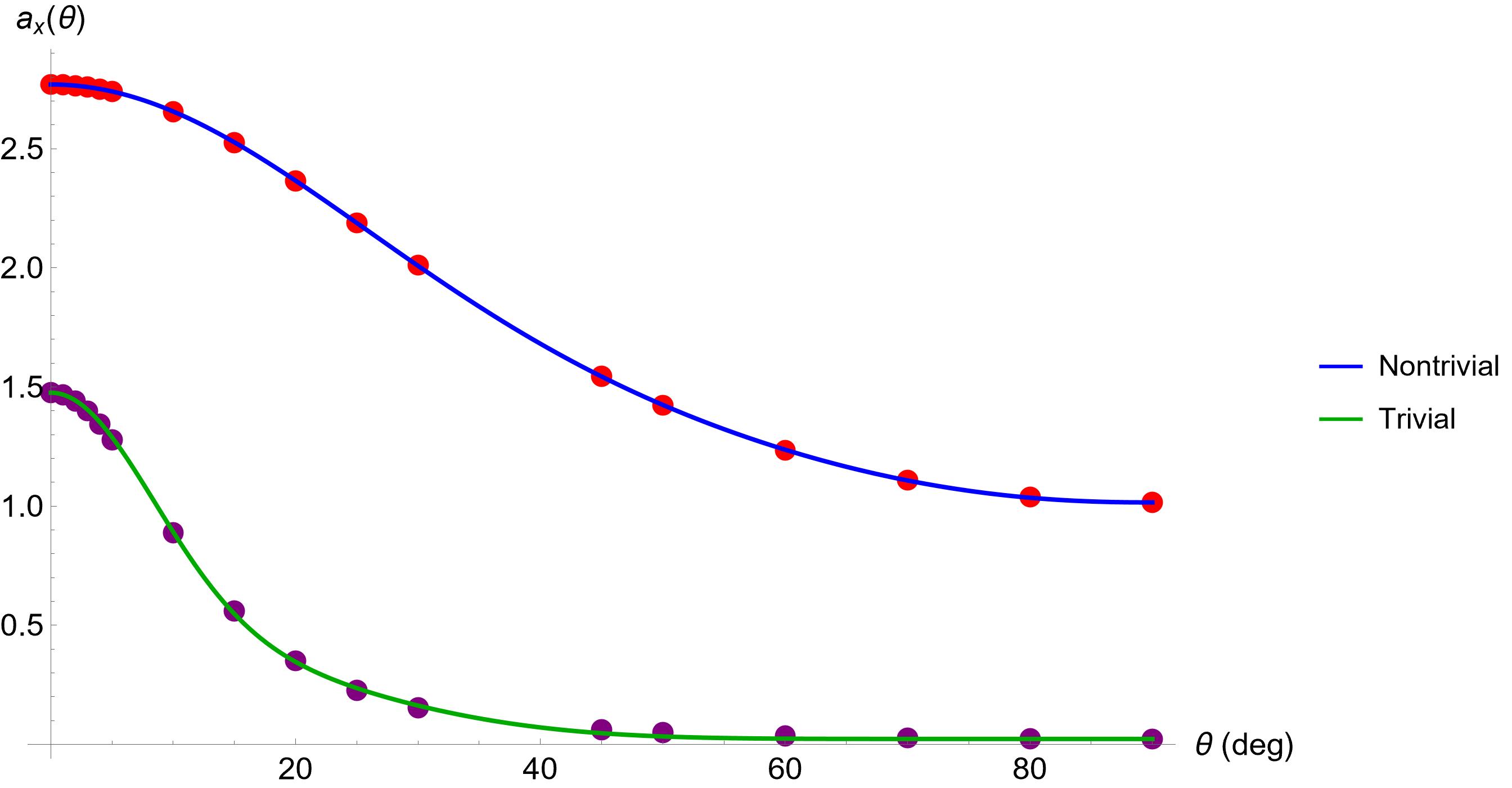}
    \hfill
    \includegraphics[width=0.48\textwidth]{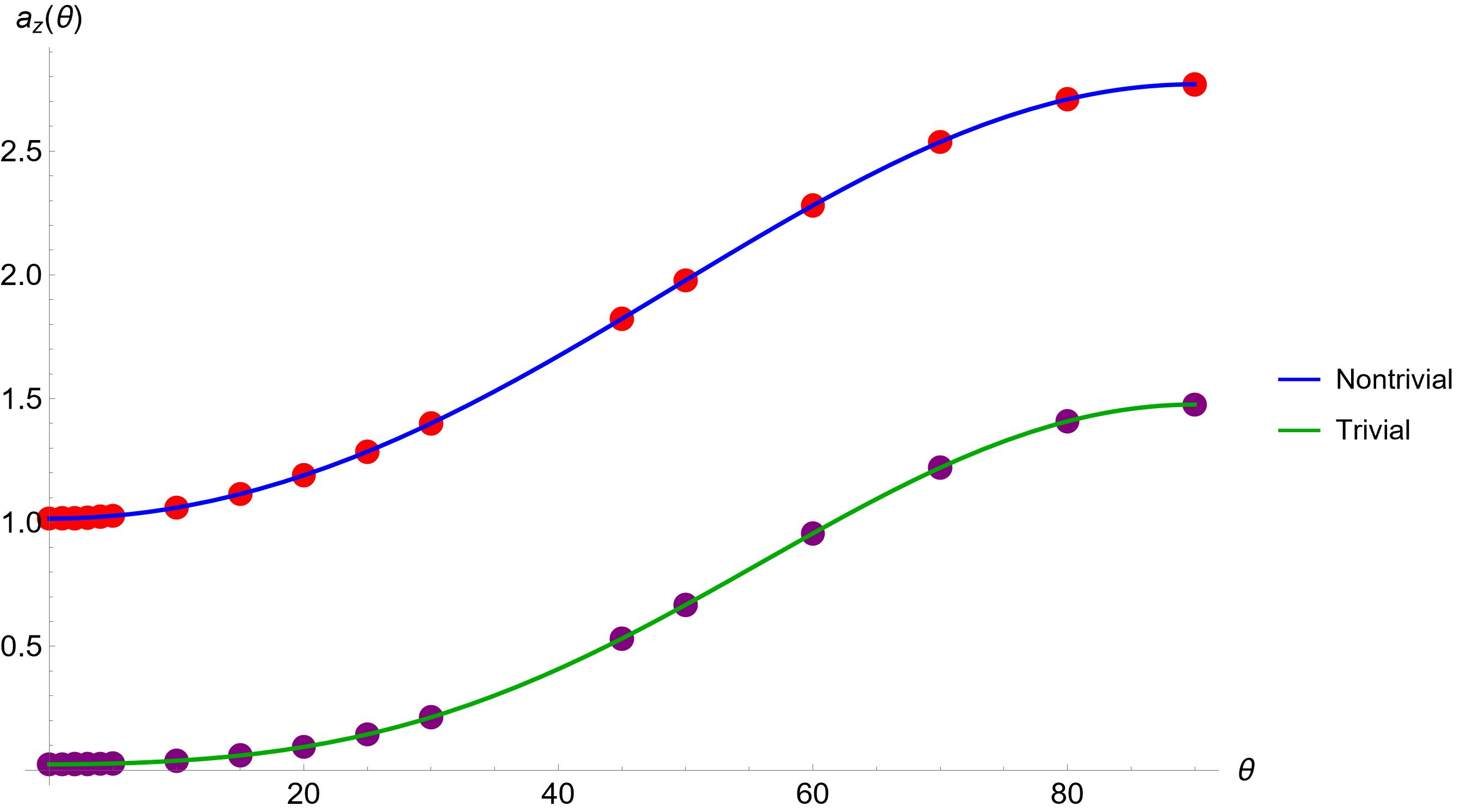}
    \caption[Fitted CMI angular coefficients]{
    Fitted angular coefficients of CMI for the representative noncritical
    pair. Left: \(a_x(\theta)\). Right: \(a_z(\theta)\). 
    }
    \label{fig:cmi_fitted_amplitudes}
\end{figure}

\begin{figure}[t]
    \centering
    \includegraphics[width=0.48\textwidth]{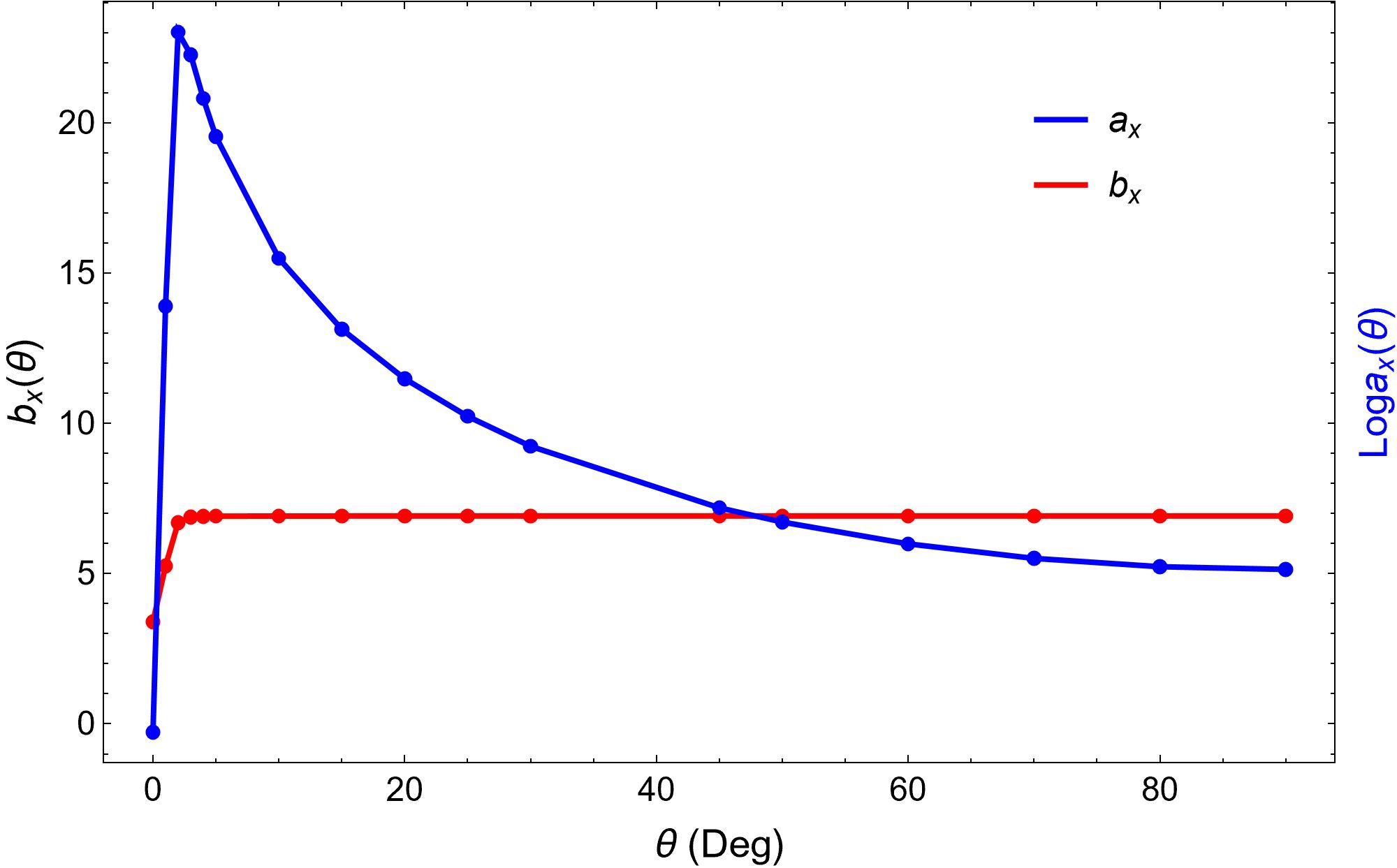}
    \hfill
    \includegraphics[width=0.48\textwidth]{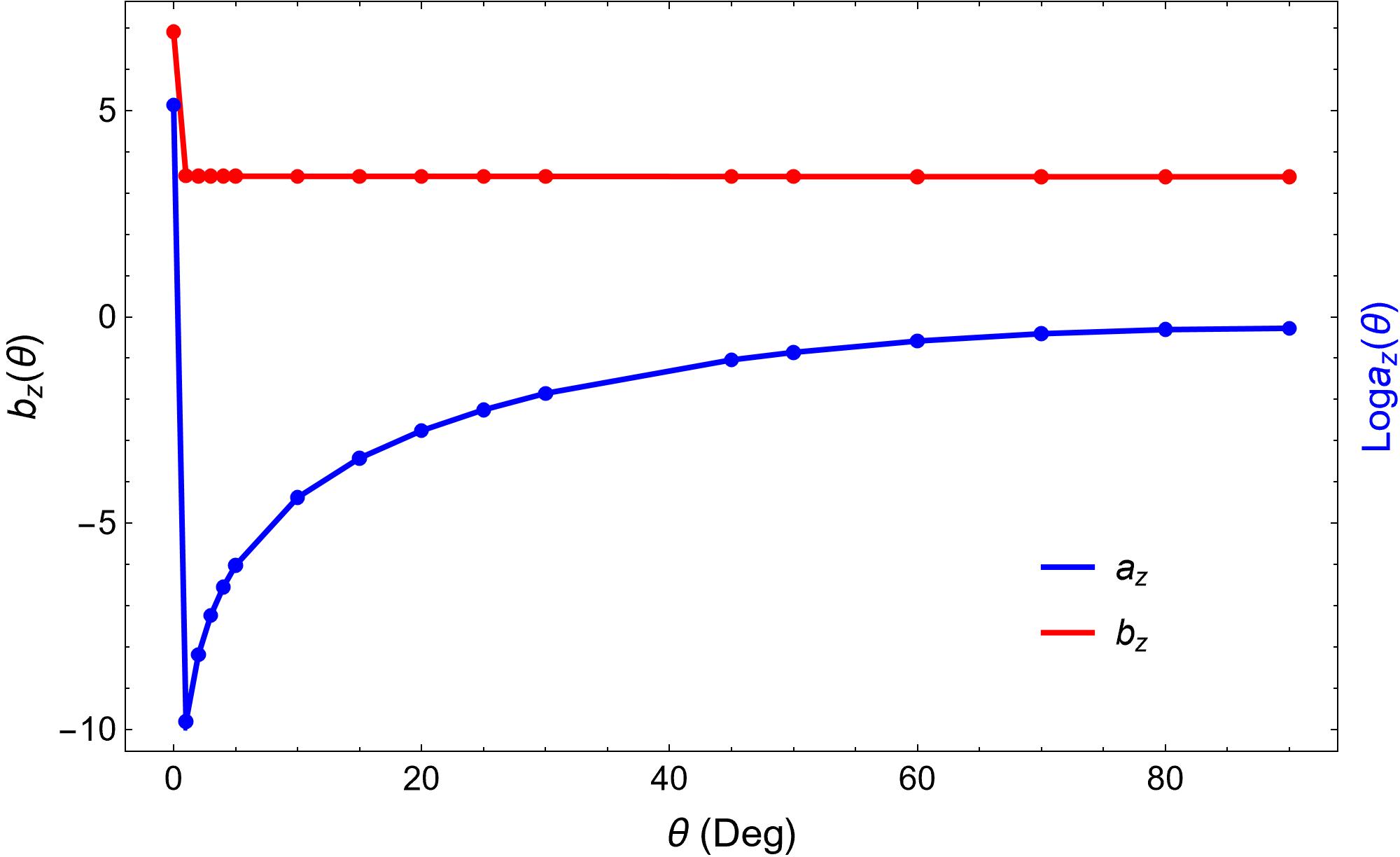}
    \caption[Critical CMI angular fitting coefficients]{
    Angular fitting coefficients for the critical CMI phase. The plots show
    \(\log a_x(\theta)\) and \(\log a_z(\theta)\), together with the
    corresponding fitted exponents \(b_x(\theta)\) and \(b_z(\theta)\).
    }
    \label{fig:cmi_critical_coefficients}
\end{figure}

\paragraph{CMI} \label{subsec:fitted_amplitude_CMI} Figure~\ref{fig:cmi_fitted_amplitudes} shows the fitted angular coefficients of the CMI in the nontrivial and trivial phases. The residual coefficient is much larger in the nontrivial phase, \(\alpha_N^{(0)}\simeq 1.016\), than in the trivial phase, \(\alpha_T^{(0)}\simeq 0.231\times10^{-1}\). The \(x\)-direction sharpness exponent also increases from \(\gamma_{x,N}^{(2)}=3.202\) to \(\gamma_{x,T}^{(2)}=25.443\). Thus the nontrivial phase retains a sizable coefficient over a wider range of angles, whereas in the trivial phase it is more strongly concentrated near the \(x\) direction. The critical solution is not fitted by the same global angular ansatz. The critical CMI data are shown in Figure~\ref{fig:cmi_critical_coefficients}, where both the coefficient and the fitted exponent vary with \(\theta\). This is consistent with the critical IR geometry, for which the large \(l\) power itself is angle dependent.

\begin{figure}[t]
    \centering
    \includegraphics[width=0.48\textwidth]{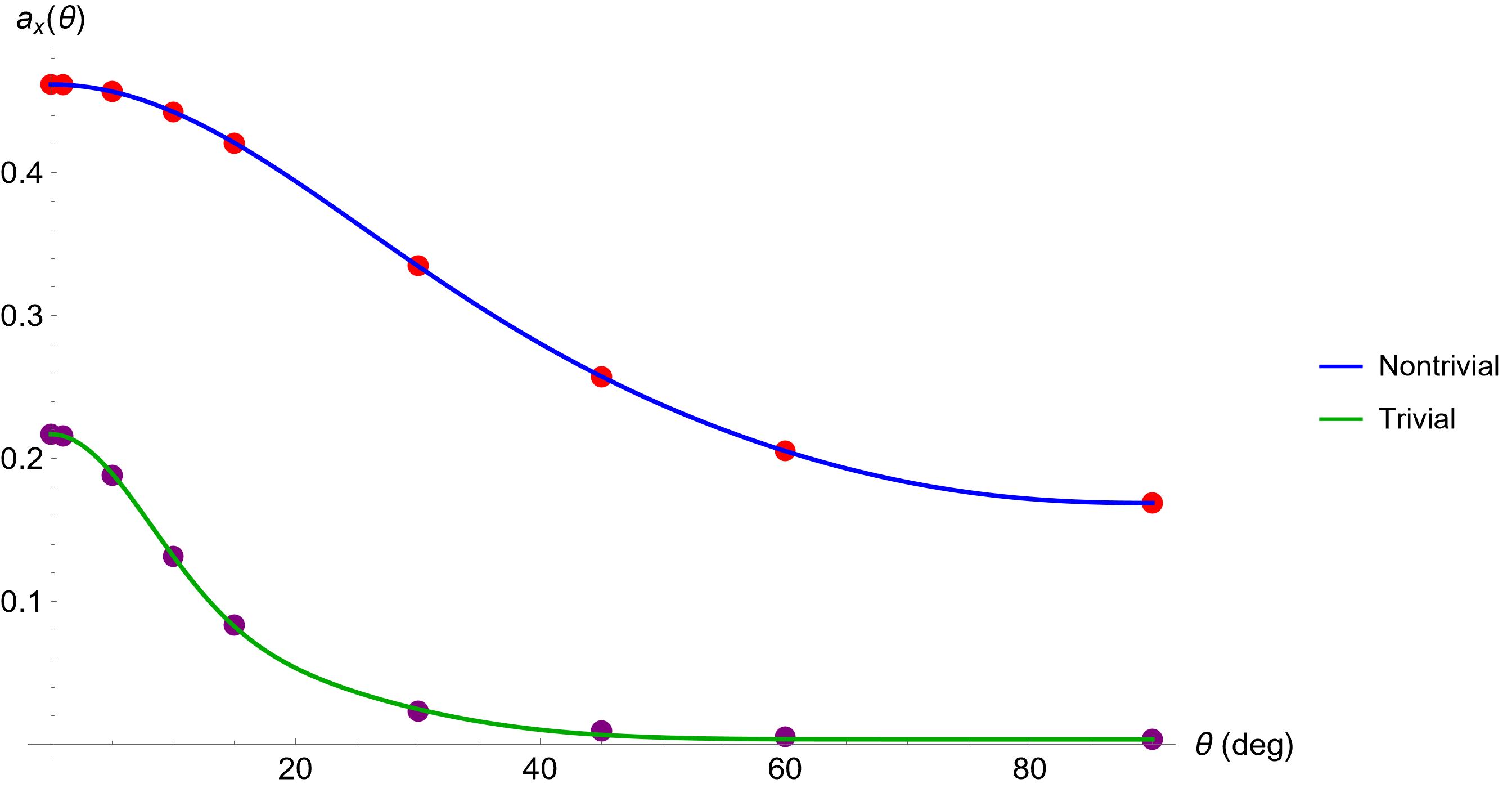}
    \hfill
    \includegraphics[width=0.48\textwidth]{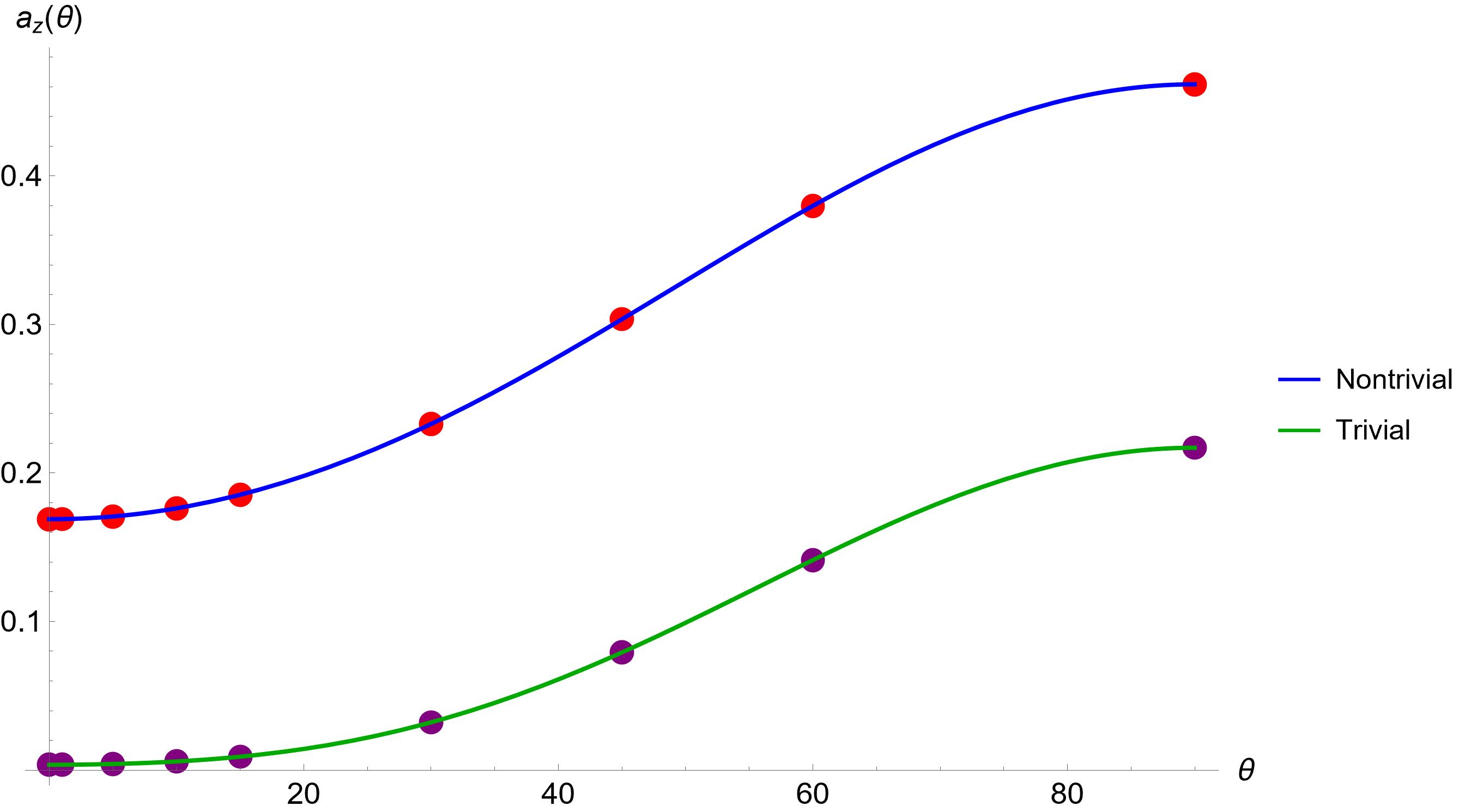}
    \caption[Fitted EWCS angular coefficients]{
    Fitted angular coefficients of EWCS for the representative noncritical
    pair. Left: \(a_x(\theta)\). Right: \(a_z(\theta)\).
    }
    \label{fig:ewcs_fitted_amplitudes}
\end{figure}

\paragraph{EWCS} \label{subsec:fitted_amplitude_EWCS} Figure~\ref{fig:ewcs_fitted_amplitudes} shows the corresponding fit for the EWCS. The same separation between the two phases is visible: the residual coefficient decreases from \(\alpha_N^{(0)}\simeq 0.169\) in the nontrivial phase to \(\alpha_T^{(0)}\simeq 0.357\times10^{-2}\) in the trivial phase. The \(x\)-direction sharpness exponent increases from \(\gamma_{x,N}^{(2)}=4.115\) to \(\gamma_{x,T}^{(2)}=26.956\). Hence the EWCS coefficient has broader angular support in the nontrivial phase and becomes strongly concentrated near the \(x\) direction in the trivial phase. The critical EWCS profile is kept as numerical angular data rather than being included in the noncritical fit. The corresponding critical coefficients and exponents are collected in Appendix~\ref{app:angular_fit_parameters}.

\begin{figure}[t]
    \centering
    \includegraphics[width=0.48\textwidth]{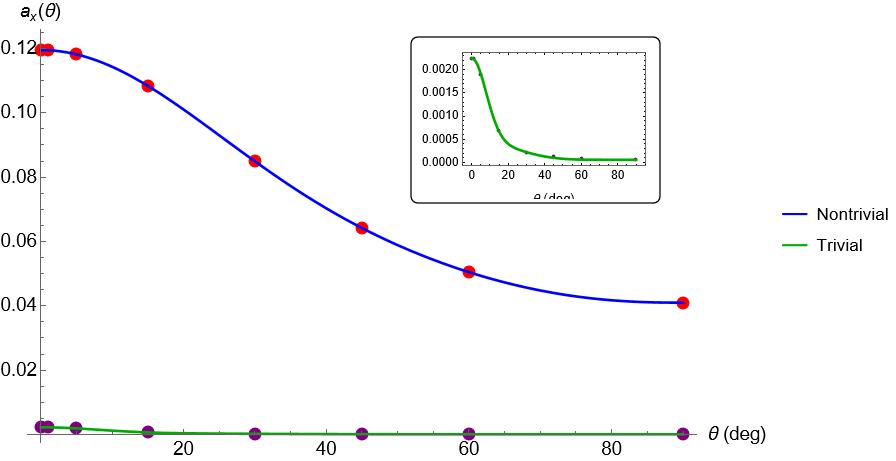}
    \hfill
    \includegraphics[width=0.48\textwidth]{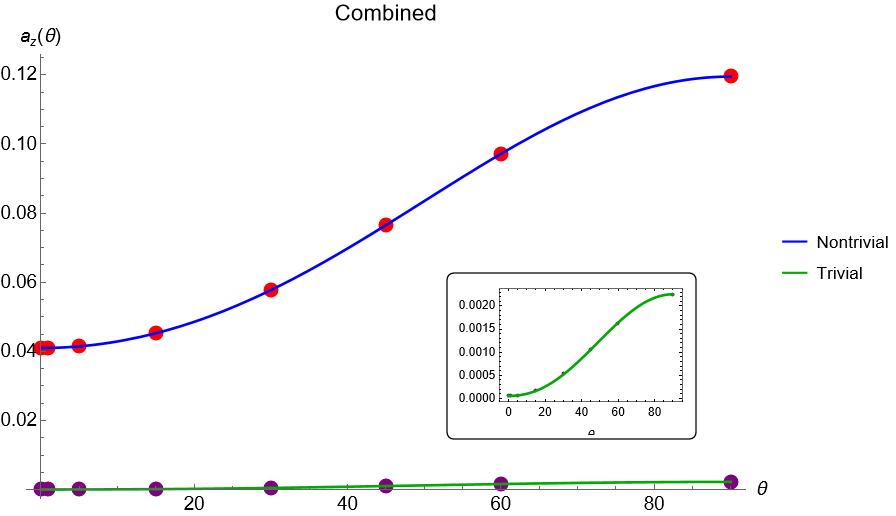}
    \caption[Fitted \(\kappa\) angular coefficients]{
    Fitted angular coefficients of \(\kappa\) for the representative
    noncritical pair. Left: \(a_x(\theta)\). Right: \(a_z(\theta)\). 
    }
    \label{fig:kappa_fitted_amplitudes}
\end{figure}

\paragraph{\texorpdfstring{\(\boldsymbol{\kappa}\)}{kappa}} \label{subsec:fitted_amplitude_kappa} Figure~\ref{fig:kappa_fitted_amplitudes} shows the fitted angular coefficients of \(\kappa\). The phase separation is especially sharp. The residual coefficient drops from \(\alpha_N^{(0)}\simeq 0.041\) in the nontrivial phase to \(\alpha_T^{(0)}\simeq 0.557\times10^{-4}\) in the trivial phase. The \(x\)-direction sharpness exponent increases from \(\gamma_{x,N}^{(2)}=4.081\) to \(\gamma_{x,T}^{(2)}=27.110\). The trivial phase therefore leaves only a narrowly concentrated angular profile.  Since the critical exponent is not approximately angle independent over the full angular range, the critical profile of \(\kappa\) is treated separately, as in the previous cases. The numerical data are reported in Appendix~\ref{app:angular_fit_parameters}.

\begin{figure}[t]
    \centering
    \includegraphics[width=0.48\textwidth]{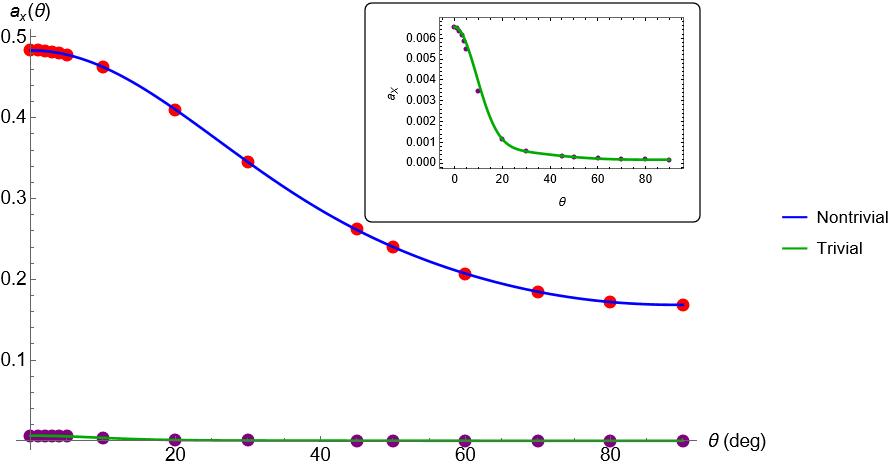}
    \hfill
    \includegraphics[width=0.48\textwidth]{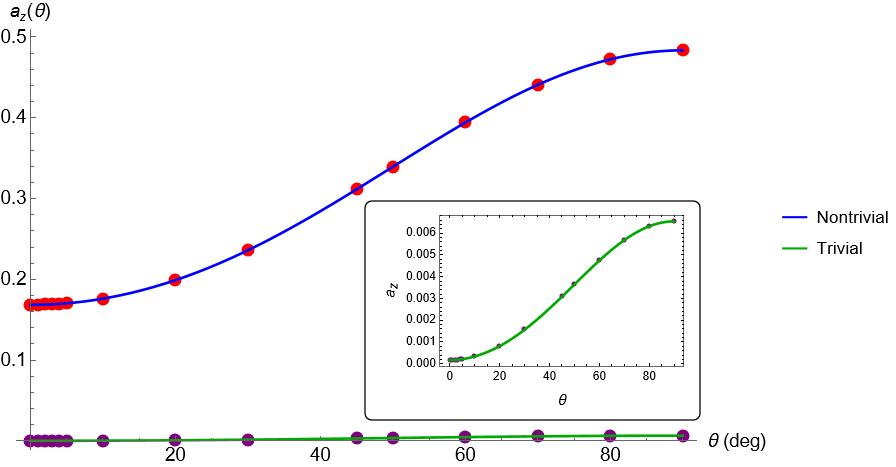}
    \caption[Fitted Markov gap angular coefficients]{
    Fitted angular coefficients of the Markov gap for the representative
    noncritical pair. Left: \(a_x(\theta)\). Right: \(a_z(\theta)\). 
    }
    \label{fig:markovgap_fitted_amplitudes}
\end{figure}

\paragraph{Markov gap} \label{subsec:fitted_amplitude_Markov_gap} Figure~\ref{fig:markovgap_fitted_amplitudes} shows the fitted angular coefficients of the Markov gap. The residual coefficient is \(\alpha_N^{(0)}\simeq 0.168\) in the nontrivial phase, but is reduced to \(\alpha_T^{(0)}\simeq 0.167\times10^{-3}\) in the trivial phase. The \(x\)-direction sharpness exponent increases from \(\gamma_{x,N}^{(2)}=3.689\) to \(\gamma_{x,T}^{(2)}=20.000\). Similar to the results above, we can see that the topologically nontrivial phase has more tripartite entanglement detected by these measures, while the tripartite entanglement is strongly suppressed away from the \(x\) direction in the trivial phase. The critical Markov gap profile is again kept as numerical data. Its angle-dependent coefficients and exponents are collected in Appendix~\ref{app:angular_fit_parameters}.

\begin{figure}[t]
    \centering
    \includegraphics[width=0.48\textwidth]{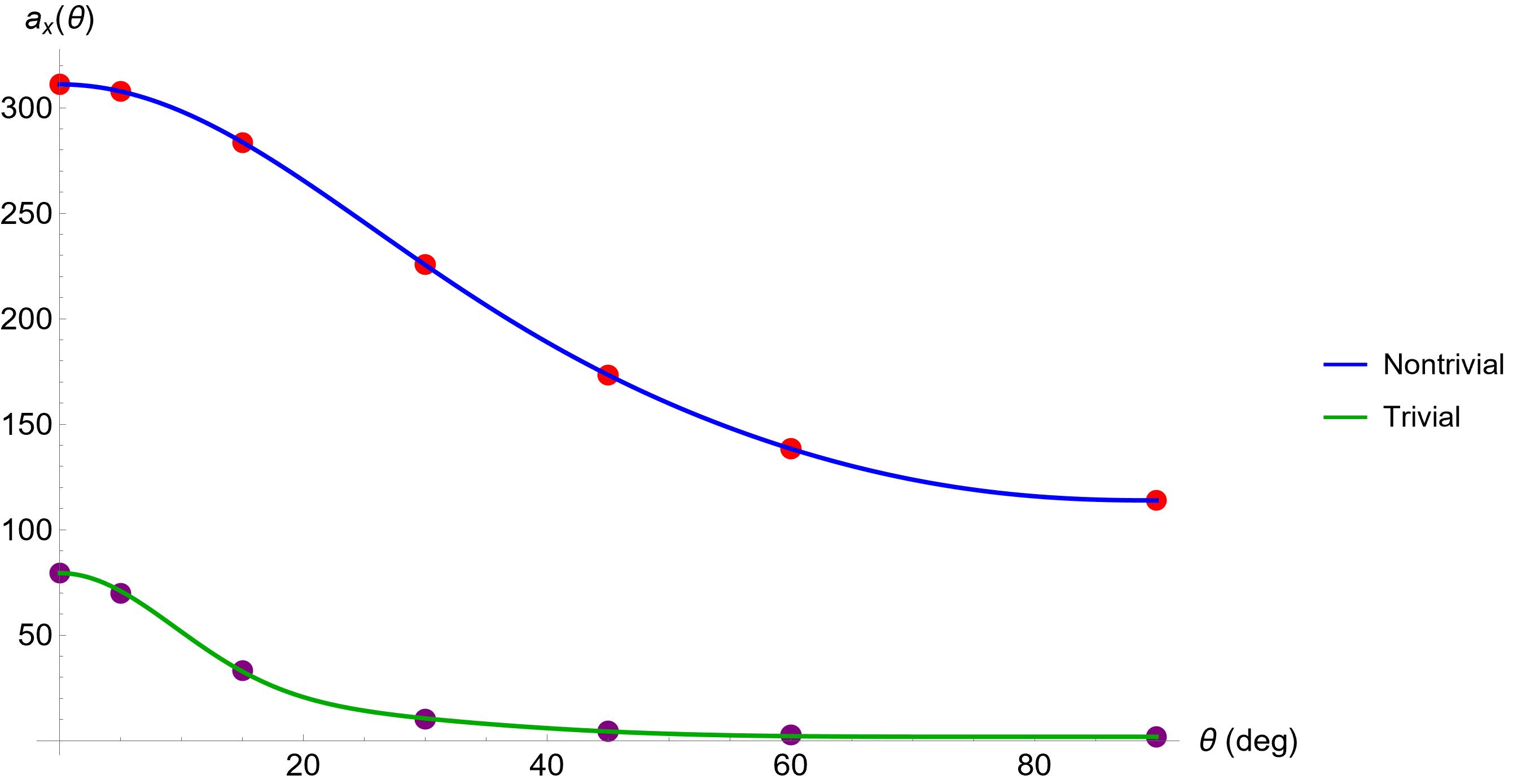}
    \hfill
    \includegraphics[width=0.48\textwidth]{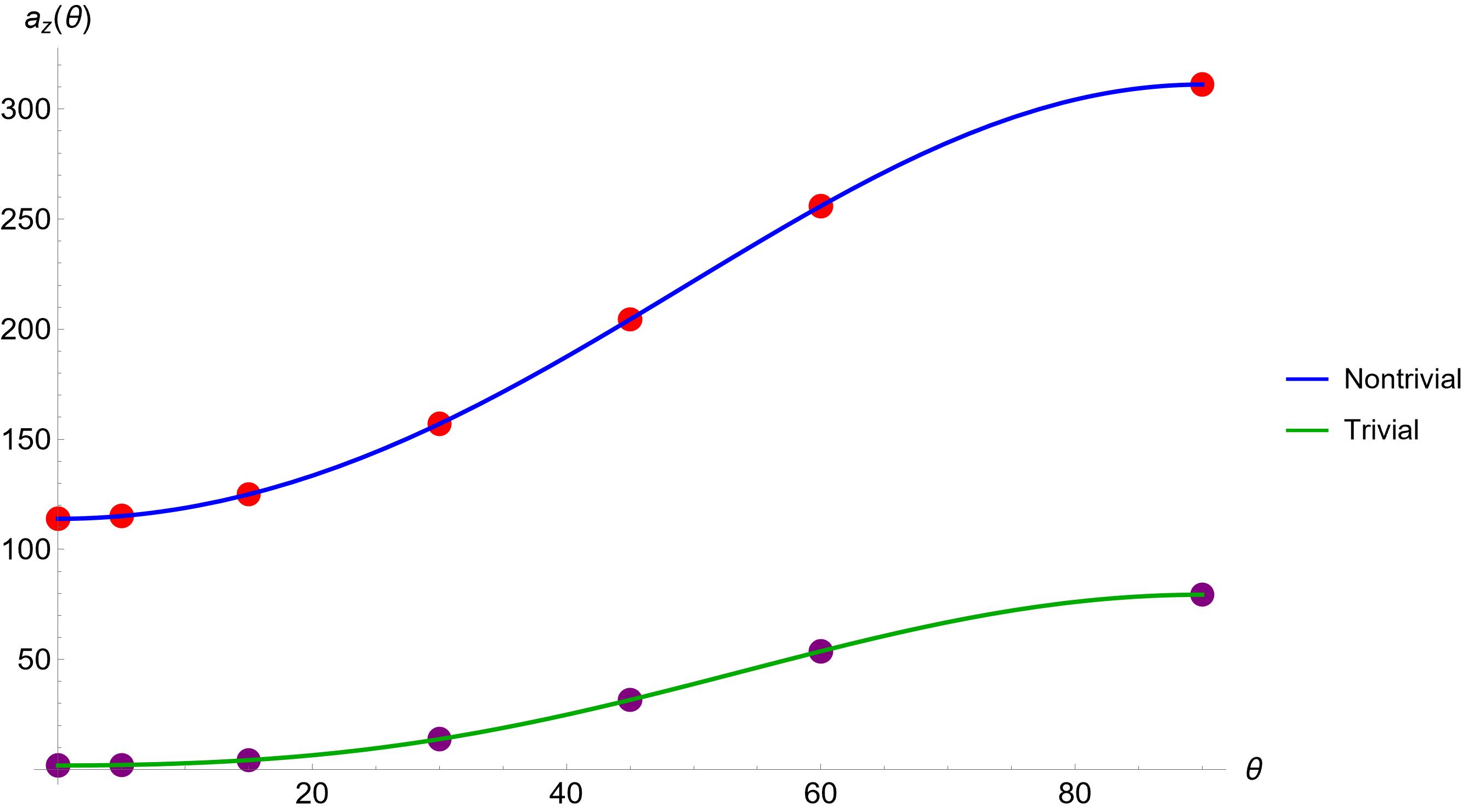}
    \caption[Fitted multi-EWCS angular coefficients]{
    Fitted angular coefficients of multi-EWCS for the representative
    noncritical pair. Left: \(a_x(\theta)\). Right: \(a_z(\theta)\).
    }
    \label{fig:multiewcs_fitted_amplitudes}
\end{figure}

\paragraph{multi-EWCS} \label{subsec:fitted_amplitude_MultiEWCS} Figure~\ref{fig:multiewcs_fitted_amplitudes} shows the fitted angular coefficients of the multi-EWCS. The nontrivial phase has a much larger residual coefficient, \(\alpha_N^{(0)}\simeq 113.861\), while the trivial phase gives \(\alpha_T^{(0)}\simeq 1.792\). The \(x\)-direction sharpness exponent increases from \(\gamma_{x,N}^{(2)}=4.045\) to \(\gamma_{x,T}^{(2)}=19.827\), showing the same narrowing of the angular profile in the trivial phase. The critical multi-EWCS profile is not described by the noncritical angular ansatz. The corresponding numerical coefficients and exponents are listed in Appendix~\ref{app:angular_fit_parameters}.

From the explicit fitting results above, we can see that the angular coefficients display a common pattern across CMI, EWCS, \(\kappa\), the Markov gap, and multi-EWCS. These quantities are constructed from different holographic data: entropy combinations, cross sections of connected entanglement wedges, multi-entropy networks, and multipartite wedge cross sections, and they capture different types of multipartite entanglement patterns as explained in Section \ref{Section2}. Nevertheless, their fitted coefficients show the same qualitative separation between the nontrivial phase and the trivial phase.

In every case listed in Table~\ref{tab:angular_fit_diagnostics}, the nontrivial phase has a larger residual coefficient \(\alpha_i^{(0)}\). Equivalently, the large \(l\) coefficient remains sizable over a wider range of angles. In the trivial phase, the residual coefficient is strongly suppressed and the \(x\)-direction sharpness exponent is much larger. The angular profile is therefore more concentrated near the \(x\) direction. This behavior is especially pronounced for \(\kappa\) and the Markov gap, but it is not limited to these two quantities.

The common behavior is a geometric statement about the large \(l\) regime of the holographic Weyl semimetal. Once the corresponding surfaces and wedge networks reach the near-horizon region, their leading angular dependence is controlled by the same anisotropic bulk geometry. In the nontrivial phase, the axial gauge field retains a finite IR value, and the angular profiles keep a larger residual component. In the trivial phase, \(A_z\) vanishes along the RG flow, and the angular profiles become more localized. The agreement among the different quantities therefore shows that the angular separation of the two phases is not tied to one particular type of entanglement structure, but follows from the common IR physics of the system.

The critical solution has to be interpreted separately. Its near-horizon geometry is anisotropic already at the level of the leading scaling powers. Consequently, the exponent \(b(\theta)\) varies with \(\theta\), and the critical data cannot be reduced to the same angle-independent-power fit used for the nontrivial and trivial phases. We therefore keep the critical angular coefficients and exponents as numerical profiles. The representative CMI profile is shown in Figure~\ref{fig:cmi_critical_coefficients}, and the remaining critical profiles are collected in Appendix~\ref{app:angular_fit_parameters}.

The angular fit thus has a specific role. For the nontrivial and trivial phases, the large \(l\) powers are the same, so the coefficients \(a_i(\theta)\) isolate the part of the entanglement geometry that still distinguishes the two IR solutions. They provide a static, nonlocal characterization of the anisotropic axial structure. These coefficients are not transport coefficients, but they are sensitive to the same IR axial data that also determine whether the anomalous Hall conductivity is finite or vanishes.
\section{Conclusions and discussion}
\label{Section6}
In this work we studied nonlocal tripartite and four-partite entanglement structures in the zero temperature holographic Weyl semimetal. We computed the CMI, EWCS, \(\kappa\), the Markov gap, multi-EWCS, and the two multi-EWCS based signals \(\Delta\) and \(g\) for strip regions, and followed their dependence on the strip width \(l\). These quantities are nonzero at finite \(l\), showing that the holographic Weyl semimetal contains nontrivial tripartite and four-partite entanglement structures. At large \(l\), the CMI, EWCS, \(\kappa\), the Markov gap, and multi-EWCS based four-partite signals all decay as powers of \(l\), rather than exponentially. This power-law decay is the expected long-distance behavior of a gapless system, where the corresponding long range entanglement is not cut off by a finite correlation length. The scaling powers of all entanglement quantities are fixed by the  IR geometry. Thus the large \(l\) behavior directly relates the multipartite entanglement structure to the IR physics of each phase.

We then used these quantities to characterize the topological quantum phase transition as \(M/b\) is varied. At fixed strip width, all the tripartite and four-partite entanglement quantities considered in this work develop clear features near the critical value \((M/b)_c\). Thus the transition is not only visible in the entanglement entropy or in the holographic \(c\)-function, but also in multipartite entanglement structures. This is consistent with the physical expectation that different phases are distinguished not only by their transport properties, but also by different entanglement patterns underlying each phase. We also studied strips whose finite direction makes an angle \(\theta\) in the \(x\)-\(z\) plane. The angular dependence of the multipartite quantities captures the anisotropy introduced by the axial source. In particular, the angular coefficients distinguish the nontrivial phase, where the IR axial structure survives, from the trivial phase, where this structure is removed along the RG flow. Thus the angular analysis shows that these multipartite entanglement quantities provide nonlocal information about the same low-energy anisotropic structure that also controls the anomalous Hall conductivity.

Several directions remain open. First, it would be useful to derive the finite-angle scaling behavior analytically from the critical IR geometry. In particular, the small angle regime should clarify how the exponent \(b(\theta)\) connects the \(x\) and \(z\) limits and why the numerical critical profiles are not captured by the same angle independent-power fit used in the nontrivial and trivial phases. Second, finite density and time-dependent backgrounds would allow one to test how the angular coefficients and multipartite entanglement structures evolve when the IR geometry is changed by charge density, quenches, or equilibration. These extensions would help determine how robust the entanglement characterization of the holographic Weyl semimetal is beyond the zero temperature static backgrounds considered here. Third, it would be interesting to search for holographic gapless systems whose IR geometries support stronger long-distance multipartite entanglement. In the present Weyl semimetal, the finite multipartite quantities remain nonzero at finite strip width but decay as powers of \(l\) and vanish as \(l\to\infty\). Other IR geometries may lead to slower decay, enhanced multipartite coefficients, or even more persistent long-distance structures. We will investigate such possibilities in forthcoming work.

\subsection*{Acknowledgments}
We thank Xin-Xiang Ju, Karl Landsteiner, Bo-Hao Liu, Yan Liu, Wen-Bin Pan, Xin-Meng Wu, Bo-Yu Xu and Yang Zhao for helpful discussions. This work was supported by the National Natural Science Foundation of China (Grant Nos. 12405078, 12575068). This work is also supported by the 2115 Talent Development Program of China Agricultural University.
\appendix

\section{Supplementary dependence on strip width in the principal directions}
\label{app:fixed_direction_supplement}
\label{app:benchmark_comparisons}

This appendix collects the supplementary fixed direction data used in subsection~\ref{subsec:symmetric_slices}. The strips are finite either along the transverse \(x\) direction or along the \(z\) direction where the Weyl nodes locate. In the main text, the \(c\)-function and the multi-EWCS are used as representative examples. Here we give the corresponding plots for CMI, EWCS, \(\kappa\), and the Markov gap.

We denote the symmetric distance from the critical point by $\Delta(M/b)=\left|M/b-(M/b)_c\right|$\footnote{This notation is distinct from the four-partite signal \(\Delta\) introduced in Section~\ref{Section2}}. The comparison between \(\Delta(M/b)=0.02\) and \(\Delta(M/b)=0.1\) tests how proximity to the critical solution affects the finite scale crossover. The data show the same pattern as in the main text: moving closer to the critical point extends the range of \(l\) influenced by the critical scaling geometry, while moving farther away makes the asymptotic noncritical behavior visible earlier. No new large \(l\) scaling pattern is generated within a fixed noncritical phase.

\begin{figure}[htbp]
    \centering

    \begin{minipage}[b]{0.48\textwidth}
        \centering
        \includegraphics[width=\linewidth]{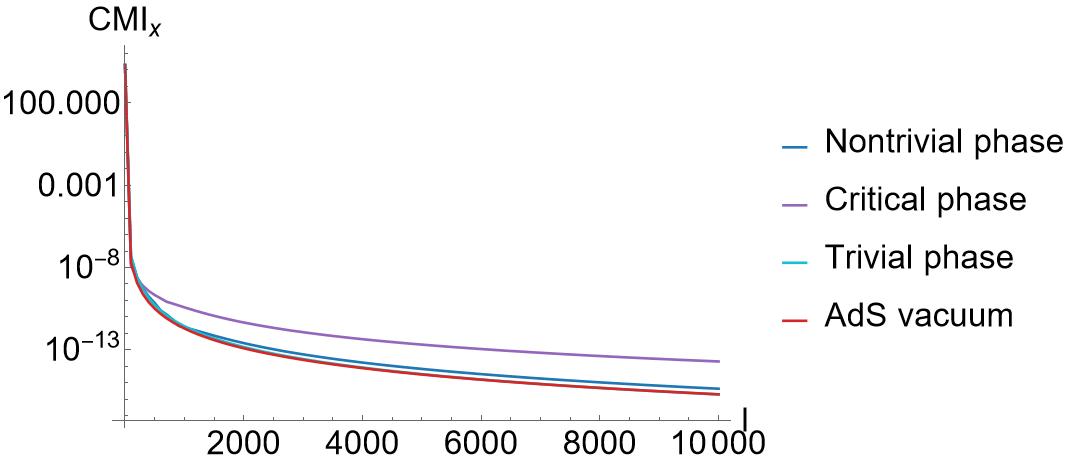}
        \small (a) \(\Delta(M/b)=0.02\): CMI\(_x(l)\)
    \end{minipage}
    \hfill
    \begin{minipage}[b]{0.48\textwidth}
        \centering
        \includegraphics[width=\linewidth]{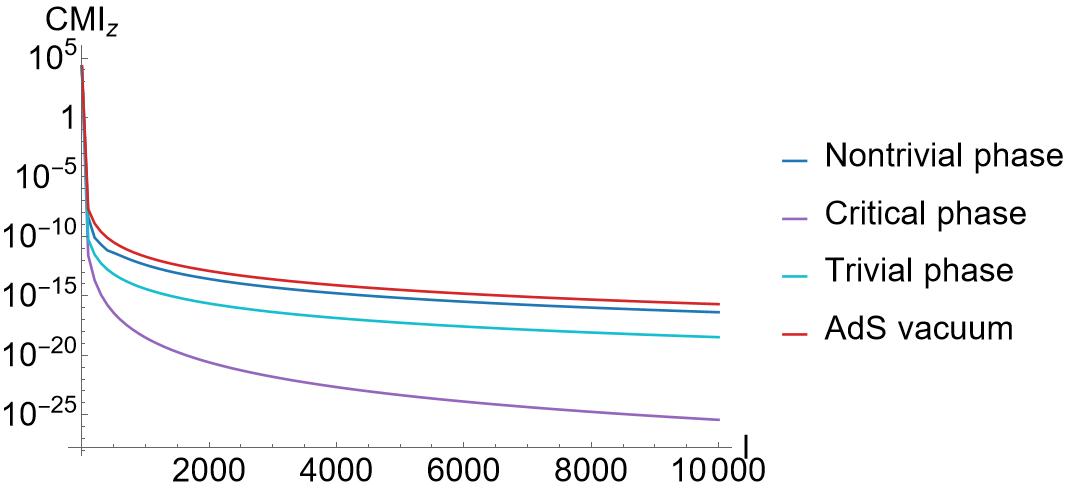}
        \small (b) \(\Delta(M/b)=0.02\): CMI\(_z(l)\)
    \end{minipage}

    \vspace{0.5em}

    \begin{minipage}[b]{0.48\textwidth}
        \centering
        \includegraphics[width=\linewidth]{Figures/DeltaMoverb=0.1-l_x-CMI.jpg}
        \small (c) \(\Delta(M/b)=0.1\): CMI\(_x(l)\)
    \end{minipage}
    \hfill
    \begin{minipage}[b]{0.48\textwidth}
        \centering
        \includegraphics[width=\linewidth]{Figures/DeltaMoverb=0.1-l_z-CMI.jpg}
        \small (d) \(\Delta(M/b)=0.1\): CMI\(_z(l)\)
    \end{minipage}

    \caption{
    Supplementary CMI data in the principal directions. Panels (a,b) show the
    slice \(\Delta(M/b)=0.02\), and panels (c,d) show the slice
    \(\Delta(M/b)=0.1\). The smaller distance from the critical point keeps
    the curves closer to the critical scaling regime over a wider range of
    \(l\), while the larger distance exposes the noncritical large \(l\)
    behavior earlier.
    }
    \label{fig:offcritical_cmi_appendix}
\end{figure}

\begin{figure}[htbp]
    \centering

    \begin{minipage}[b]{0.48\textwidth}
        \centering
        \includegraphics[width=\linewidth]{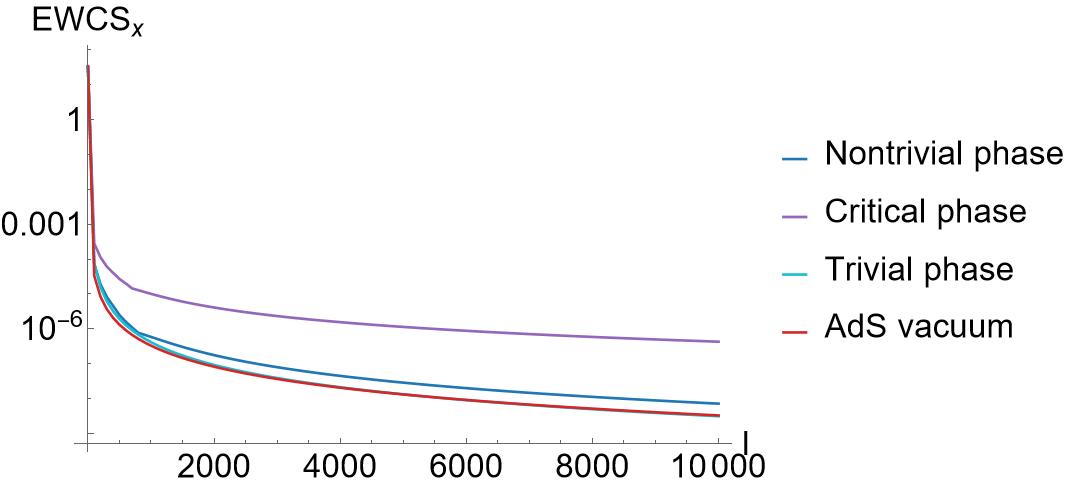}
        \small (a) \(\Delta(M/b)=0.02\): EWCS\(_x(l)\)
    \end{minipage}
    \hfill
    \begin{minipage}[b]{0.48\textwidth}
        \centering
        \includegraphics[width=\linewidth]{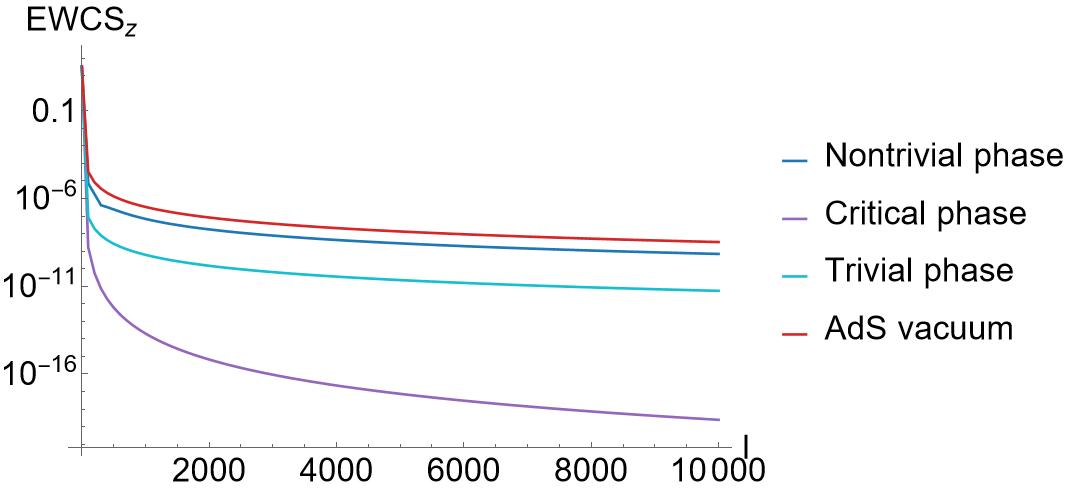}
        \small (b) \(\Delta(M/b)=0.02\): EWCS\(_z(l)\)
    \end{minipage}

    \vspace{0.5em}

    \begin{minipage}[b]{0.48\textwidth}
        \centering
        \includegraphics[width=\linewidth]{Figures/DeltaMoverb=0.1-l_x-EWCS.jpg}
        \small (c) \(\Delta(M/b)=0.1\): EWCS\(_x(l)\)
    \end{minipage}
    \hfill
    \begin{minipage}[b]{0.48\textwidth}
        \centering
        \includegraphics[width=\linewidth]{Figures/DeltaMoverb=0.1-l_z-EWCS.jpg}
        \small (d) \(\Delta(M/b)=0.1\): EWCS\(_z(l)\)
    \end{minipage}

    \caption{
    Supplementary EWCS data in the principal directions. The comparison shows the same finite scale effect as the CMI data: the slice closer to the critical point retains the influence of the critical scaling geometry over a longer interval in \(l\), whereas the slice farther away reaches the noncritical large \(l\) regime more rapidly. Without loss of generality, an appropriate ratio of $l_{\text{strip}}$ to $l_{\text{gap}}$ has been chosen to guarantee that the entanglement wedges between different regions remain connected as the scale increases.
    }
    \label{fig:offcritical_ewcs_appendix}
\end{figure}

\begin{figure}[htbp]
    \centering

    \begin{minipage}[b]{0.48\textwidth}
        \centering
        \includegraphics[width=\linewidth]{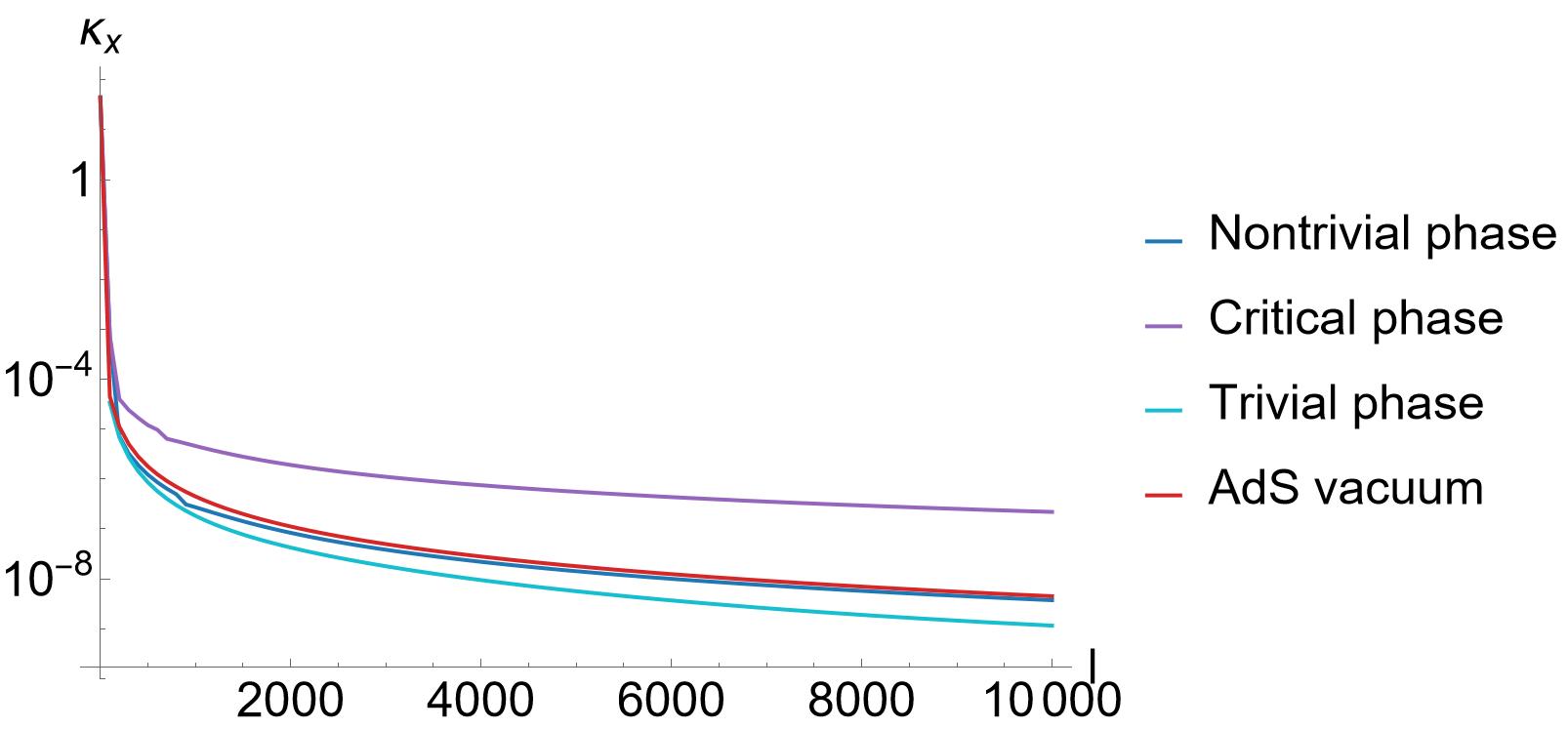}
        \small (a) \(\Delta(M/b)=0.02\): \(\kappa_x(l)\)
    \end{minipage}
    \hfill
    \begin{minipage}[b]{0.48\textwidth}
        \centering
        \includegraphics[width=\linewidth]{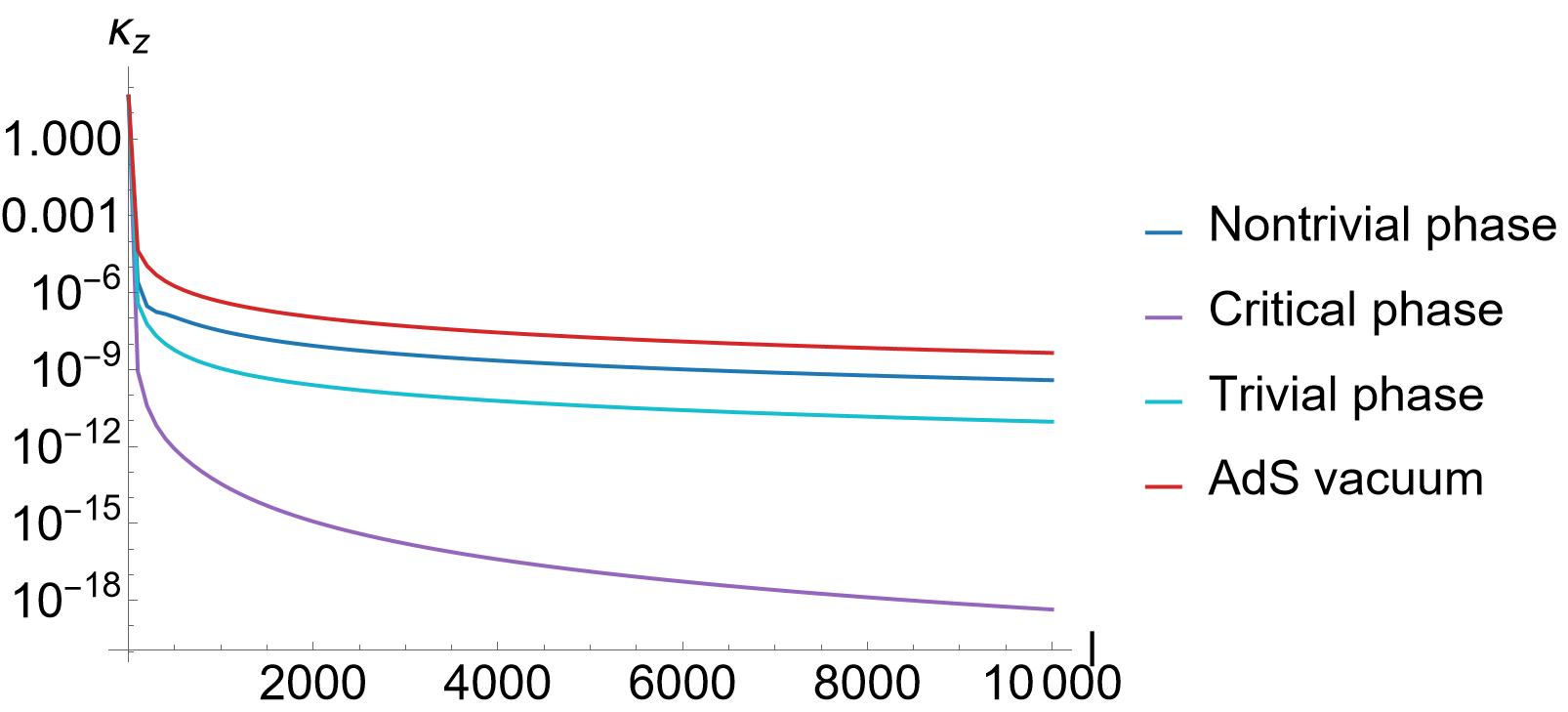}
        \small (b) \(\Delta(M/b)=0.02\): \(\kappa_z(l)\)
    \end{minipage}

    \vspace{0.5em}

    \begin{minipage}[b]{0.48\textwidth}
        \centering
        \includegraphics[width=\linewidth]{Figures/DeltaMoverb=0.1-l_x-kappa.jpg}
        \small (c) \(\Delta(M/b)=0.1\): \(\kappa_x(l)\)
    \end{minipage}
    \hfill
    \begin{minipage}[b]{0.48\textwidth}
        \centering
        \includegraphics[width=\linewidth]{Figures/DeltaMoverb=0.1-l_z-kappa.jpg}
        \small (d) \(\Delta(M/b)=0.1\): \(\kappa_z(l)\)
    \end{minipage}

    \caption{
    Supplementary \(\kappa\) data in the principal directions. The same
    crossover pattern persists after the reducible pairwise entropy
    contributions are subtracted. Thus the finite scale influence of the
    nearby critical geometry is also visible in the tripartite sector selected
    by the multi-entropy construction.
    }
    \label{fig:offcritical_kappa_appendix}
\end{figure}

\begin{figure}[htbp]
    \centering

    \begin{minipage}[b]{0.48\textwidth}
        \centering
        \includegraphics[width=\linewidth]{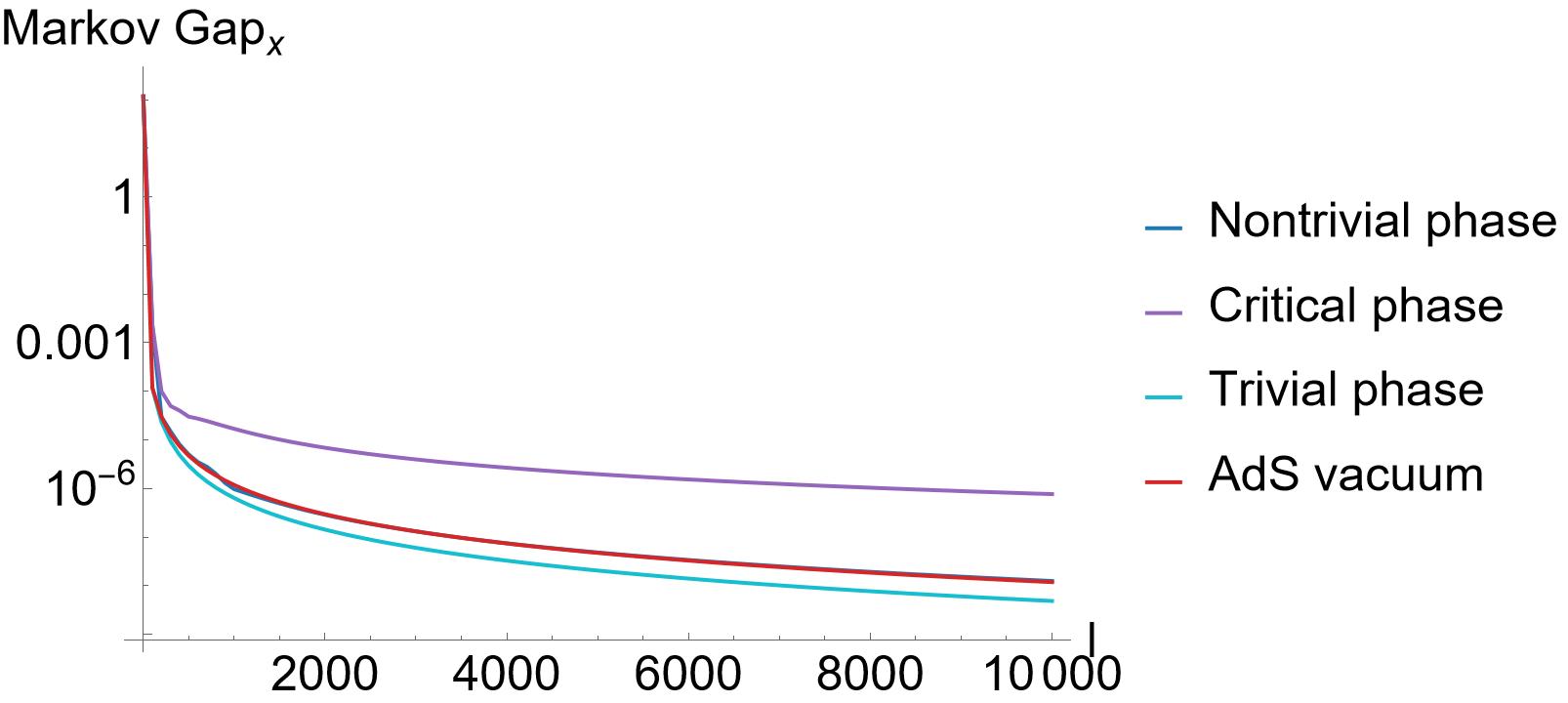}
        \small (a) \(\Delta(M/b)=0.02\): Markov gap\(_x(l)\)
    \end{minipage}
    \hfill
    \begin{minipage}[b]{0.48\textwidth}
        \centering
        \includegraphics[width=\linewidth]{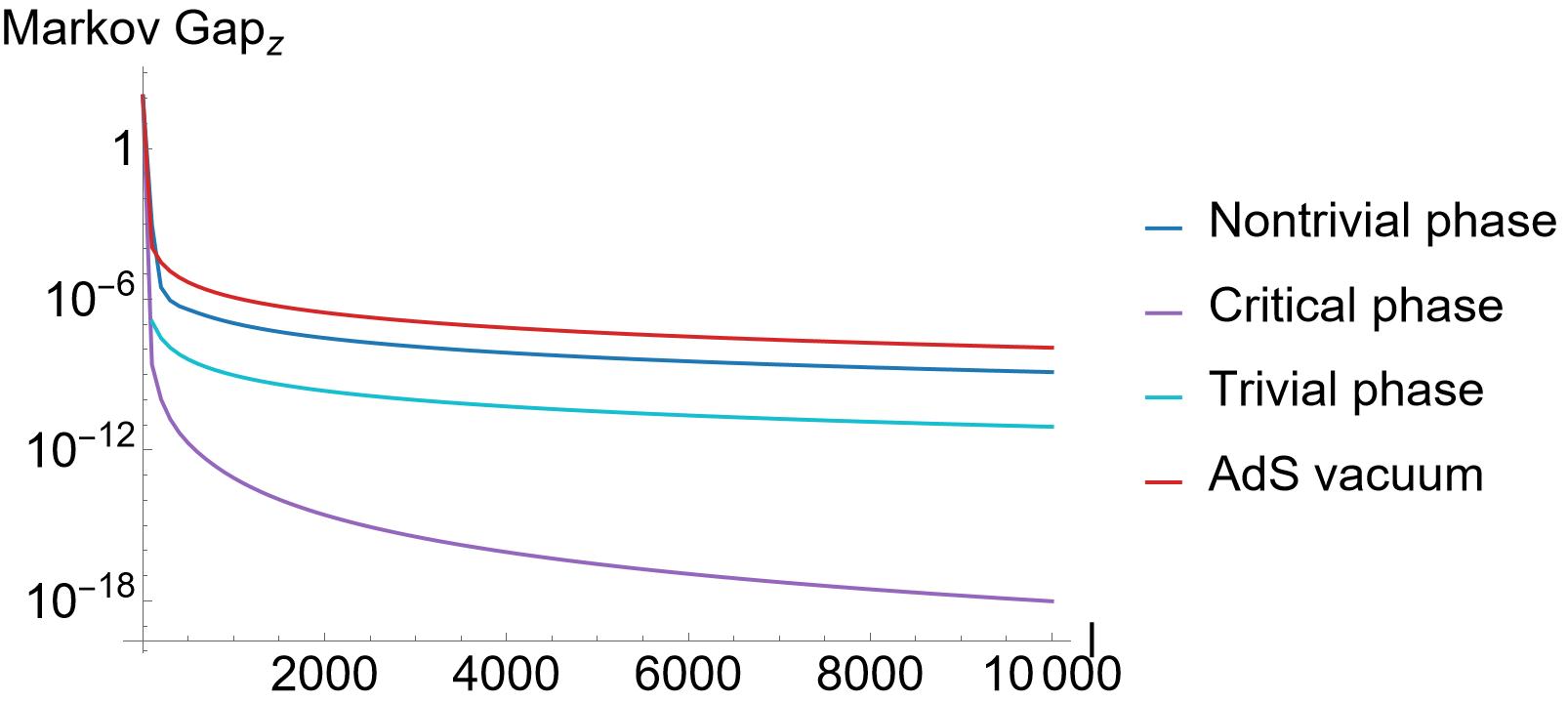}
        \small (b) \(\Delta(M/b)=0.02\): Markov gap\(_z(l)\)
    \end{minipage}

    \vspace{0.5em}

    \begin{minipage}[b]{0.48\textwidth}
        \centering
        \includegraphics[width=\linewidth]{Figures/DeltaMoverb=0.1-l_x-MarkovGap.jpg}
        \small (c) \(\Delta(M/b)=0.1\): Markov gap\(_x(l)\)
    \end{minipage}
    \hfill
    \begin{minipage}[b]{0.48\textwidth}
        \centering
        \includegraphics[width=\linewidth]{Figures/DeltaMoverb=0.1-l_z-MarkovGap.jpg}
        \small (d) \(\Delta(M/b)=0.1\): Markov gap\(_z(l)\)
    \end{minipage}

    \caption[Supplementary Markov gap comparison in the principal directions]{
    Supplementary Markov gap data in the principal directions. The comparison
    confirms that changing \(\Delta(M/b)\) mainly shifts the crossover scale in
    \(l\). It does not change the asymptotic large \(l\) character of the
    corresponding noncritical phase. Without loss of generality, an appropriate ratio of $l_{\text{strip}}$ to $l_{\text{gap}}$ has been chosen to guarantee that the entanglement wedges between different regions remain connected as the scale increases.
    }
    \label{fig:offcritical_markovgap_appendix}
\end{figure}

\section{Angular fit parameters and critical angular profiles}
\label{app:angular_fit_parameters}

This appendix gives the full angular fit data used in Section~\ref{Section5}. The main text lists only the subset of parameters that most directly diagnoses the noncritical phase comparison. Here we give the complete five-parameter fits for the representative noncritical pair, together with the critical angular profiles not shown in the main text.

The holographic \(c\)-function is not included in the fit table because, in the noncritical phases, it approaches an IR plateau rather than a decaying power law. The multi-EWCS based signals \(\Delta\) and \(g\) are also not fitted separately. As explained in subsection~\ref{subsec:raw_angular_response_01}, they are built from the multi-EWCS and lower partition data, so their angular dependence is not an independent large \(l\) datum. We therefore use multi-EWCS as the representative four-partite wedge quantity in the angular
coefficient analysis.

For completeness, we repeat the fitting ansatz used in subsection~\ref{subsec:fitting_protocol_physical_meaning}:
\begin{align}
a_x(\theta)
&=
\alpha^{(0)}_x
+
\alpha^{(1)}_x
\left[
\beta^{(1)}_x
(\cos^2\theta)^{\gamma^{(1)}_x}
+
\left(
1-\beta^{(1)}_x
\right)
(\cos^2\theta)^{\gamma^{(2)}_x}
\right],
\label{eq:appendix_ax_fit}
\\
a_z(\theta)
&=
\alpha^{(0)}_z
+
\alpha^{(1)}_z
\left[
\beta^{(1)}_z
(\sin^2\theta)^{\gamma^{(1)}_z}
+
\left(
1-\beta^{(1)}_z
\right)
(\sin^2\theta)^{\gamma^{(2)}_z}
\right].
\label{eq:appendix_az_fit}
\end{align}
The parameters \(\alpha^{(0)}\) and \(\alpha^{(1)}\) encode the residual coefficient and the total angular variation, while \(\gamma^{(1)}\) and \(\gamma^{(2)}\) characterize the angular sharpness.

The full fit parameters are listed in Table~\ref{tab:full_angular_fit_parameters}. The phase labels \(N\) and \(T\) denote the topologically nontrivial and topologically trivial phases, represented by \(M/b=(M/b)_c\mp0.1\), respectively. For the critical solution we do not impose the noncritical ansatz. The critical geometry has anisotropic leading IR powers, and the fitted exponent \(b_d(\theta)\) is not approximately angle independent over the full angular range. The CMI profile is shown in the main text as a representative case; the remaining critical profiles are collected below.

\begin{table}[p]
\centering
\small
\setlength{\tabcolsep}{4pt}
\resizebox{\textwidth}{!}{
\begin{tabular}{llccccc}
\toprule
measure / channel
& phase
& \(\alpha^{(0)}\)
& \(\alpha^{(1)}\)
& \(\beta^{(1)}\)
& \(\gamma^{(1)}\)
& \(\gamma^{(2)}\) \\
\midrule
CMI, \(x\)
& \(N\)
& \(1.016\)
& \(1.754\)
& \(0.469\)
& \(0.961\)
& \(3.202\) \\
CMI, \(z\)
& \(N\)
& \(1.016\)
& \(1.754\)
& \(0.804\)
& \(0.993\)
& \(1.805\) \\
CMI, \(x\)
& \(T\)
& \(0.231\times10^{-1}\)
& \(1.453\)
& \(0.333\)
& \(4.322\)
& \(25.443\) \\
CMI, \(z\)
& \(T\)
& \(0.231\times10^{-1}\)
& \(1.453\)
& \(0.230\)
& \(0.949\)
& \(1.740\) \\
\midrule
EWCS, \(x\)
& \(N\)
& \(0.169\)
& \(0.293\)
& \(0.645\)
& \(1.195\)
& \(4.115\) \\
EWCS, \(z\)
& \(N\)
& \(0.169\)
& \(0.293\)
& \(0.817\)
& \(0.998\)
& \(1.847\) \\
EWCS, \(x\)
& \(T\)
& \(0.357\times10^{-2}\)
& \(0.213\)
& \(0.379\)
& \(4.674\)
& \(26.956\) \\
EWCS, \(z\)
& \(T\)
& \(0.357\times10^{-2}\)
& \(0.213\)
& \(0.228\)
& \(0.938\)
& \(1.713\) \\
\midrule
\(\kappa\), \(x\)
& \(N\)
& \(0.041\)
& \(0.785\times10^{-1}\)
& \(0.626\)
& \(1.195\)
& \(4.081\) \\
\(\kappa\), \(z\)
& \(N\)
& \(0.041\)
& \(0.785\times10^{-1}\)
& \(0.775\)
& \(0.993\)
& \(1.835\) \\
\(\kappa\), \(x\)
& \(T\)
& \(0.557\times10^{-4}\)
& \(0.220\times10^{-2}\)
& \(0.197\)
& \(3.357\)
& \(27.110\) \\
\(\kappa\), \(z\)
& \(T\)
& \(0.557\times10^{-4}\)
& \(0.220\times10^{-2}\)
& \(0.726\)
& \(0.959\)
& \(1.803\) \\
\midrule
Markov gap, \(x\)
& \(N\)
& \(0.168\)
& \(0.315\)
& \(0.568\)
& \(1.117\)
& \(3.689\) \\
Markov gap, \(z\)
& \(N\)
& \(0.168\)
& \(0.315\)
& \(0.780\)
& \(0.992\)
& \(1.826\) \\
Markov gap, \(x\)
& \(T\)
& \(0.167\times10^{-3}\)
& \(0.637\times10^{-2}\)
& \(0.100\)
& \(1.800\)
& \(20.000\) \\
Markov gap, \(z\)
& \(T\)
& \(0.167\times10^{-3}\)
& \(0.637\times10^{-2}\)
& \(0.766\)
& \(0.964\)
& \(1.846\) \\
\midrule
multi-EWCS, \(x\)
& \(N\)
& \(113.861\)
& \(197.234\)
& \(0.636\)
& \(1.187\)
& \(4.045\) \\
multi-EWCS, \(z\)
& \(N\)
& \(113.861\)
& \(197.234\)
& \(0.806\)
& \(0.994\)
& \(1.828\) \\
multi-EWCS, \(x\)
& \(T\)
& \(1.792\)
& \(77.583\)
& \(0.261\)
& \(3.001\)
& \(19.827\) \\
multi-EWCS, \(z\)
& \(T\)
& \(1.792\)
& \(77.583\)
& \(0.210\)
& \(0.877\)
& \(1.550\) \\
\bottomrule
\end{tabular}}
\caption[Full noncritical angular fit parameters]{
Full noncritical angular fit parameters for CMI, EWCS, \(\kappa\), the Markov
gap, and multi-EWCS. The labels \(N\) and \(T\) denote the topologically
nontrivial and topologically trivial phases. The \(x\) channel is fitted with
powers of \(\cos^2\theta\), and the \(z\) channel with powers of
\(\sin^2\theta\). The diagnostic subset of this table is shown in
Table~\ref{tab:angular_fit_diagnostics}.
}
\label{tab:full_angular_fit_parameters}
\end{table}

\begin{figure}[p]
    \centering
    \includegraphics[width=0.48\textwidth]{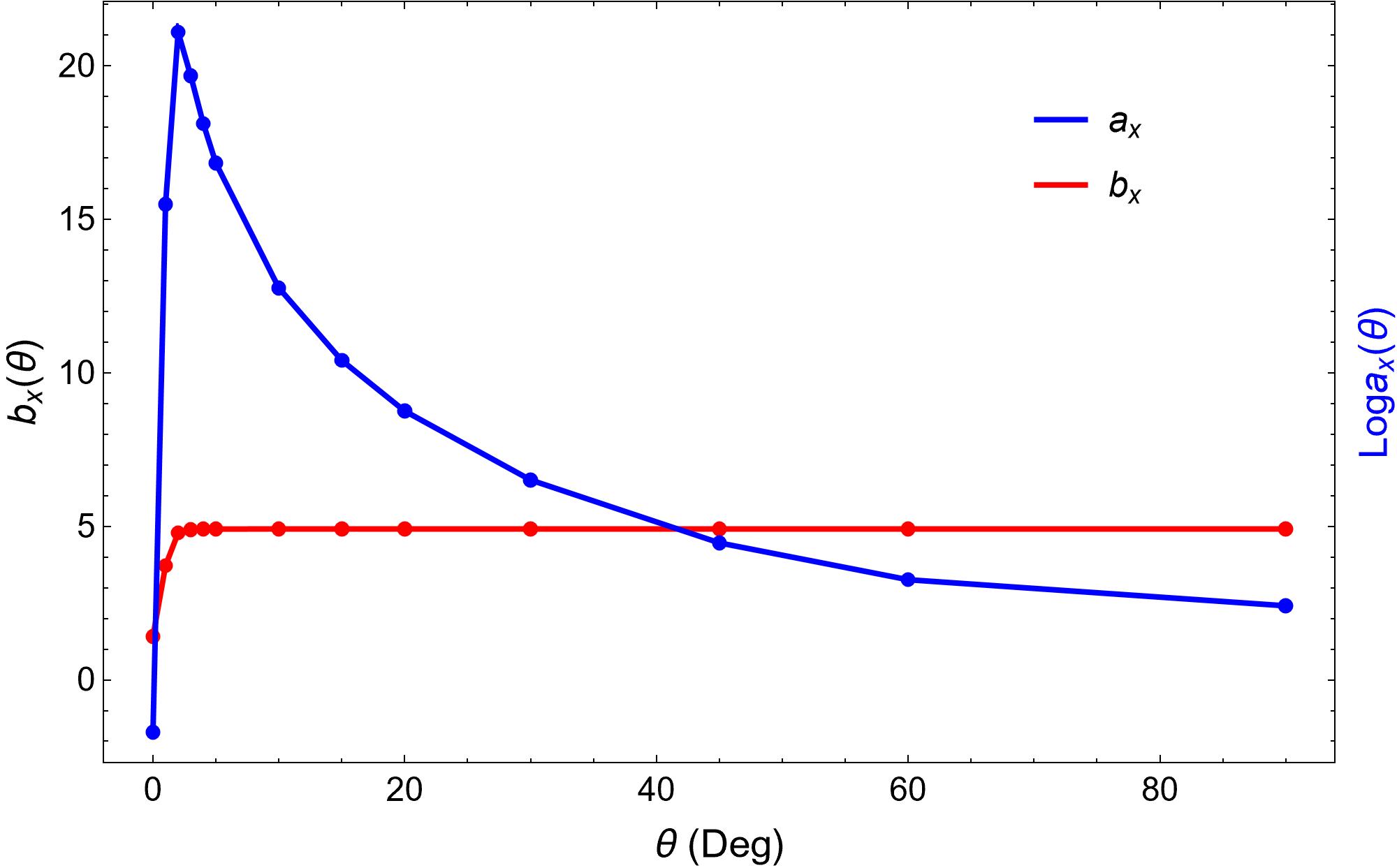}
    \hfill
    \includegraphics[width=0.48\textwidth]{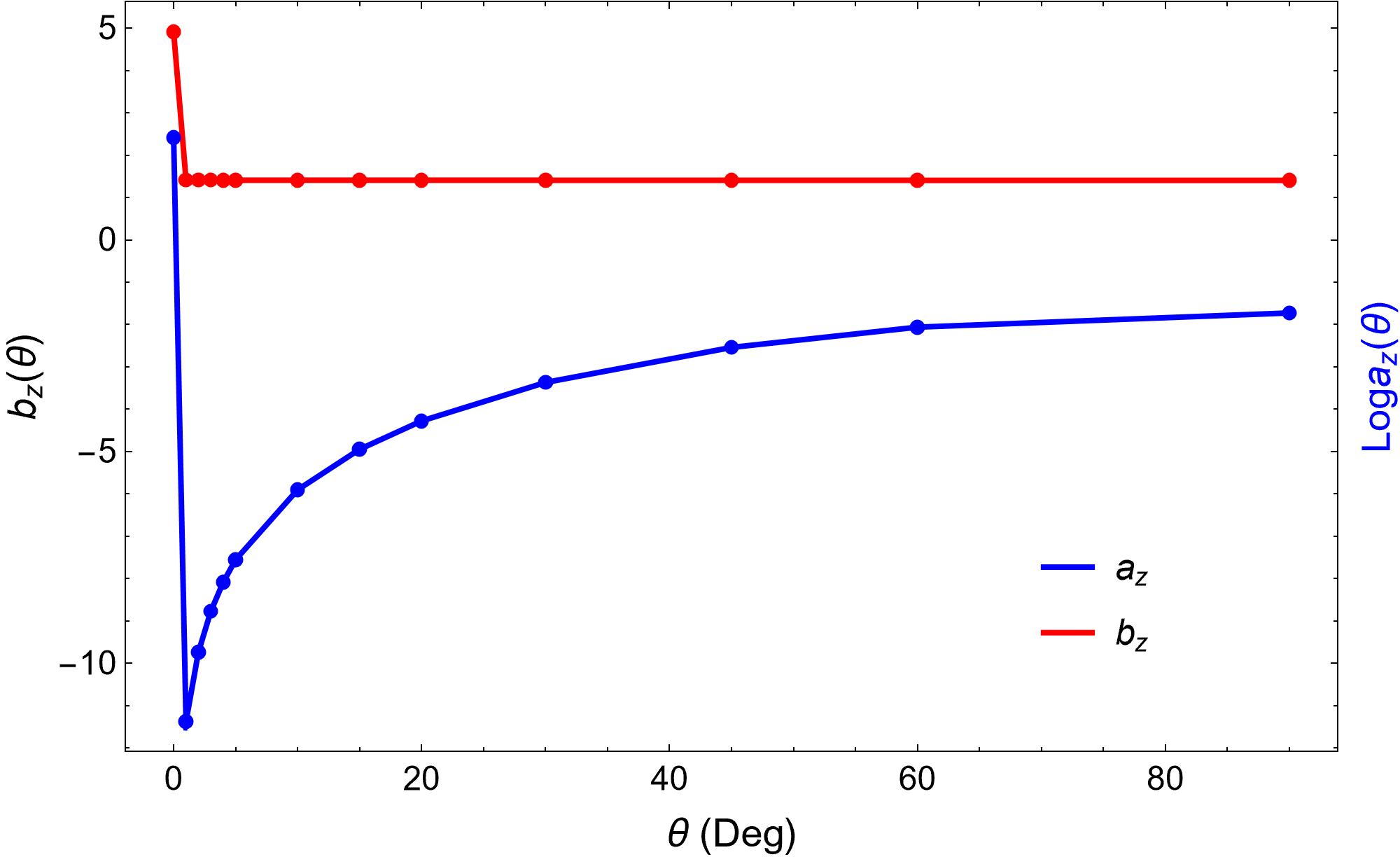}
    \caption{
    Critical angular profile of the EWCS. The plots show
    \(\log a_x(\theta)\) and \(\log a_z(\theta)\), together with the fitted
    exponents \(b_x(\theta)\) and \(b_z(\theta)\).
    }
    \label{fig:ewcs_critical_coefficients}
\end{figure}

\begin{figure}[p]
    \centering
    \includegraphics[width=0.48\textwidth]{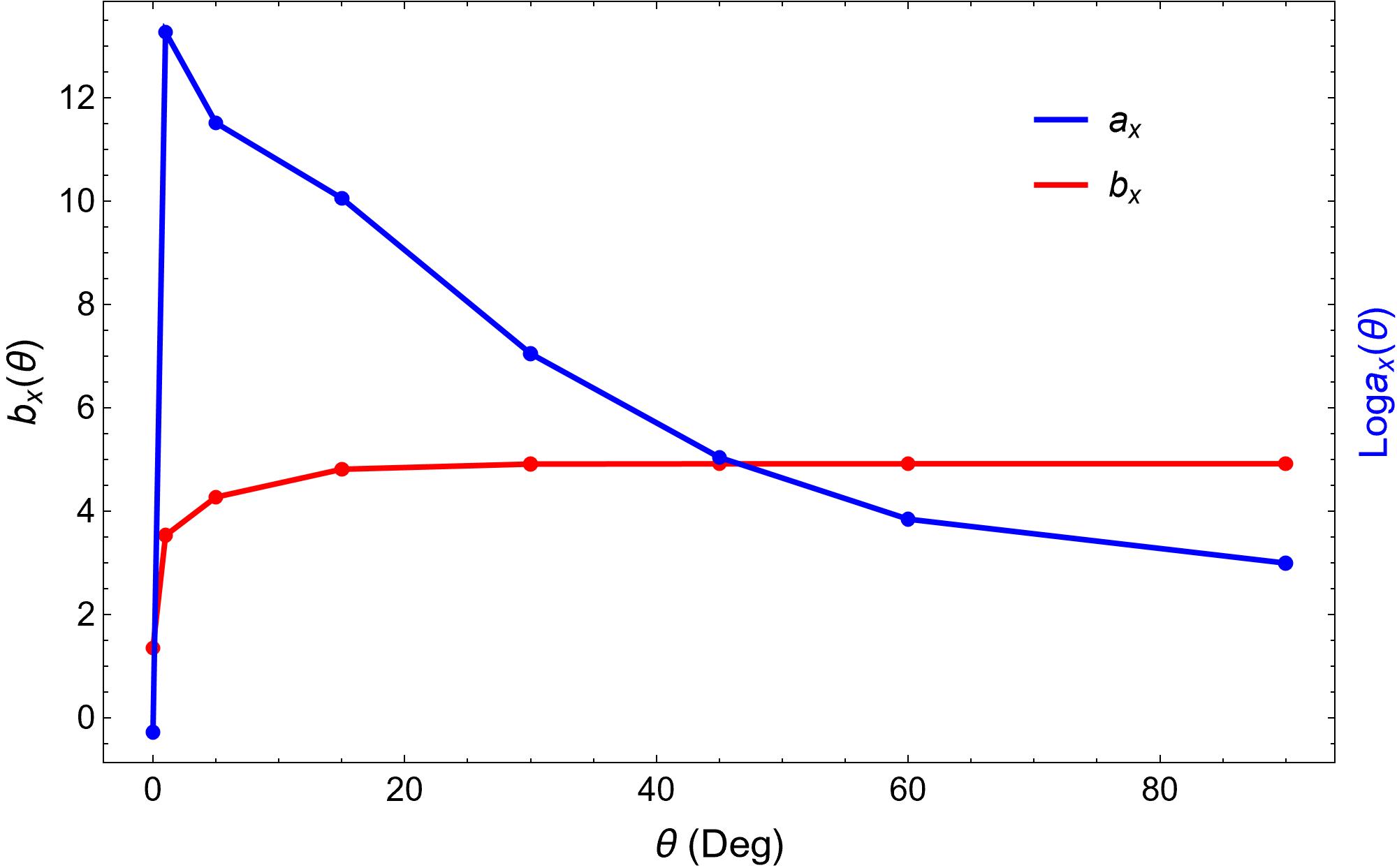}
    \hfill
    \includegraphics[width=0.48\textwidth]{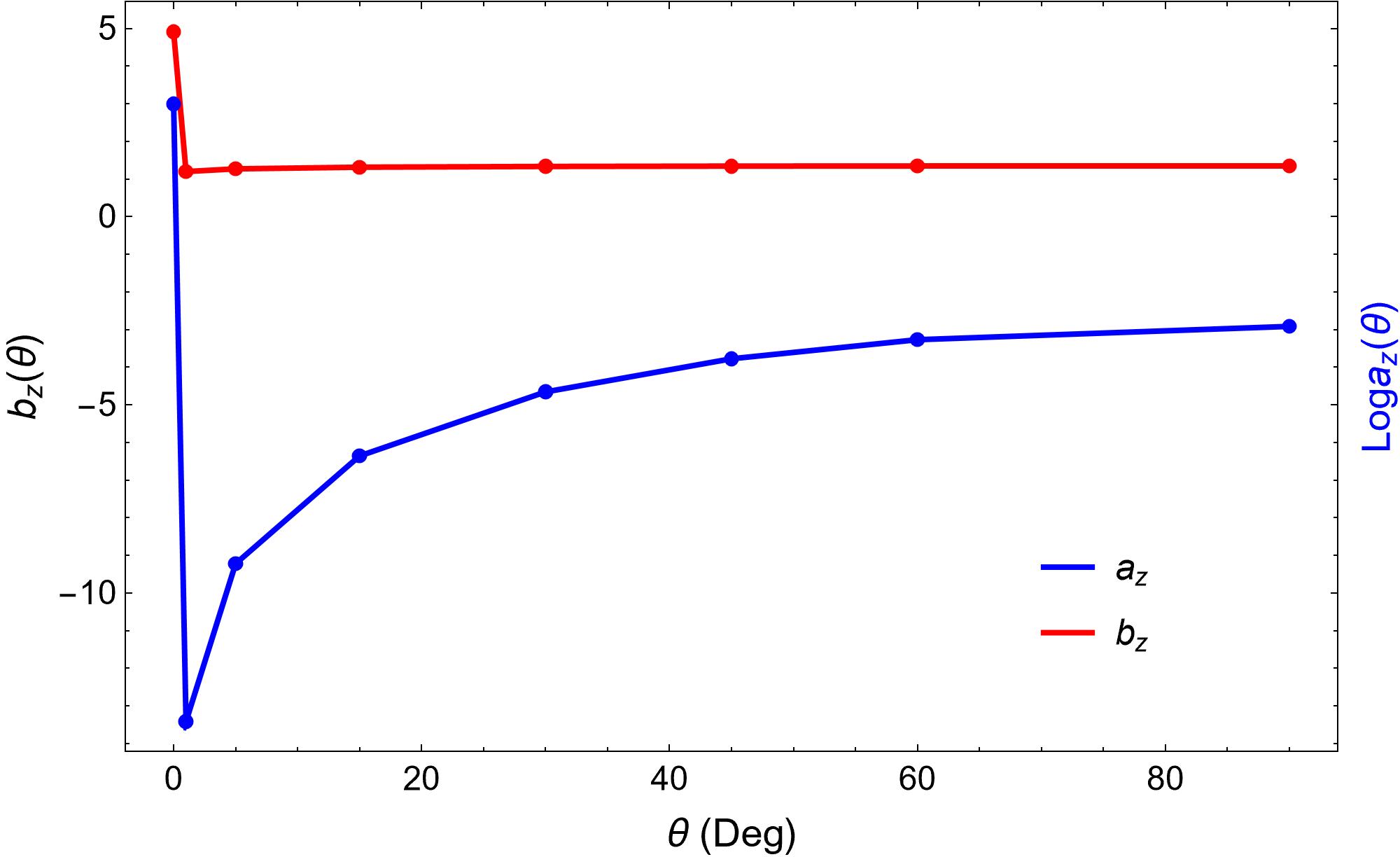}
    \caption{
    Critical angular profile of \(\kappa\). The plots show
    \(\log a_x(\theta)\) and \(\log a_z(\theta)\), together with the fitted
    exponents \(b_x(\theta)\) and \(b_z(\theta)\).
    }
    \label{fig:kappa_critical_coefficients}
\end{figure}

\begin{figure}[p]
    \centering
    \includegraphics[width=0.48\textwidth]{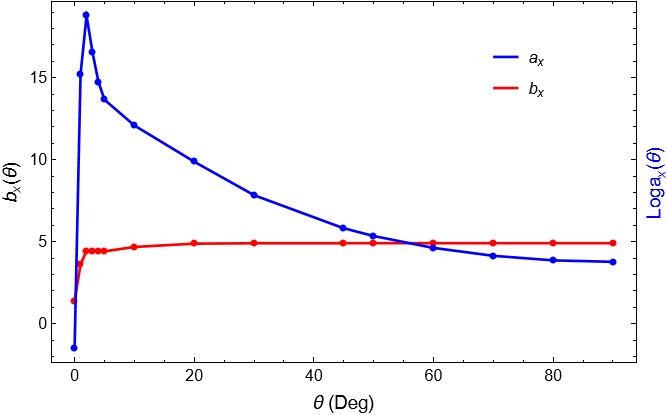}
    \hfill
    \includegraphics[width=0.48\textwidth]{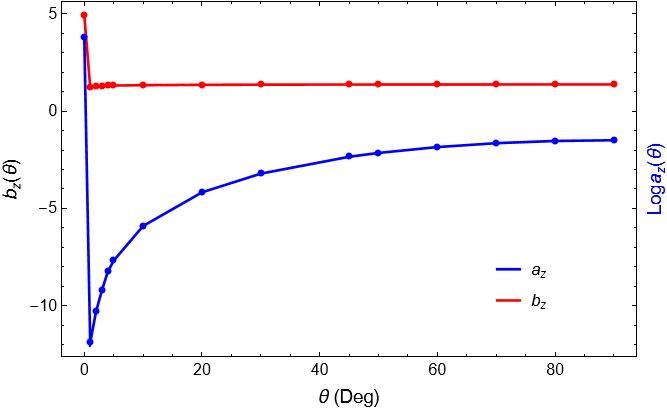}
    \caption[Critical Markov gap angular profile]{
    Critical angular profile of the Markov gap. The plots show
    \(\log a_x(\theta)\) and \(\log a_z(\theta)\), together with the fitted
    exponents \(b_x(\theta)\) and \(b_z(\theta)\).
    }
    \label{fig:markovgap_critical_coefficients}
\end{figure}

\begin{figure}[p]
    \centering
    \includegraphics[width=0.48\textwidth]{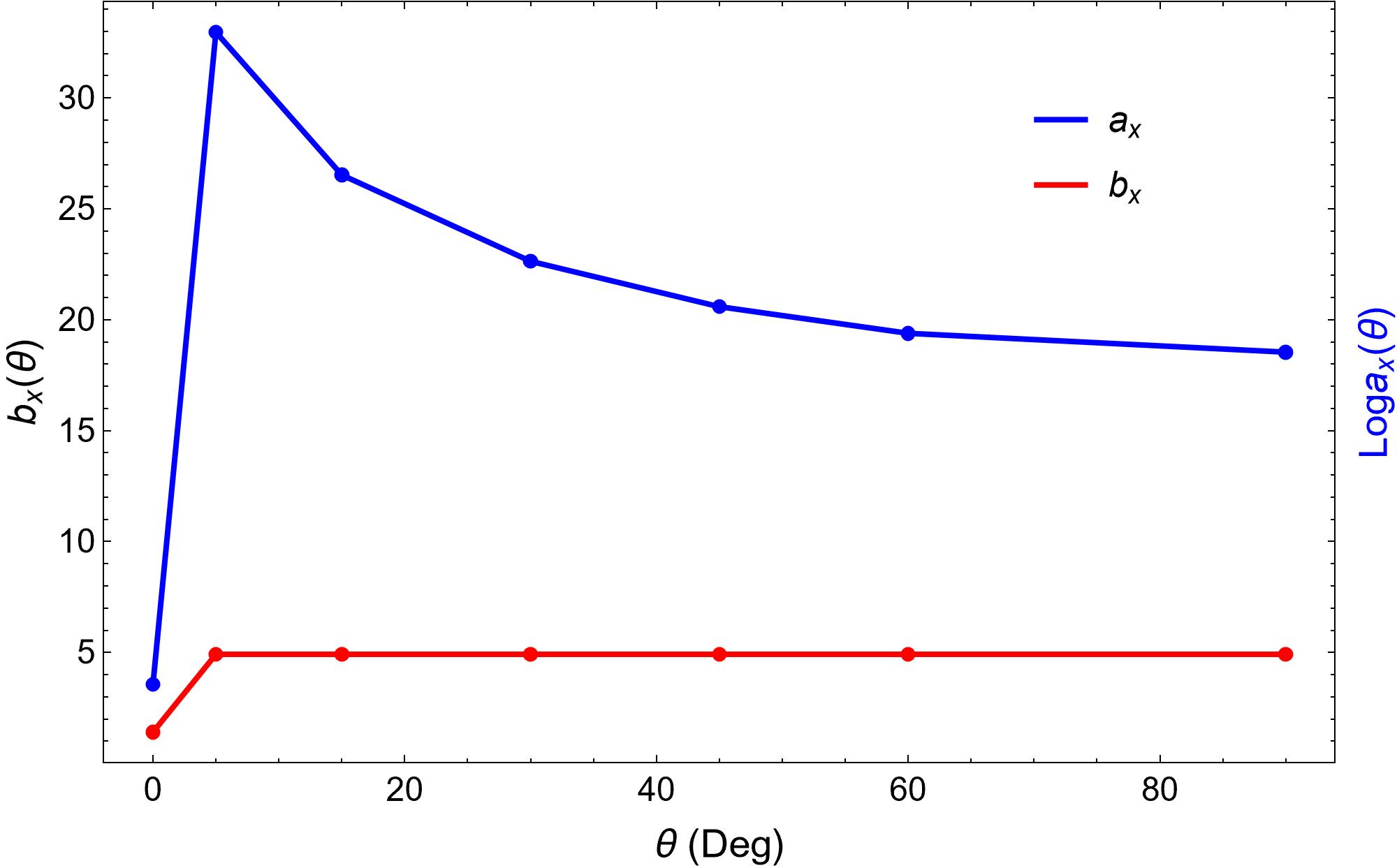}
    \hfill
    \includegraphics[width=0.48\textwidth]{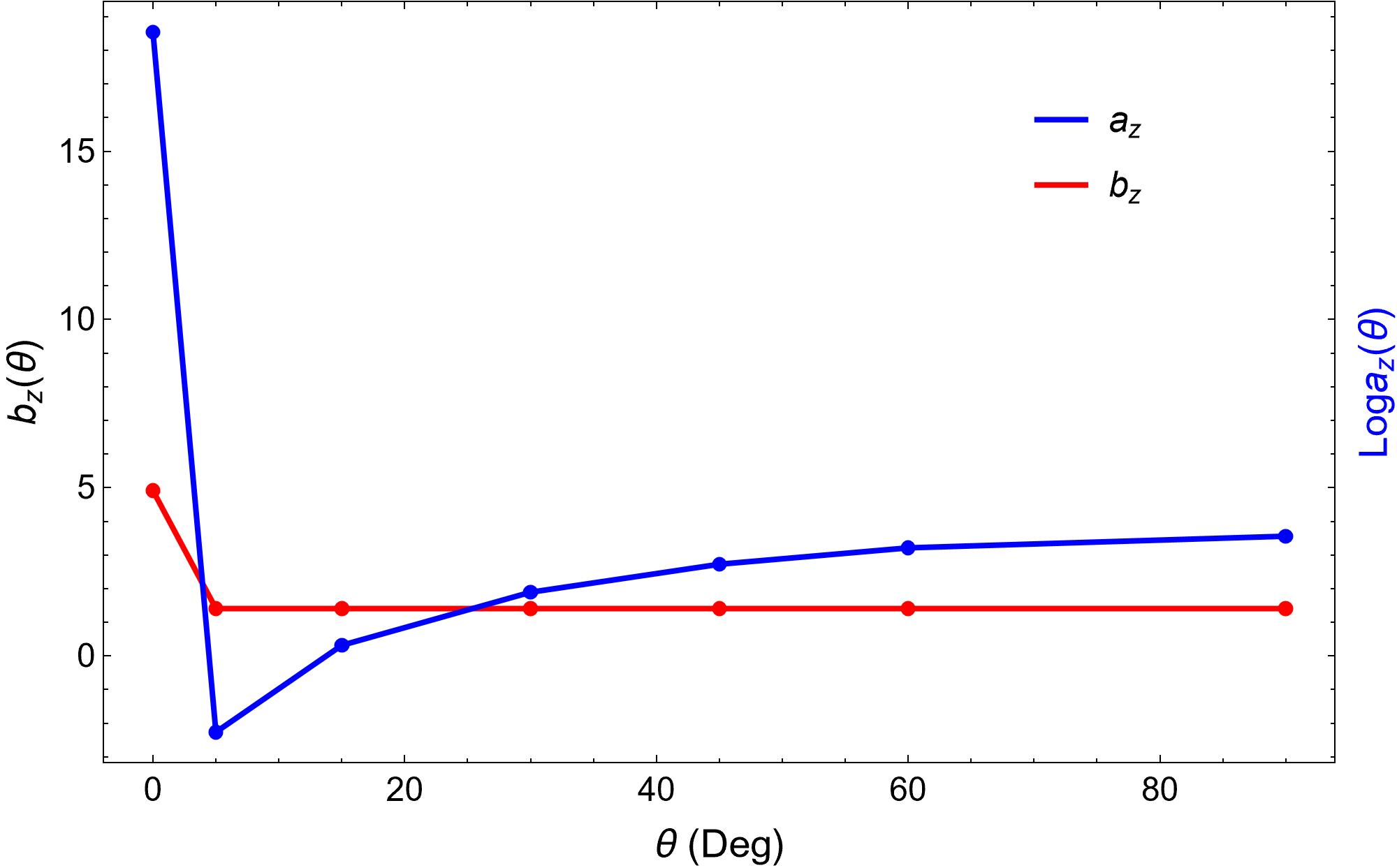}
    \caption{
    Critical angular profile of the multi-EWCS. The coefficients vary
    continuously between the transverse and longitudinal limits, but the
    fitted exponents are not angle independent near the small angle regime.
    }
    \label{fig:multiewcs_critical_coefficients}
\end{figure}

\bibliographystyle{elsarticle-num}
\bibliography{biblio}

\end{document}